\documentclass[acmsmall]{acmart}
\AtBeginDocument{%
  }

\setcopyright{acmlicensed}
\copyrightyear{2018}
\acmYear{2018}
\acmDOI{XXXXXXX.XXXXXXX}

\acmJournal{JACM}
\acmVolume{37}
\acmNumber{4}
\acmArticle{111}
\acmMonth{8}

\usepackage{hyperref}
\usepackage{booktabs}
\usepackage{multirow}
\usepackage{makecell}
\usepackage{longtable}
\usepackage{supertabular}
\usepackage{subfigure}
\usepackage{algorithmic}
\usepackage[linesnumbered, ruled]{algorithm2e}
\usepackage[table]{xcolor}

\usepackage{colortbl}

\begin{document}

\title{Learning to Hash for Recommendation: A Survey}

\author{Fangyuan Luo}
\affiliation{%
  \institution{the College of Computer Science, Beijing University of Technology}
  \city{Beijing}
  \country{China}
}
\email{luofangyuan@bjut.edu.cn}

\author{Yankai Chen}
\authornote{Corresponding authors.}
\affiliation{%
  \institution{University of Illinois Chicago}
  \city{Chicago}
  \country{United States}
}
\affiliation{%
  \institution{MBZUAI, McGill University}
  \city{Montreal}
  \country{Canada}
}
\email{yankaichen@acm.org}

\author{Jun Wu}
\authornotemark[1]
\affiliation{%
  \institution{School of Computer Science and Technology, Beijing Jiaotong University}
  \city{Beijing}
  \country{China}
}
\email{wuj@bjtu.edu.cn}

\author{Tong Li}
\affiliation{%
  \institution{the College of Computer Science, Beijing University of Technology}
  \city{Beijing}
  \country{China}
}
\email{litong@bjut.edu.cn}

\author{Philip S. Yu}
\affiliation{%
  \institution{the Department of Computer Science, University of Illinois Chicago}
  \city{Chicago}
  \country{United States}
}
\email{psyu@uic.edu}

\author{Xue Liu}
\affiliation{%
  \institution{the Department of Mathematics and Statistics, MBZUAI, McGill University}
  \city{Montreal}
  \country{Canada}
}
\email{Steve.Liu@mbzuai.ac.ae}








\renewcommand{\shortauthors}{F. Luo et al.}

\begin{abstract}
  With the explosive growth of users and items, Recommender Systems are facing unprecedented challenges in terms of retrieval efficiency and storage overhead. Learning to Hash techniques have emerged as a promising solution to these issues by encoding high-dimensional data into compact hash codes. As a result, hashing-based recommendation methods (HashRec) have garnered growing attention for enabling large-scale and efficient recommendation services. This survey provides a comprehensive overview of state-of-the-art HashRec algorithms. Specifically, we begin by introducing the common two-tower architecture used in the recall stage and by detailing two predominant hash search strategies. Then, we categorize existing works into a three-tier taxonomy based on: (i) learning objectives, (ii) optimization strategies, and (iii) recommendation scenarios. Additionally, we summarize widely adopted evaluation metrics for assessing both the effectiveness and efficiency of HashRec algorithms. Finally, we discuss current limitations in the field and outline promising directions for future research. We index these HashRec methods at the repository \href{https://github.com/Luo-Fangyuan/HashRec}{https://github.com/Luo-Fangyuan/HashRec}. 
\end{abstract}

\ccsdesc[500]{Information systems~Recommender systems}

\keywords{Recommender Systems, Learning to Hash, Survey}


\maketitle

\section{Introduction}
\begin{figure}[!t]
    \centering
    \includegraphics[width=0.65\textwidth]{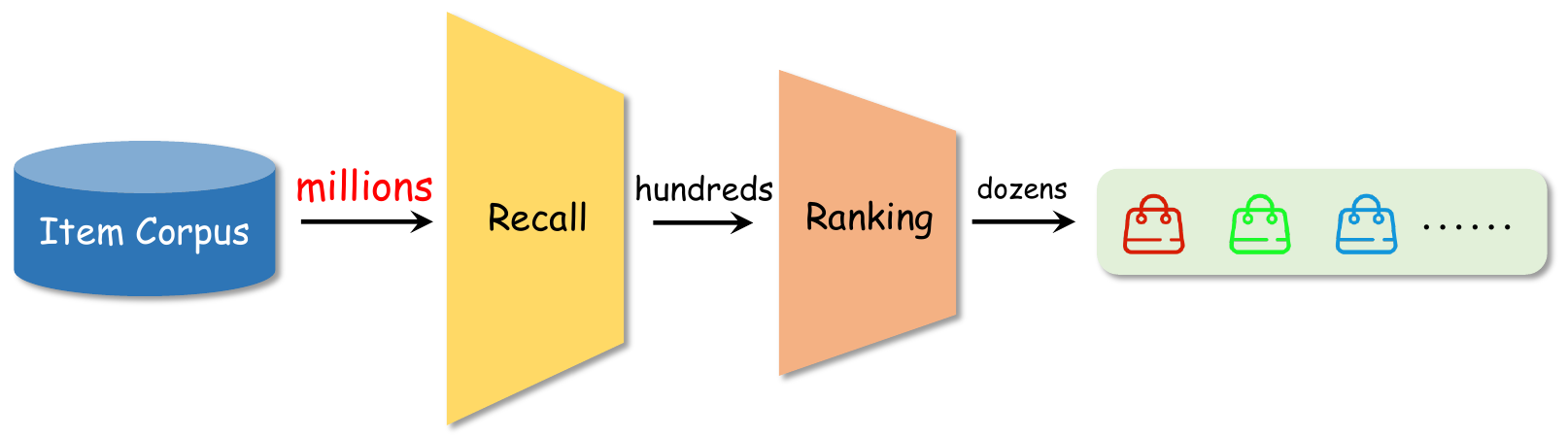}
    \caption{The architecture of industry recommender systems, which includes recall and ranking phase. Specifically, the recall model retrieves hundreds of candidate items from millions of item corpus, and the ranking model presents a list which contains dozens of items sorted in descending order of preference scores. It is obvious that the recall model suffers from serious efficiency issue.}
    \label{fig:RS_architeure}
\end{figure}

Recommender Systems (RS) \cite{Zhang19RS_Survey,Alhijawi20RS_Survey} play an important role in helping users find relevant and personalized items, which can significantly alleviate the information overload issue.
The core idea is to discover a short list of items that are likely to be interacted by users, which is widely applied in modern Internet services, e.g., E-commerce platforms (Taobao \cite{Gong20Taobao}, Amazon \cite{Linden03Amazon}), video-sharing platforms (Kuaishou \cite{Chang23Kuaishou}, TikTok \cite{Zannettou24Tiktok}), social media (Weibo \cite{Alsini21Weibo}, RED \cite{liu24RED}), and so on. With the rapid growth of users and items\footnote{Take the Wechat platform as an example, there are 1385 million monthly active users at the end of 2024. Data source from \href{https://static.www.tencent.com/uploads/2025/03/19/5894f24eb4ade2dea94826d62bd7b11b.pdf}{link}.}, it is challenging to rapidly identify a small number of items from a large item corpus due to stringent response requirements of Internet services. Therefore, the ``recall-and-ranking'' recommendation architecture has garnered widespread adoption \cite{DNN2016YouTube}, which is shown in Fig. \ref{fig:RS_architeure}. Specifically, the recall model first takes users' history as input, and retrieves hundreds of candidate items that are broadly relevant to the user with high precision. Subsequently, the ranking model leverages rich user and item features to compute a preference score for each candidate item. This process produces a finely-tailored list of several dozen recommendations by their preference scores.
In contrast, the ranking model only need to score hundreds of candidate items, while the recall model must evaluate and score every item in the entire item corpus. Given the sheer volume of users and items during the recall stage, efficiency has become an urgent problem that needs to be solved immediately \cite{Paun20RS_Efficiency}.


\begin{figure}[!h]
\centering
\subfigure{\label{fig:sign}
    \includegraphics[width=0.25\textwidth]{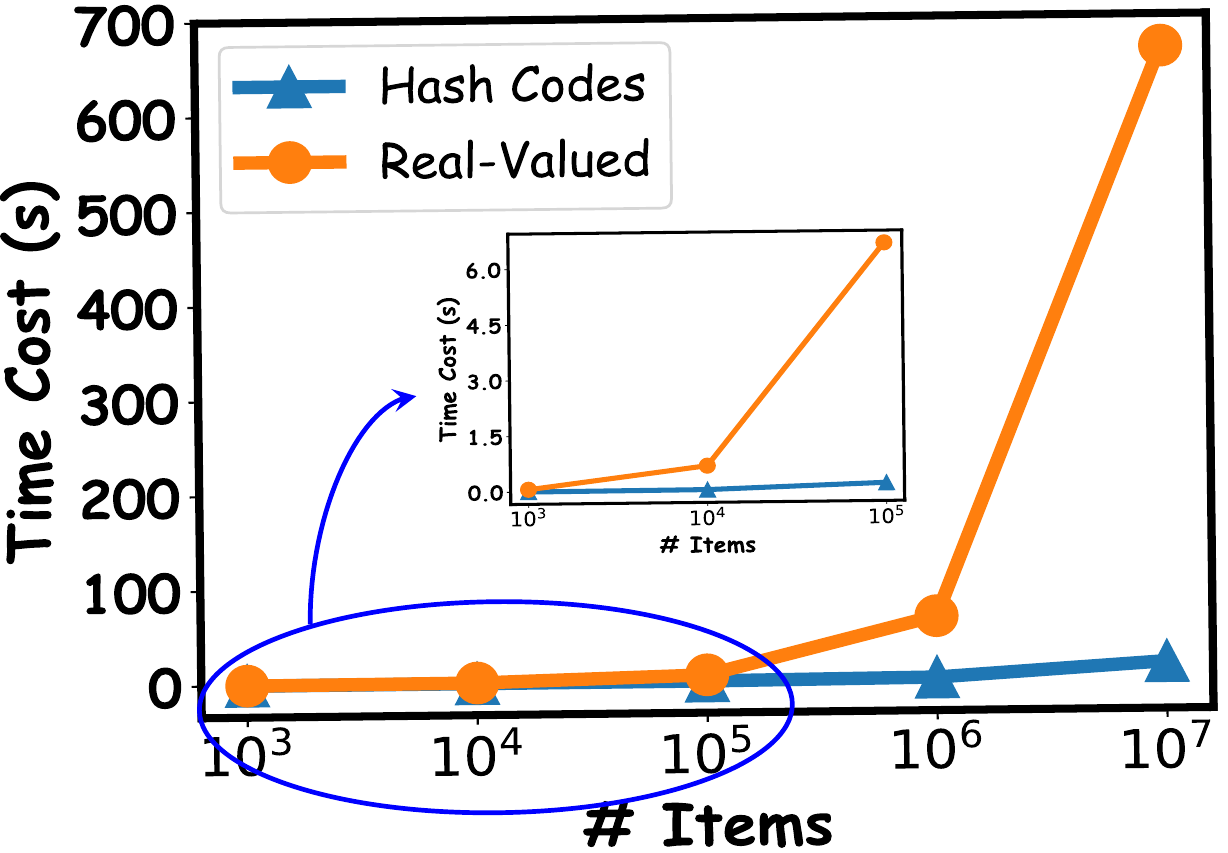}}\hspace{10mm}
\subfigure{\label{fig:tanh}
    \includegraphics[width=0.25\textwidth]{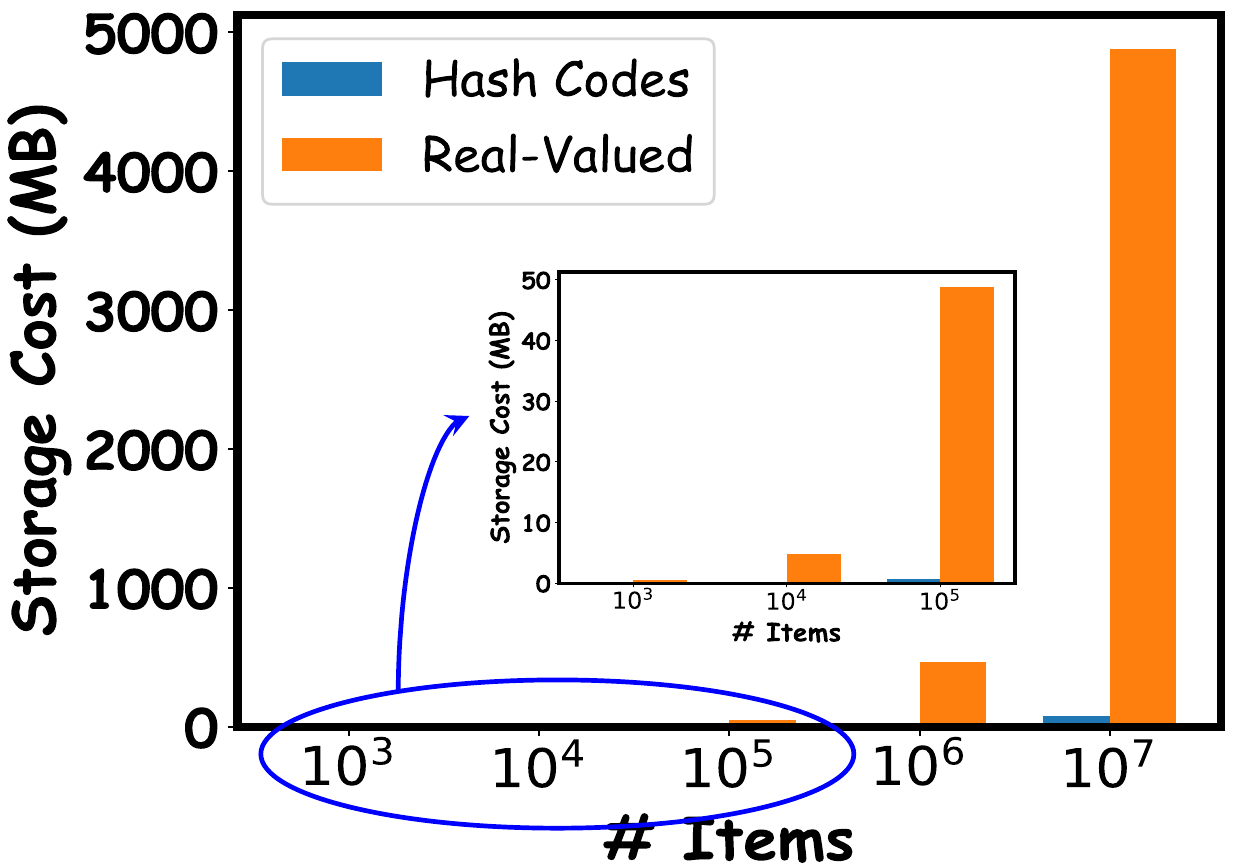}}
\caption{Comparisons on inference time (the left panel) and storage cost (the right panel) between bit operations (hash codes) and floating-point calculation (real-valued) varying the number of items. The experiments are conducted by randomly generating hash codes and real-valued vectors with a length of 64 on $10^3$ users and $10^3$-$10^7$ items, and we report the average results over 10 repetitions.}
\label{fig:time_storage_comparison}
\end{figure}

To address this issue, the two-tower recall model is widely adopted to balance accuracy and efficiency \cite{DNN2016YouTube}. It consists of a user tower and an item tower. Specifically, the user tower is responsible for producing the real-valued representations (a.k.a. embedding) of users, while the item tower is tasked with generating the real-valued representations of items. In the offline phase, the two-tower recall model generates real-valued representations for both users and items by leveraging historical interactions, content features, and side information. After training, the real-valued representations are stored in the database. In the online phase, the target user's representation can be directly picked up according to his/her unique ID when he/she logs into the platform. And then, the similarity between the target user and items can be calculated by inner product or cosine similarity of their embeddings. In this sense, the recall can be regarded as a similarity search problem, i.e., finding items similar to the target user. Although the inference cost can be reduced by precomputing the representations, the computational complexity remains high. Given $m$ users and $n$ items, the computational complexity for generating $K$ preferred items for all users is $\mathcal{O}(mnf+mnlogK)$ \cite{Zhang2016DCF,Lian17DCMF,Zhang18DRMF,Tan20HashGNN}, where $f$ is the dimension of the real-valued representations. Therefore, these methods are still computationally expensive and lead to low-efficiency issues when either $m$ or $n$ is large. Intuitive results can be observed from Fig. \ref{fig:time_storage_comparison}. As the number of items increases, the storage cost and inference time increase dramatically, based on their real-valued representations.

One promising solution for improving efficiency is Learning to Hash (L2H), whose core idea is mapping data points in the vector space into Hamming space, and then the real-valued representations can be replaced by compact hash codes (a.k.a binary codes). L2H methods have been intensively studied in many fields, such as Computer Vision (CV), telecommunication, computer graphics, etc., for several decades \cite{Knuth98L2H}. Existing L2H methods can be categorized into two types: data-independent and data-dependent. As a representative of data-independent method, Locality-Sensitive Hashing (LSH) has attracted widespread attention \cite{Gionis99LSH}. The key component of LSH is a hash function, which generate identical code for nearby data points in the original vector space with high probability. Therefore, a lot of researchers have devoted efforts to designing random hash functions that satisfy the locality sensitivity property for various distance measures \cite{Broder97LSH1,Broder97LSH2,Charikar02LSH3,Datar04LSH4,Motwani07LSH5,ODonnell14LSH6,Dasgupta11FLSH}. Other researchers develop better search schemes \cite{Gan12LSH7,Lv07LSH8}, or provide a similarity estimator with small variance \cite{Li06LSH9,Ji12LSH10,Li12LSH11,Ji13LSH12}, or design faster computation of hash functions \cite{Li06LSH13,Shrivastava14LSH14}. Although LSH can provide asymptotic theoretical properties leading to performance guarantees, they are suffering from long encoding problem. To this end, some researchers suggest leveraging machine learning techniques to produce more effective hash codes \cite{Cayton07DD1}. Their goal is to learn data-dependent and task-specific hash functions that generate compact hash codes for improved search performance \cite{He11DD2}. To achieve this, various approaches have been proposed, including kernel-based methods \cite{Kulis12KLSH,Liu12DD3}, supervised learning methods \cite{Liu12DD3,Kulis09DD4}, semi-supervised learning methods \cite{Wang12SemiHash}. Furthermore, with the rapid advancement of deep learning \cite{Matiur18Deep_Survey,Hu21Deep_Survey}, deep L2H methods for generating hash codes have become highly popular and demonstrate superior performance compared to traditional methods \cite{Singh22DeepHash}.

Inspired by L2H techniques, which have demonstrated a remarkable ability to efficiently handle large-scale data, a surge of research efforts has been devoted to applying L2H in RS (\textbf{HashRec}). HashRec methods aim to improve recall efficiency and reduce memory overhead by mapping users and items into Hamming space, and then the user-item preference can be efficiently calculated in a Hamming space by bit operation rather than in a vector space by floating-point calculation. In this way, HashRec methods not only reduce the computational overhead but also can be scalable and suitable for practical application. However, there lacks a comprehensive survey that systematically categorizes and presents the existing studies. To fill this gap, we conduct an extensive survey to categorize HashRec methods from the perspectives of learning objectives, optimization strategies and recommendation scenarios. Furthermore, we examine the strengths and limitations of each category, highlighting the unique challenges associated with them. Additionally, we provide a detailed analysis of the current trends and future directions in the field of HashRec, identifying key areas for further exploration and innovation. 

\subsection{Contribution of This Paper}
Although there have been several reviews \cite{Wang16L2H,Wang18L2H,Luo23DeepHash,Singh22DeepHash} on L2H, they primarily concentrate on methods within CV and do not cover HashRec approaches. Due to the uniqueness of the recommendation, directly applying L2H techniques from other domains remains challenging. Specifically, CV models typically process dense input data and treat individual instances as independent, with the primary goal being content-based similarity search between homogeneous entities. However, the input data of RS are extremely sparse (e.g., one-hot ID and categorical features of users/items), and the main task is personalized ranking based on heterogeneous user-item relationships. This key distinction necessitates different modeling approaches: CV models focus on preserving instance-level feature similarity, whereas recommendation models need to capture interactive preferences. Consequently, the unique challenges also differ. CV concerns itself with fine-grained similarity preservation, while recommendation systems aim to address unique issues like data sparsity, cold-start scenarios, and the inherent heterogeneity between users and items, making direct transfer of L2H techniques from CV to recommendation largely ineffective. Given these unique characteristics of RS, a comprehensive survey dedicated to HashRec is essential to systematically review and analyze relevant methodologies.



We summarize the contribution of this survey as follows.
\begin{itemize}
\item To the best of our knowledge, we take the pioneering step in providing a systematic survey of HashRec methods, a promising yet underexplored area. 
\item We propose a novel three-tier taxonomy of HashRec from the perspectives of learning objectives, optimization strategies, and recommendation scenarios, and provide a comprehensive discussion of existing HashRec methods, highlighting their strengths and weaknesses.
\item We discuss important yet unresolved problems in this area, and propose promising directions on this topic that can inspire further research in this potential field.
\item We release a GitHub repository to encompass all reviewed papers and multiple datasets, aiming to facilitate the deeper understanding of HashRec models. 
\end{itemize} 

\begin{figure}[!h]
    \centering
    \includegraphics[width=0.5\textwidth]{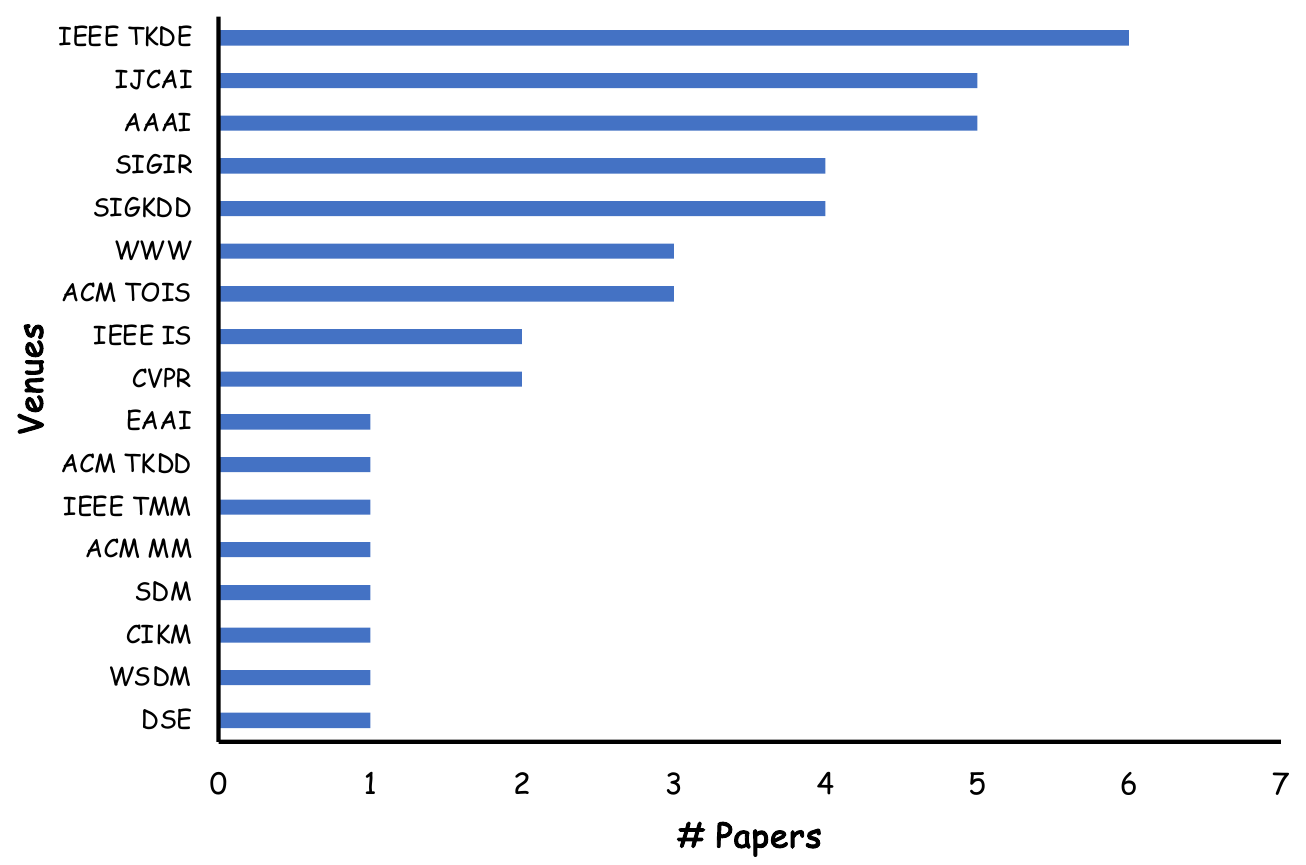}
    \caption{The number of HashRec papers published in relevant journals and conferences. We believe it would help researchers in this field to identify appropriate venues where HashRec papers are published.}
    \label{fig:paper_collection}
\end{figure}

\subsection{Paper Collection} 
We adopt DBLP\footnote{https://dblp.org} and Google Scholar\footnote{https://scholar.google.com/} as the main search engine to collect the related papers. Then, we search the most popular related conferences and journals, such as SIGKDD, SIGIR, AAAI, IJCAI, CIKM, SDM, WWW, IEEE TKDE, etc, to collect recent researches. Fig. \ref{fig:paper_collection} illustrates the statistics of collected papers. Specifically, these research articles were searched for by using a combination of major keywords, such as ``hash + recommend'', ``hash + collaborative filtering'', ``discrete + recommend'', and ``discrete + collaborative filtering''. To avoid omissions of relevant work, we further read the references of each paper that included the relevant search keywords.

\subsection{Organization of This Paper}
The rest of this paper is organized as follows. We explain key notations and definitions. Then in Section \ref{sec:background}, we introduce the preliminary. We then introduce the HashRec methods in Section \ref{sec:HashRec}, followed by a comprehensive evaluation metric introduction in Section \ref{sec:metrics}. Finally, we discuss current challenges and future research directions in Section \ref{sec:discussion}, and conclude the paper in Section \ref{sec:conclusion}. 


\section{Preliminary}\label{sec:background}
In this section, we first give a brief introduction to the two-tower model, which is commonly adopted during the recall stage. Then, we detail search strategies utilized in L2H techniques. In addition, the related notations adopted in this survey are summarized in Table \ref{tab:notations}.

\begin{table}[!t]
\centering
\caption{Main notations used in the paper.}
\label{tab:notations}
\begin{tabular}{ll}
\hline
\textbf{Symbols} & \textbf{Description} \\
\hline
$\mathbf{Y}$       & the user-item rating/preference matrix           \\
$\mathbf{B} \in \{\pm 1\}^{f\times m}$       & hash codes of $m$ users             \\
$\mathbf{D} \in \{\pm 1\}^{f\times n}$       & hash codes of $n$ items         \\
$\mathbf{P}\in \mathbb{R}^{f\times m}$       & real-valued representations of $m$ users             \\
$\mathbf{Q}\in \mathbb{R}^{f\times n}$       & real-valued representations of $n$ items         \\
$\mathcal{U}$, $\mathcal{I}$ & user set, item set  \\
$\mathcal{I}_u^+$ & the positive item set of user $u$ \\
$\mathcal{I}_u^-$ & the negative item set of user $u$ \\
$\Omega$  & the 2–tuple index set of observed entries \\
$\Omega_u$  & the set of observed entries for user $u$ \\
$\Omega_i$  & the set of observed entries for item $i$ \\
$f$ & the length of hash codes/dimension of representations\\
$m$ & the number of users \\
$n$ & the number of items \\
\hline
\end{tabular}
\end{table}


\subsection{Two-Tower Model}
\begin{figure}[!t]
    \centering
    \includegraphics[width=0.45\textwidth]{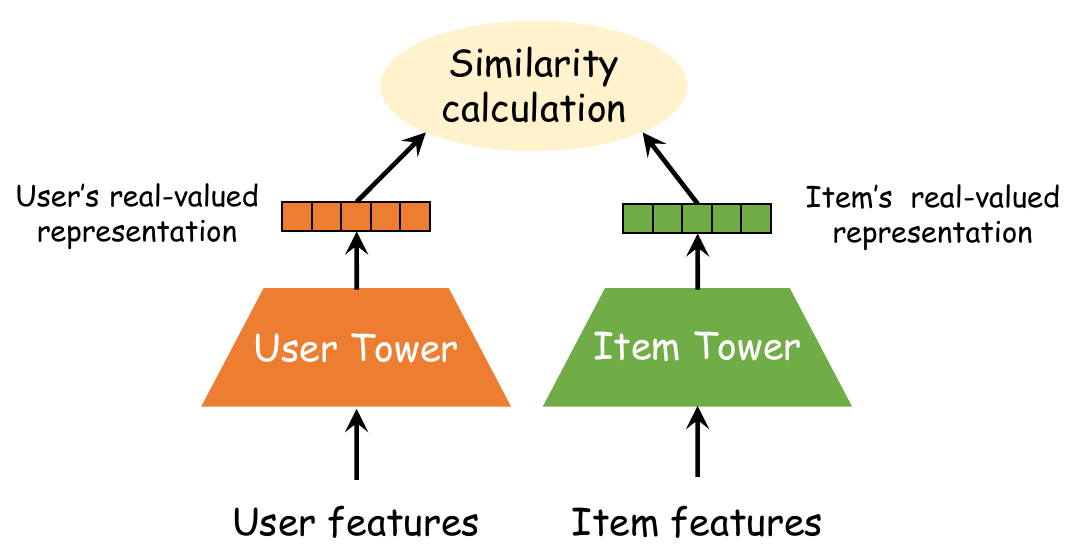}
    \caption{The architecture of two-tower recall model, where the user tower and item tower produce users' real-valued representations and items' real-valued representations respectively.}
    \label{fig:twotower_architeure}
\end{figure}

As shown in Fig. \ref{fig:twotower_architeure}, the two-tower model consists of a user tower and an item tower, where the user tower produces the users' latent representations, and the item tower produces the items' latent representations. Then, the similarity between them can be calculated by inner product of their representations. According to the complexity of model architecture, existing two-tower models can be divided into shallow models, deep models, and Graph Neural Network (GNN)-based models.

\subsubsection{Shallow Models} 
Early-stage research primarily focused on heuristic methods for Collaborative Filtering (CF) \cite{Adomavicius05CF}, which analyze the interdependence between users and items, with the goal of 
identifying unobserved user-item associations. Here, we introduce two classical shallow two-tower models.
\begin{itemize}
\item \textbf{Matrix Factorization (MF)} \cite{Koren09MF}, a classical representative of shallow two-tower model to build RS, has attracted much more attention. Specifically, MF factorizes a partially observed user-item interaction matrix into users' latent preference matrix and items' latent preference matrix, where the interaction matrix can be explicit (e.g., rating values) that directly reflect user preference on the rated item, or be implicit (e.g., click, purchase, share) indicating whether the user has interacted with the item. Subsequently, the user's preference over the item can be estimated by the inner product of their embeddings. 


\item \textbf{SimpleX} \cite{Mao21SimpleX} is built as a unified model that integrates MF and user behavior modeling. Specifically, it comprises a behavior aggregation layer to obtain a user's preference vector from the historically interacted items, and then fuses with the user embedding vector via a weighted sum. For users with a different size of interacted items, either padding or chunking can be applied accordingly. Then, the final user's preference vector can be represented as the weighted sum of the preference representations and behavior representations. Finally, the preference of the user over the item is defined as the cosine similarity.
\end{itemize}

\subsubsection{Deep Models}
With the rapid growth of Deep Neural Networks (DNN), deep two-tower approaches have been extensively studied to improve recall quality at scale. Unlike shallow two-tower models, deep two-tower methods represent each object as a low-dimensional real-valued vector using DNNs. 
\begin{itemize}
    \item \textbf{DSSM} \cite{DSSM13} takes high-dimensional raw text features as input and first processes them through a word hashing layer, which significantly reduces the dimensionality of bag-of-words term vectors. This method is based on letter $n$-grams: each word is augmented with start and end markers and then decomposed into a set of letter $n$-grams. The word is subsequently represented as a vector of these $n$-grams. The resulting hashed features are then passed through multiple layers of non-linear projection. The $tanh(\cdot)$ function is used as the activation function in the output layer. Finally, the semantic relevance between a user and an item is quantified using cosine similarity.
    \item \textbf{YouTubeDNN} \cite{DNN2016YouTube} is a classic two-tower model for video recommendation. The user tower takes features such as the user's watch history sequence (pooled via weighted average or RNN) and demographic features, while the item tower uses embedding features of the video ID. The two towers output user and video embedding vectors, respectively, and the matching score is computed via inner product. A key design feature of this model is the use of negative sampling, which frames the recommendation task as an extreme multi-class classification problem and employs sampled softmax loss for efficient training, making it highly optimized for large-scale industrial scenarios.
\end{itemize}

\subsubsection{GNN-Based Models}
GNNs, as a specialized form of neural networks for structured data, have achieved state-of-the-art performance in RS \cite{Wu23GNN_Survey}. The core idea of GNNs is to iteratively aggregate the feature information from neighbors and integrate the aggregated information with the current central node representation during the propagation process \cite{Wu21GNN_Survey,Zhou20GNN_Survey}. 
\begin{itemize}
\item \textbf{NGCF} \cite{Wang19NGCF} largely follows the standard GCN \cite{Kipf17GCN}, which includes the nonlinear function and feature transformation matrices. It encodes the interaction between users and items into message being passed via their element-wise product, making the message dependent on the affinity between them. After propagating with $L$ layers, the final embedding for users and items are obtained by concatenating representations in different layers. Finally, the user's preference towards the target item is calculated by inner product.




\item \textbf{LightGCN} \cite{He2020LightGCN} is the light-weight version of NGCF. It argues that the feature transformation and nonlinear activation function are not as useful for CF. Therefore, it only contains the neighborhood aggregation and propagates the embeddings on the user-item interaction graph to refine them. 
After $L$ layers, the embeddings obtained at each layer are conducted average pooling to form the final representation of a user or an item.
Likewise, the model prediction is defined as the inner product of user and item final representations.

\end{itemize}

\subsection{Hamming Similarity}
The Hamming similarity calculates the similarity of hash codes between users and items.
Let $\mathbf{B}\in \{-1, 1\}^{f\times m}$ and $\mathbf{D}\in \{-1, 1\}^{f\times n}$ denote the hash codes of users and items respectively\footnote{Sometimes, the hash codes are represented by 0/1, i.e., $\mathbf{B}\in \{0, 1\}^{f\times m}$ and $\mathbf{D}\in \{0, 1\}^{f\times n}$}. The details of how to obtain hash codes will be illustrated in the Section \ref{sec:HashRec}. Then, the similarity calculation can be conducted efficiently by bit operation \cite{Zhang2016DCF}. 
\begin{align}\label{Eq:HammingSimilarity}
\hat{y}_{ui} &= \frac{1}{f}\sum_{k=1}^f \mathbb{I}(b_{uk} = d_{ik})=\frac{1}{2f}(\sum_{k=1}^f\mathbb{I}(b_{uk} = d_{ik}) + f - \sum_{k=1}^f\mathbb{I}(b_{uk} \neq d_{ik}), \nonumber\\
&=\frac{1}{2f}(f + \sum_{k=1}^f b_{uk}d_{ik})=\frac{1}{2} + \frac{1}{2f}\mathbf{b}_u^{\mathsf{T}}\mathbf{d}_i,
\end{align}
where $\mathbf{b}_u \in \{-1, 1\}^f$ and $\mathbf{d}_i \in \{-1, 1\}^f$ denote the hash codes of user $u$ and item $i$, $b_{uk}$ and $d_{ik}$ is the $k$-th bit of $\mathbf{b}_u$ and $\mathbf{d}_i$. To perform similarity search on these computed hash codes, there are two basic strategies. One is hash table lookup, and the other is hash code ranking. 

\begin{figure}[!t]
    \centering
    \includegraphics[width=0.6\textwidth]{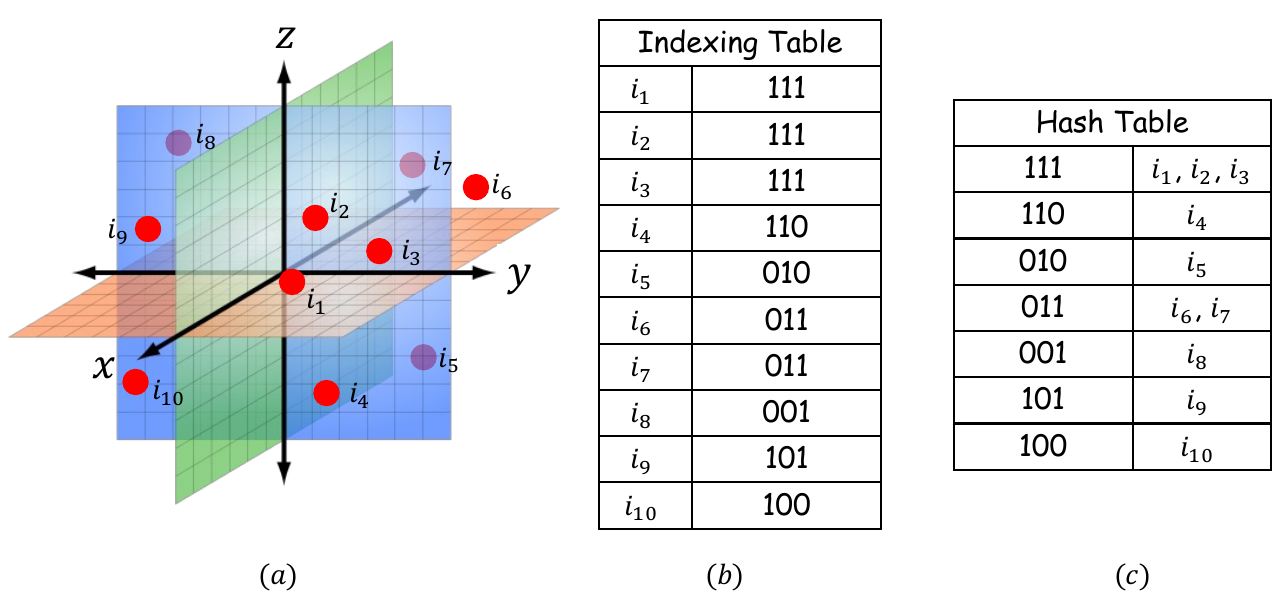}
    \caption{Illustration of Hash Table Lookup, where we suppose that each data point is mapped into 3-dimensional Hamming space.}
    \label{fig:Hash_Table}
\end{figure}

\subsubsection{Hash Table Lookup}
The main idea of hash table lookup for accelerating the search is reducing the number of similarity computations. The search procedure of hash table lookup is illustrated in Fig. \ref{fig:Hash_Table}. Given ten items $\{i_1, i_2, \cdots, i_{10}\}$ and their corresponding hash codes, as show in Fig. \ref{fig:Hash_Table}(a) and Fig. \ref{fig:Hash_Table}(b). According to the indexing table, we can obtain the hash table which is shown in Fig. \ref{fig:Hash_Table}(c). Specifically, the hash table consists of buckets, each indexed by a hash code, and each item is placed into its corresponding bucket according to its hash code. Then, given the target user and its hash code, the items lying in the same bucket as that of the user are retrieved as the candidate items. Normally, this is followed by a reranking step, i.e., reranking the candidate items according to the similarity between the target user and candidates. It is worth noting that the hash table lookup method, such as LSH essentially aims to maximize the probability of collision of similar items while minimizing the probability of collision of the items that are far away. It is the primary difference from conventional hashing algorithms, e.g., hash map in data structures, in computer science which avoid collision (i.e., mapping two items into some same buckets).

\subsubsection{Hash Code Ranking}
Hash code ranking performs an exhaustive search which computes the similarity between the target user and each item in the dataset. The similarity computation leverages the CPU instruction $\_\_$\emph{popcnt}, which is specifically optimized for calculating Hamming similarity between hash codes. The Hamming similarity measures the number of the identical bits between two hash codes, providing a straightforward yet effective way to quantify the similarity between items. After computing the Hamming similarity for all items, hash code ranking retrieves the items that exhibit the highest similarities to the target user, designating them as candidate items. These candidate items represent the most likely matches based on their similarity to the target user. Since the computational complexity of similarity calculation between hash codes is lower than that of real-valued representations-owing to the simplicity of bit operations inherent in hash code comparisons-hash code ranking can significantly improve recall efficiency.

\begin{figure*}[!t]
    \centering
    \includegraphics[width=0.8\textwidth]{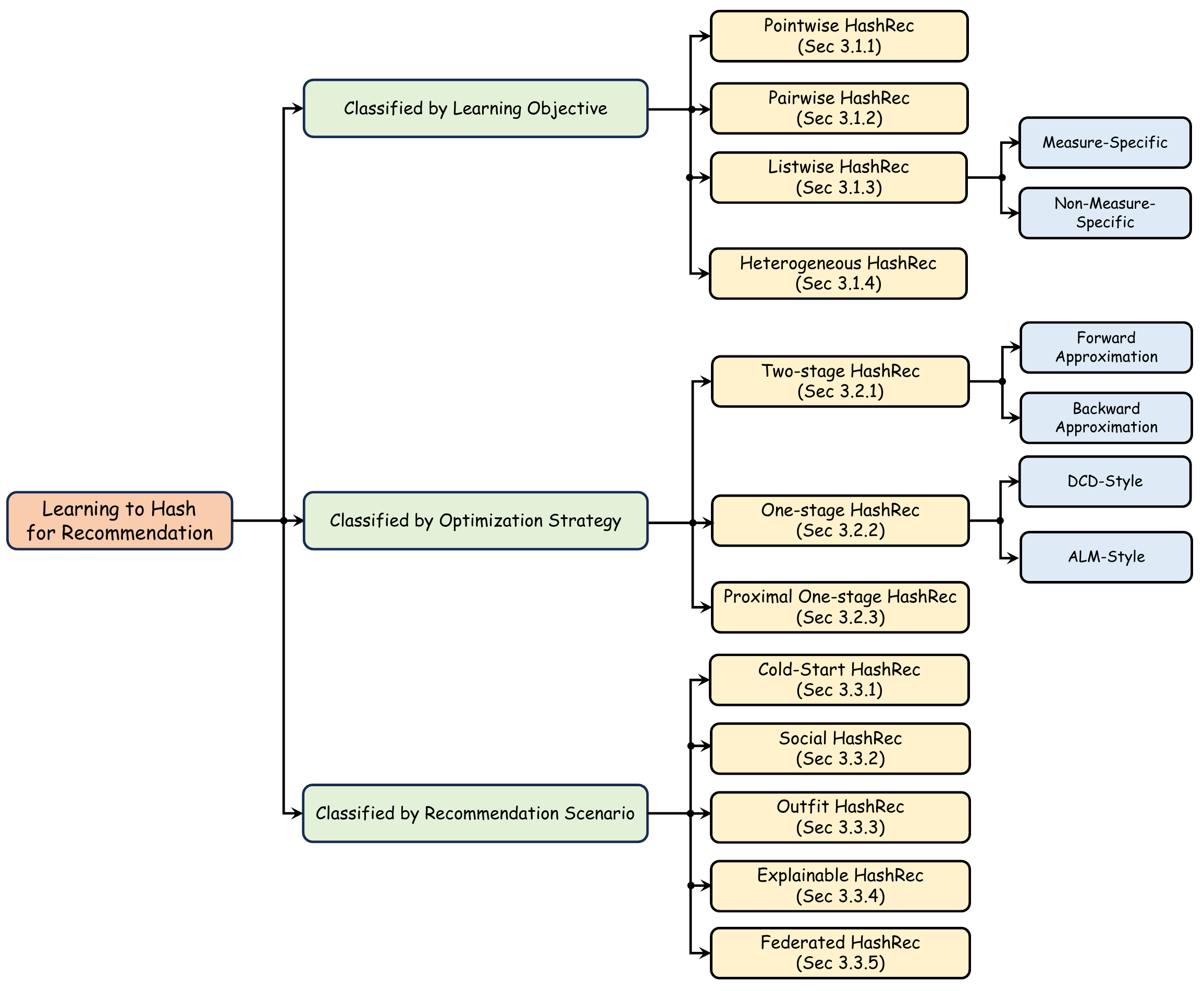}
    \caption{Taxonomy of HashRec Methods.}
    \label{fig:classfication}
\end{figure*}

\section{Learning to Hash for Recommendation}\label{sec:HashRec}

Fig. \ref{fig:classfication} summarizes a general taxonomy of current HashRec algorithms, which is classified from learning objective, optimization strategy and recommendation scenario. 
We introduce the three primary categories in Section \ref{sec:obj}, Section \ref{sec:opt}, and Section \ref{sec:recommendation_task}, respectively. Moreover, we give a brief summary of existing HashRec methods in Table \ref{tab:summary}, including its venue, recommendation scenario, learning objective, optimization strategy, and metrics used in its paper.

{
\setlength{\tabcolsep}{4pt}
\tiny
\begin{longtable}{cccccc}
    \caption{Summary of HashRec methods, where ``UI CF'', ``CR'', ``OR'', ``SR'', ``ER'', and ``FR'' indicate the User-Item CF, Cold-Start Recommendation, Outfit Recommendation, Social Recommendation, Explainable Recommendation, and Federated Recommendation, respectively.} \\
    \label{tab:summary} \\
    \hline
    \textbf{Method} & \textbf{Venue} & \textbf{\makecell[c]{Recommendation\\ Scenario}} & \textbf{\makecell[c]{Learning\\ Objective}} & \textbf{\makecell[c]{Optimization \\Strategy}} & \textbf{Metric} \\
    \hline
    \endfirsthead
    \hline
    \textbf{Method} & \textbf{Venue} & \textbf{\makecell[c]{Recommendation\\ Task}} & \textbf{\makecell[c]{Learning\\ Objective}} & \textbf{\makecell[c]{Optimization \\Strategy}} & \textbf{Metric} \\ 
    \hline
    \endhead
    \hline
    \multicolumn{6}{r@{}}{Continued}
    \endfoot
    \hline
    \endlastfoot
    BCCF \cite{Zhou2012BCCF}  & SIGKDD2012   & UI CF  & Pointwise,Pairwise & Two-Stage & DCG, Precision \\
    CH \cite{Liu14CH}   & CVPR2014  & UI CF  & Pointwise & Two-Stage & Precision, NDCG,MAP, PH2 \\
    PPH \cite{Zhang2014PPH}  & SIGIR2014 & UI CF  & Pointwise & Two-Stage & NDCG \\
    DCF \cite{Zhang2016DCF}  & SIGIR2016 & UI CF  & Pointwise      & One-Stage      & NDCG  \\
    DCMF \cite{Lian17DCMF} & SIGKDD2017   & CR  & Pointwise      & One-Stage      & NDCG, MPR  \\
    DPR \cite{Zhang17DPR}  & AAAI2017  & UI CF  & Pairwise      & One-Stage      & AUC \\
    DFM \cite{Liu18DFM}  & IJCAI2018 & UI CF  & Pointwise      & One-Stage      & NDCG \\
    DSFCH \cite{Liu18DSFCH} & DSE2018 & OR  & Pointwise      & One-stage      & AUC \\
    DRMF \cite{Zhang18DRMF}  & SIGKDD2018   & UI CF  & Pairwise      & One-Stage      & NDCG \\
    DDL \cite{Zhang2018DDL}  & WSDM2018  & CR  & Pointwise      & One-Stage      & Accuracy, MRR \\
    DTMF \cite{Guo2019DTMF}  & IJCAI2019 & SR  & Pointwise      & One-Stage      & NDCG \\
    CIGAR \cite{Kang19CIGAR} & CIKM2019  & UI CF  & Pairwise      & Two-Stage      & HR, MRR  \\
    CCCF \cite{Liu19CCCF}  & SIGIR2019 & UI CF  & Pointwise      & One-Stage      & NDCG \\
    DSR \cite{Liu2019DSR}  & AAAI2019  & SR  & Pointwise      & One-Stage      & NDCG  \\
    FHN \cite{Lu19FHN}  & CVPR2019  & OR  & Pairwise      & Two-Stage      & AUC, NDCG, FITB \\
    DGCN-BinCF \cite{Wang19DGCNBinCF} & IJCAI2019 & UI CF  & Pairwise      & Two-Stage      & NDCG, Recall, MAP \\
    ABinCF \cite{Wang19ABinCF} & AAAI2019  & UI CF  & Pointwise      & Two-Stage       & Precision, NDCG  \\
    NeuHash-CF \cite{Hansen20NeuHash-CF} & SIGIR2020 & CR  & Pointwise      & Two-Stage      & NDCG, MRR \\
    HashGNN \cite{Tan20HashGNN} & WWW2020   & UI CF  & Heterogeneous      & Two-Stage      & HR, NDCG \\
    SDMF \cite{Wu2020SDMF}  & IEEE IS2020 & UI CF  & Pointwise      & One-Stage      & NDCG \\
    MFDCF \cite{Xu20MFDCF} & AAAI2020  & CR  & Pointwise       & One-Stage      & Accuracy \\
    CGH \cite{Zhang20CGH}  & IEEE IS2020 & CR  & Pointwise      & Two-Stage       & Accuracy \\
    DPH \cite{Zhang22DPH}   & IEEE TKDE2020 & CR  & Pairwise      & One-Stage      & Accuracy, MRR \\
    VH$_{\rm{PHD}}$ \cite{Hansen2021VHPHD} & WWW2021   & UI CF  & Pointwise      & Two-Stage      & NDCG, MRR \\
    DMF \cite{Lian2021DMF}   & IEEE TKDE2021 & CR & Pointwise      & One-Stage       & NDCG, AUC, Recall \\
    DLCF \cite{Liu21DLCF}  & SDM2021   & UI CF  & Listwise      & One-Stage      & NDCG  \\
    SDSR \cite{Luo21SDSR} & AAAI2021  & SR  & Pointwise       & One-Stage      & NDCG \\
    BiGeaR \cite{Chen22BiGeaR} & SIGKDD2022   & UI CF  & Pairwise      & Two-Stage       & NDCG, Recall \\
    BIHGH \cite{Guan22BIHGH} & ACM MM2022 & OR  & Pairwise      & Two-Stage      & AUC, MRR, NDCG \\
    DLPR \cite{Luo22DLPR}  & IJCAI2022 & UI CF  & Listwise      & Proximal One-Stage      & MAP, NDCG, Recall, MRR \\
    HCFRec \cite{Wang22HCFRec} & IJCAI2022 & UI CF  & Pointwise       & Two-Stage      & NDCG, MAP \\
    EDCF \cite{Zhu23EDCF} & IEEE TKDE2022 & ER  & Pointwise      & Two-Stage       & Accuracy, NDCG, Recall, Precision, F1 \\
    DLACF \cite{Hu23DLACF} & EAAI2023 & SR & Pointwise & Two-Stage & \makecell[c]{Precision, Recall, F1-Macro, F1-Micro, NDCG}\\
    FHN+ \cite{Lu23FHN} & IEEE TMM2023  & OR  & Pairwise      & Two-Stage      & AUC, NDCG, FITB \\
    BGCH \cite{Chen23BGCH} & WWW2023   & UI CF  & Heterogeneous      & Two-Stage      & NDCG, Recall \\
    MDCF \cite{Xu23MDCF} & IEEE TKDE2023 & CR  & Pointwise      & One-Stage       & Accuracy, NDCG \\
    DLFM \cite{Luo24DLFM} & ACM TKDD2024 & UI CF & Listwise & Proximal One-Stage  & NDCG, Recall, Accuracy \\
    LightFR \cite{Zhang23LightFR} & ACM TOIS2024 & FR & Pointwise &One-Stage & HR, NDCG\\
    DFMR \cite{Yang24DFMR} & ACM TOIS2024 & FR & Pointwise &One-Stage & \makecell[c]{NDCG, Recall, Precision} \\
    TSGNH \cite{Xu24TSGNH} & IEEE TKDE2024 & SR  & Pointwise      & Two-Stage       & HR, NDCG  \\
    BGCH+ \cite{Chen24BGCH+} & IEEE TKDE2024 & UI CF & Pairwise      & Two-Stage      & NDCG, Recall \\
    BiGeaR-SS \cite{Chen25BiGeaR-SS}  & ACM TOIS2025  & UI CF & Pairwise & Two-Stage & NDCG, Recall \\
\end{longtable}
}

\subsection{Learning Objective}\label{sec:obj}
According to the learning paradigm in Learning to Rank (LTR) \cite{Liu11LTR,Li11LTR}, existing methods can be divided into four types: (i) the pointwise HashRec methods, (ii) the pairwise HashRec methods, (iii) the listwise HashRec methods, and (iv) the heterogeneous HashRec methods. 

\subsubsection{The Pointwise HashRec}
The pointwise HashRec methods take a single user-item pair as the input, which aims to optimize their correlation. In the explicit recommendation scenario where the relevance between users and items is an integer value, the recommendation task is formulated as a regression problem, and the Mean Square Error (MSE) is widely adopted as the learning objective \cite{Zhou2012BCCF,Liu14CH,Liu18DFM,Zhang2018DDL,Liu2019DSR,Guo2019DTMF,Hansen20NeuHash-CF,Xu20MFDCF,Liu18DSFCH,Wu2020SDMF,Xu23MDCF,Zhang20CGH,Luo21SDSR,Zhang23LightFR,Hu23DLACF,Yang24DFMR}, which aims to predict the exact relevance degree of each user to an item. Specifically, the learning objective is as follows:
\begin{align}\label{Eq:MSE}
\mathcal{L} &= \frac{1}{|\Omega|}\sum_{(u,i)\in \Omega} (y_{ui} - \hat{y}_{ui})^2= \frac{1}{|\Omega|}\sum_{(u,i)\in \Omega} (y_{ui} - \frac{1}{2}-\frac{1}{2f}\mathbf{b}_u^{\mathsf{T}}\mathbf{d}_i)^2, \nonumber\\
&s.t., \mathbf{B}\in \{-1, +1\}^{f\times m}, \mathbf{D}\in \{-1, +1\}^{f\times n},
\end{align}
where $|\cdot|$ represents the size of a set. $\mathbf{B}\in \{-1, +1\}^{f\times m}$ and $\mathbf{D}\in \{-1, +1\}^{f\times n}$ are discrete constraints, causing that the problem in Eq. (\ref{Eq:MSE}), is a discrete optimization problem. The optimization strategy will be elaborated in Section \ref{sec:opt}. 
\textbf{PPH} \cite{Zhang2014PPH} proposed the constant feature norm constraint $(\sum_{u=1}^m(||\mathbf{b}_u||_F^2-\frac{y_{max}}{2})^2 + \sum_{i=1}^n(||\mathbf{d}_i||_F^2-\frac{y_{max}}{2})^2)$, which can avoid that the inner product deviates from similarity. $y_{max}$ is the maximum value of all ratings. 
\textbf{DCF} \cite{Zhang2016DCF} introduced the balanced and decorrelated constraints ($\mathbf{B}\mathbf{1} = 0, \mathbf{D}\mathbf{1}=0, \mathbf{B}\mathbf{B}^{\mathsf{T}}=m\mathbf{I}, \mathbf{D}\mathbf{D}^{\mathsf{T}}=n\mathbf{I}$) that yield compact yet informative hash codes. 
$\mathbf{B}\mathbf{1} = 0$ and $\mathbf{D}\mathbf{1}=0$ are balanced constraints, which require that each bit to split the dataset as balanced as possible to maximize the information entropy of the bit. $\mathbf{B}\mathbf{B}^{\mathsf{T}}=m\mathbf{I}$ and $\mathbf{D}\mathbf{D}^{\mathsf{T}}=n\mathbf{I}$ are decorrelation constraints, which enforce each bit independent. Then, the bits will be uncorrelated and the variance is maximized by removing the redundancy among the bits. 
\textbf{DCMF} \cite{Lian17DCMF} and \textbf{DMF} \cite{Lian2021DMF} introduced an interaction regularization $\sum_{u\in U}\sum_{i\in I}(\mathbf{b}_u^{\mathsf{T}}\mathbf{d}_i)^2$, penalizing non-zero predicted preference, which should be imposed for better recommendation performance.
\textbf{DFM} \cite{Liu18DFM} proposed to binarize factorization machine that learns hash codes for each feature:
\begin{align}
\mathcal{L} &= \sum_{(\mathbf{x},y)\in\Omega}(y-w_0-\sum_{i=1}^{n_f}w_ix_i-\sum_{i=1}^{n_f}\sum_{j=i+1}^{n_f}<\mathbf{b}_i,\mathbf{b}_j>x_ix_j)^2 +\alpha\sum_{i=1}^{n_f}w_i^2,\nonumber\\
&s.t.,  \mathbf{B}\in \{-1, +1\}^{f\times n_f},\mathbf{B}\mathbf{1} = 0, \mathbf{B}\mathbf{B}^{\mathsf{T}}=n_f\mathbf{I},
\end{align}
where $n_f$ denotes the number of features, $\mathbf{w}\in \mathbb{R}^{n_f}$ is the model bias parameter. $\mathbf{x}\in\mathbb{R}^{n_f}$ is the feature representation of rich side-information, concatenated by one-hot user ID, item ID, user and item content features. It is worth noting that the preference score calculation between user $u$ and item $i$ involves real-valued content features, thereby achieving better recommendation performance.
\textbf{CCCF} \cite{Liu19CCCF} represents each user/item with a set of hash codes, associated with a sparse real-valued weight vector. Each element of the weight vector encodes the importance of the corresponding hash code to the user/item. Therefore, the similarity between the user $u$ and item $i$ is calculated as $\hat{y}_{ui} = \sum_{k=1}^Gw_{ui}^{(k)}(\mathbf{b}_u^{(k)})^{\mathsf{T}}\mathbf{d}_i^{(k)}$, 
where $w_{ui}^{(k)}=\eta_u^{(k)}\cdot\xi_i^{(k)}$ is the importance weight of the $k$-th components with respect to user $u$ and item $i$. Following the setting of compositional matrix approximation, user weight $\eta_u^{(k)}$ can be instantiated with an Epanechnikov kernel \cite{KernelSmoothing} $\eta_u^{(k)}=\frac{3}{4}(1-d(u,u_t')^2)\mathbb{I}(d(u,u_t')<h)$. 
$h>0$ is a bandwidth parameter, and $d(u, u_t')=\arccos(\frac{<\mathbf{b}_u, \mathbf{b}_{u_t'}>}{||\mathbf{b}_u||\cdot ||\mathbf{b}_{u_t'}||})$ is a distance function to measure the similarity between $u$ and its anchor point $u_t'$. Item weight $\xi_i^{(k)}$ follows the analogous formulation.
\textbf{VH$_{\rm{PHD}}$} \cite{Hansen2021VHPHD} proposed the projected Hamming dissimilarity $\delta_H^P(\mathbf{b}_u, \mathbf{d}_i) = \rm{SUM}(\mathbf{b}_u~XOR (\mathbf{b}_u~AND~\mathbf{d}_i))$, effectively allowing a binary importance weighting of the hash code.
$\rm{SUM}$ can be computed using the $\_\_popcnt$ bit string instruction. In particular, this method requires no additional storage and no computational overhead.


In the implicit recommendation scenario where the relevance between users and items is a binary variable, the recommendation task is transformed into the classification problem, and the Binary Cross Entropy (BCE) loss has attracted significant attention from researchers. BCE loss estimates the probability that one user likes one item by maximizing the likelihood $P = \prod_{(u,i)\in \Omega}\hat{y}_{ui}^{y_{ui}} \cdot (1 - \hat{y}_{ui})^{1 - y_{ui}}$.
To maximize the likelihood $P$, some researchers reformulate it as a minimization problem by taking the negative log likelihood, thereby obtaining the BCE loss.
\begin{align}\label{Eq:BCE}
&\mathcal{L} = -\sum_{(u,i)\in \Omega}\bigg(y_{ui}\log(\hat{y}_{ui}) + (1 - y_{ui})\log(1 - \hat{y}_{ui})\bigg), \nonumber\\
&s.t., \mathbf{B}\in \{-1, +1\}^{f\times m}, \mathbf{D}\in \{-1, +1\}^{f\times n},\nonumber\\
& \quad\ \mathbf{B}\mathbf{1} = 0, \mathbf{D}\mathbf{1}=0, \mathbf{B}\mathbf{B}^{\mathsf{T}}=m\mathbf{I}, \mathbf{D}\mathbf{D}^{\mathsf{T}}=n\mathbf{I}.
\end{align}
\textbf{ABinCF} \cite{Wang19ABinCF} proposed an adversarial binary collaborative filtering learning objective:
\begin{align}
D^* &= \max_{\phi}\sum_{u=1}^m(\mathbb{E}_{k\sim p_{ture}(i|u)}[\mathrm{In}\sigma(s_{\phi}(u,i))] + \mathbb{E}_{k\sim p_{\theta}(i|u)}[\mathrm{In}(1-\sigma(s_{\phi}(u,i)))]) + \lambda||\phi||_{\mathcal{L}_2}, \nonumber\\
G^* &= \min_{\theta}\sum_{u=1}^m\sum_{k\sim p_{\theta}(i|u)}\mathrm{In}p_{\theta}(i|u)\mathrm{In}(1-\sigma(s_{\phi}(u,i))), \nonumber\\
& s.t., \theta = H(\theta^*)
\end{align}
where $\theta^*$ is real-valued and $H(\cdot)$ is a hash function, $\lambda$ is a coefficient of the regularization. $s_{\phi}(u,i)=b_i+\_\_popcount(f-2\mathrm{XOR}(\mathbf{b}_u, \mathbf{d}_i))$ is the scoring function.
\textbf{DMF} \cite{Lian2021DMF} adopt the logistic loss as the learning objective. Due to its non-linearity, the learning objective function can be reduced to an inhomogeneous Binary Quadratic Programming (BQP) problem by seeking an upper variational quadratic bound, enabling the application of Discrete Coordinate Descent (DCD) algorithm, which will be detailed in Section \ref{subsubsec:one-stage}:
\begin{align}
\mathcal{L} &= \log(1 + e^{-y_{ui}\mathbf{b}_u^{\mathsf{T}}\mathbf{d}_i})= \log(1 + e^{\mathbf{b}_u^{\mathsf{T}}\mathbf{d}_i})-\frac{1+y_{ui}}{2}\mathbf{b}_u^{\mathsf{T}}\mathbf{d}_i, \nonumber\\
&\leq \lambda(\hat{y}_{ui})((\mathbf{b}_u^{\mathsf{T}}\mathbf{d}_i)^2-\hat{y}_{ui}^2)-\frac{1}{2}(y_{ui}\mathbf{b}_u^{\mathsf{T}}\mathbf{d}_i+\hat{y}_{ui})+\log(1+e^{\hat{y}_{ui}}),
\end{align}
where $\lambda(x)=\frac{1}{4x}\tanh(x/2)=\frac{1}{2x}(\sigma(x)-\frac{1}{2})$ and the equality holds on only if $\hat{y}_{ui}=\mathbf{b}_u^{\mathsf{T}}\mathbf{d}_i$.
\textbf{HCFRec} \cite{Wang22HCFRec} assumed that the observed rating data obey the Poisson distribution, such that $\mathcal{L} = \sum_{u,i}-\frac{1}{y_{ui}!}\exp(y_{ui}\log(\hat{y}_{ui})-\hat{y}_{ui})$, 
where $\hat{y}_{ui}$ is Hamming similarity between user $u$ and item $i$, which is calculated via Eq. (\ref{Eq:HammingSimilarity}).
\textbf{EDCF} \cite{Zhu23EDCF} and \textbf{TSGNH} \cite{Xu24TSGNH} adopt the Eq. (\ref{Eq:BCE}) as the learning objective, and normalize $y_{ui}$ to the range of 0 and 1 with $\frac{y_{ui}}{y_{max}}$. 

\textbf{Pros and cons of the pointwise HashRec.} From the Eq. (\ref{Eq:MSE}) and Eq. (\ref{Eq:BCE}), we observe that both the MSE and BCE loss functions have a linear computational complexity with respect to the size of dataset $|\Omega|$, which is denoted as $\mathcal{O}(|\Omega|)$. It means that they are efficient, particularly in handling large-scale datasets where the number of user-item samples can be abundant. Nevertheless, both of them possess a notable deficiency that they do not consider the inter-dependency between items. 
It indicates that the loss functions do not take into account how the ranking of one item relative to others might impact the overall performance or relevance of the ranking.

\subsubsection{The Pairwise HashRec} Different from the pointwise HashRec methods, the pairwise counterparts focus on the relative order between two items. In this sense, it is closer to the concept of ``ranking'' than the pointwise approaches. Specifically, the pairwise methods aim to determine which item in a pair is preferred. In most cases, Bayesian Personalized Ranking (BPR) \cite{Rendle09BPR} is adopted as the learning objective \cite{Wang19DGCNBinCF,Chen22BiGeaR,Guan22BIHGH,Chen24BGCH+,Lu23FHN,Chen25BiGeaR-SS}. BPR is an elegant framework that leverages bayesian inference to model user preferences within a probabilistic formulation, naturally accommodating the relative nature of pairwise comparisons. By maximizing the likelihood of observed preference orderings, BPR guides the pairwise HashRec methods towards constructing representations inherently tailored to ranking tasks, thereby enhancing their effectiveness in accurately arranging items according to user preferences. 
\begin{align}\label{Eq:BPR}
&\mathcal{L} = \frac{1}{|\Omega_s|}\sum_{(u, i, j)\in \Omega_s}\mathrm{In}\sigma(\hat{y}_{ui}-\hat{y}_{uj}), \nonumber\\
&s.t., \mathbf{B}\in \{-1, +1\}^{f\times m}, \mathbf{D}\in \{-1, +1\}^{f\times n},\nonumber\\
& \quad\ \mathbf{B}\mathbf{1} = 0, \mathbf{D}\mathbf{1}=0, \mathbf{B}\mathbf{B}^{\mathsf{T}}=m\mathbf{I}, \mathbf{D}\mathbf{D}^{\mathsf{T}}=n\mathbf{I},
\end{align}
where $\Omega_s := \{(u,i,j)|i\in I_u^+ \wedge j\in I_u^-\}$ and the semantics of $(u,i,j)\in \Omega_s$ is that user $u$ is assumed to prefer item $i$ over item $j$. $I_u^+$ and $I_u^-$ represent the positive item set and negative item set of user $u$. In general, the negative samples are usually obtained through sampling techniques \cite{Ma24NegativeSampling}. And $\sigma(x) = 1/(1+e^{-x})$ denotes the sigmoid function.
\textbf{BCCF} \cite{Zhou2012BCCF}, \textbf{DPR} \cite{Zhang17DPR}, and \textbf{DPH} \cite{Zhang22DPH} transform the Eq. (\ref{Eq:BPR}) into a pairwise least square loss as the learning objective, enabling the application of DCD optimization algorithm. Specifically, it is defined as:
\begin{align}\label{Eq:least_square_loss}
&\mathcal{L} = \frac{1}{|\Omega_s|}\sum_{(u, i, j)\in \Omega_s}(1-(\hat{y}_{ui}-\hat{y}_{uj}))^2, \nonumber\\
&s.t., \mathbf{B}\in \{-1, +1\}^{f\times m}, \mathbf{D}\in \{-1, +1\}^{f\times n},\nonumber\\
& \quad\ \mathbf{B}\mathbf{1} = 0, \mathbf{D}\mathbf{1}=0, \mathbf{B}\mathbf{B}^{\mathsf{T}}=m\mathbf{I}, \mathbf{D}\mathbf{D}^{\mathsf{T}}=n\mathbf{I}.
\end{align}
\textbf{CIGAR} \cite{Kang19CIGAR} and \textbf{FHN} \cite{Lu19FHN} introduce auxiliary real-valued embeddings $\mathbf{p}_u \in \mathbb{R}^f$ and $\mathbf{q}_i \in \mathbb{R}^f$. Then, the learning objective is formulated as $\mathcal{L} = -\sum_{(u,i,j)\in \Omega_s}\mathrm{In}\sigma_{\alpha}(<\mathrm{tanh}(\beta\mathbf{p}_u), \mathrm{tanh}(\beta\mathbf{q}_i)>-<\mathrm{tanh}(\beta\mathbf{p}_u), \mathrm{tanh}(\beta\mathbf{q}_j)>)$.
where $\mathbf{b}_u=\mathrm{sign}(\mathbf{p}_u)=\lim_{\beta\rightarrow \infty}\mathrm{tanh}(\beta\mathbf{p}_u)$ and $\mathbf{d}_i=\mathrm{sign}(\mathbf{q}_i)=\lim_{\beta\rightarrow \infty}\mathrm{tanh}(\beta\mathbf{q}_i)$. $\sigma_{\alpha}(x) = \sigma(\alpha x)=1/(1+\exp(-\alpha x))$. And $<\cdot, \cdot>$ represents the inner product. \textbf{BiGeaR-SS} \cite{Chen25BiGeaR-SS} proposes the graph layer-wise quantization by computing both quantized embeddings and embedding scalers. The preference scores are calculated by the inner product of the weighted concatenation of quantized embeddings in each layer, where the weight increases linearly from lower layers to higher layers, primarily for computational simplicity and stability. 



\textbf{Pros and cons of the pairwise HashRec.} Pairwise HashRec methods are designed to generate recommendation lists that are tailored to the target user's specific preferences by taking into account how the user might compare and contrast different options. Consequently, they exhibit a closer alignment with the ultimate goal of RS. Despite their effectiveness, two major challenges remain. First, the learning objectives only consider the relative order between two items, but the position of items in the final recommendation list can hardly be derived. Additionally, the number of item pairs varies greatly among users, wit some having hundreds of pairs while others have only a few dozen. The imbalance in the training data poses a challenge for accurately assessing the overall effectiveness of RS.

\subsubsection{The Listwise HashRec} Listwise HashRec methods take the entire set of items associated with a user as input and predict their labels. Existing listwise HashRec methods can be categorized into two types based on whether the learning objective is directly related to the evaluation metric. The first type is measure-specific, where the objective is explicitly related to the evaluation metric (e.g., the differentiable approximation of evaluation metric) \cite{Luo22DLPR,Luo24DLFM}. The second type is non-measure-specific, where the objective is not explicitly tied to the evaluation metric \cite{Liu21DLCF}. 

\textbf{Measure-Specific Listwise HashRec.} In the measure-specific listwise HashRec methods, the learning objective is generally the differentiable approximation of the evaluation metric. According to the results on web search \cite{Donmez09} and text retrieval \cite{Yilmaz10}, when targeting at less informative metrics such as Precision and Recall, optimizing more informative metrics, like Normalized Discounted Cumulative Gain (NDCG) or Average Precision (AP), can perform even better \cite{Li21MetricOpt}. Let's take NDCG as an example. Formally, given the ground truth list $\mathbf{y}_u = [y_{u1}, y_{u2}, \cdots, y_{un}]$ and predicted list $\hat{\mathbf{y}}_u = [\hat{y}_{u1}, \hat{y}_{u2}, \cdots, \hat{y}_{un}]$ of user $u$. And we define $\pi^u$ as a permutation over $\hat{\mathbf{y}}_u$, which records the indices of elements sorted in descending order. To represent the ranking list of $\mathbf{y}_u$ determined by $\pi^u$, let's further define a permutation matrix $\mathbf{Z}^u\in\{0,1\}^{n\times n}=\left\{
             \begin{array}{lr}
             1, \quad j = \pi^u_i & \\
             0, \quad otherwise. &
             \end{array}
            \right.$ associated with $\pi^u$.
Based on $\mathbf{Z}^u$, the specific ranking list of $\mathbf{y}_u$ is defined by $\mathbf{Z}^u\mathbf{y}_u$. Then, the NDCG@K for user $u$ can be formulated as:
\begin{align}\label{Eq:lossNDCG}
NDCG@K &= \frac{DCG@K(\mathbf{y}_u, \mathbf{Z}^u)}{DCG@K(\mathbf{y}_u, \mathbf{Z}^*)} =\sum_{k=1}^K \frac{\mathbf{Z}^u_k(E_2(\mathbf{y}_u - \mathbf{1}))}{\log_2(k+1)DCG@K(\mathbf{y}_u, \mathbf{Z}^*)},
\end{align}
where $\mathbf{Z}^*$ is an ideal permutation matrix over $\mathbf{y}_u$, $E_2(\mathbf{y}_u) \triangleq [2^{y_{u1}}, 2^{y_{u2}}, \cdots, 2^{y_{un}}]^{\mathsf{T}}$ and $\mathbf{1}$ is an all-one vector with length of $n$. Due to the existence of $\mathbf{Z}^u$, Eq. (\ref{Eq:lossNDCG}) is non-differentiable. Inspired by \cite{Grover19NeuralSort}, some researchers \cite{Luo22DLPR,Luo24DLFM} replace $\mathbf{Z}^u \in \{0,1\}^{n\times n}$ with its soften version $\widetilde{\mathbf{Z}}^u \in [0,1]^{n\times n}$ to smooth the Eq. (\ref{Eq:lossNDCG}). Concretely, the $k$-th row of $\widetilde{\mathbf{Z}}^u$ is defined as $\widetilde{\mathbf{Z}}^u_{k} = \rm{softmax}[\tau^{-1}((|\mathcal{I}_u|+1-2k)\mathbf{\hat{y}}^u-\mathbf{A}^u\mathbf{1})]$, 
where $\tau>0$ is a temperature parameter that controls the degree of approximation, and $\mathbf{A}^u$ is a matrix whose elements are defined as $A_{ij}^u=\mathrm{abs}(\hat{y}_{ui}-\hat{y}_{uj})$. $\rm{abs}(\cdot)$ denotes the absolute value of an inputted number. By replacing $\mathbf{Z}^u$ as its smooth version, the learning objective is formulated as:
\begin{align}
&\mathcal{L} = 1 - \sum_{u=1}^m \sum_{k=1}^k \frac{\widetilde{\mathbf{Z}}^u_{k} (E_2( \mathbf{y}_u)-\mathbf{1})} {\log_2(k+1) DCG@K(\mathbf{y}_u,\mathbf{Z}^*)}, \nonumber\\
&s.t., \mathbf{B}\in \{-1, +1\}^{f\times m}, \mathbf{D}\in \{-1, +1\}^{f\times n}, \nonumber\\
& \quad\ \mathbf{B}\mathbf{1} = 0, \mathbf{D}\mathbf{1}=0, \mathbf{B}\mathbf{B}^{\mathsf{T}}=m\mathbf{I}, \mathbf{D}\mathbf{D}^{\mathsf{T}}=n\mathbf{I}.
\end{align}
Specifically, the proximity between the learning objective and the original evaluation metric is ensured by \cite{Luo24DLFM,Luo22DLPR}, thereby providing a robust foundation for the approximation. Other evaluation metrics can be similarly transformed into learning objectives.

\textbf{Non-Measure-Specific Listwise HashRec.} In non-measure-specific listwise HashRec methods, PlackettLuce model \cite{Cao07ListNET,Wu18Sqlrank} is a widely used permutation probability model, representing each user as a probability distribution over permutations of relevant items. Formally, let $\pi^u$ denote a particular permutation of the $n$ items of user $u$, which is a random variable and takes values from the set of all possible permutations. $\pi^u_1$ denotes the item with the highest relevance, and $\pi^u_n$ is the lowest ranked. Then, the probability of generating the ranking permutation matrix $\pi = [\pi^1, \pi^2, \cdots, \pi^m]^{\mathsf{T}} \in \mathbb{Z}^{m\times n}$ can be formulated as $P(\pi|\mathbf{B}, \mathbf{D}) = \prod_{u=1}^m\prod_{i=1}^n \frac{e^{\phi(\mathbf{b}_u^{\mathsf{T}}\mathbf{d}_{\pi^u_i})}}{\sum_{l=j}^ne^{\phi(\mathbf{b}_u^{\mathsf{T}}\mathbf{d}_{\pi^u_l})}}$,
where $\phi(\cdot)$ is an increasing and strictly positive function. It is noteworthy that the rating data may include many items with the same ratings for each user in explicit scenarios, resulting in permutations that coincide with the rating. To address this limitation, Wu et al. \cite{Wu18Sqlrank} adopt a stochastic queuing process to shuffle the ordering of observed items with the same ratings. Specifically, let $S(\Omega, \mathbf{Y})$ denote the set of valid permutations. Then, the probability of generating the observed ratings $\mathbf{Y}$ can be formulated as:
\begin{align}\label{Eq:PL}
P(\mathbf{Y}|\mathbf{B}, \mathbf{D}) = \sum_{\pi \in S(\Omega, \mathbf{Y})}P(\pi|\mathbf{B}, \mathbf{D}) = \sum_{\pi \in S(\Omega, \mathbf{Y})}\prod_{u=1}^m P_u(\pi_u, \mathbf{b}_u, \mathbf{D}) = \sum_{\pi \in S(\Omega, \mathbf{Y})}\prod_{u=1}^m\prod_{i=1}^{n_u} \frac{e^{\phi(\mathbf{b}_u^{\mathsf{T}}\mathbf{d}_{\pi^u_i})}}{\sum_{l=j}^{n_u}e^{\phi(\mathbf{b}_u^{\mathsf{T}}\mathbf{d}_{\pi^u_l})}},
\end{align}
where $n_u$ denotes the number of interacted items for user $u$. Based on Eq. (\ref{Eq:PL}), the learning objective is as follows:
\begin{align}
&\mathcal{L} = -\log \sum_{\pi \in S(\Omega, \mathbf{Y})} P(\pi|\mathbf{B}, \mathbf{D}), \nonumber\\
&s.t., \mathbf{B}\in \{-1, +1\}^{f\times m}, \mathbf{D}\in \{-1, +1\}^{f\times n},\nonumber\\
& \quad\ \mathbf{B}\mathbf{1} = 0, \mathbf{D}\mathbf{1}=0, \mathbf{B}\mathbf{B}^{\mathsf{T}}=m\mathbf{I}, \mathbf{D}\mathbf{D}^{\mathsf{T}}=n\mathbf{I}.
\end{align}

\textbf{Pros and cons of the listwise HashRec.} The learning objective of listwise HashRec methods is inherently more aligned with the ultimate goal of ranking tasks, as they consider the entire list as input rather than individual items or pairs. Consequently, listwise HashRec methods demonstrate a superior capability in capturing the complex interactions and dependencies among items, achieving better performance. However, despite these advantages in recommendation accuracy, listwise HashRec methods often face a higher computational complexity. This is primarily due to the fact that they need to evaluate and optimize the entire ranking list during training stage, which can involve time-consuming sort operations. As a result, the time and computational resources required for training can be significantly greater, especially for large-scale datasets.


\subsubsection{The Heterogeneous HashRec}
The heterogeneous HashRec methods are trained end-to-end by jointly optimizing two losses, i.e., reconstruction loss from reconstructing observed links, and ranking loss from preserving the relative ordering of hash codes.
\textbf{HashGNN} \cite{Tan20HashGNN} combined BCE loss and hinge loss, which is formulated as:
\begin{align}
\mathcal{L} &= -\sum_{(u,i)\in\Omega}y_{ui}\log(\sigma(<\mathbf{b}_u, \mathbf{d}_i>))+(1-y_{ui})\log(1-\sigma(<\mathbf{b}_u, \mathbf{d}_i>)) \nonumber\\
&+ \lambda \sum_{(u,i,j)\in \Omega_s}\max(0, -\sigma(<\mathbf{b}_u, \mathbf{d}_i>)+\sigma(<\mathbf{b}_u, \mathbf{d}_j>)+\alpha),
\end{align}
where $\lambda$ is the trade-off parameter to balance the importance between entropy and ranking loss.
\textbf{BGCH} \cite{Chen23BGCH} consists of BCE loss and BPR loss, which harness the regularization effect to each other. 
\begin{align}
\mathcal{L} &= -\sum_{(u,i)\in\Omega}y_{ui}\log(\sigma(<\mathbf{b}_u, \mathbf{d}_i>))+(1-y_{ui})\log(1-\sigma(<\mathbf{b}_u, \mathbf{d}_i>)) \nonumber\\
&-\lambda_1\sum_{(u,i,j)\in \Omega_s} \mathrm{In} (\sigma(<\mathbf{b}_u, \mathbf{d}_i>)-\sigma(<\mathbf{b}_u, \mathbf{d}_j>)).
\end{align}

\textbf{Pros and cons of the heterogeneous HashRec.} Compared with pointwise HashRec and pairwise HashRec methods, heterogeneous HashRec methods simultaneously consider both the correlation between users and items and the relative relationship between item pairs achieving higher recommendation performance. However, heterogeneous HashRec methods can be regarded as a multi-objective optimization problem, requiring careful tuning of weight parameters to balance performance. In addition, although heterogeneous HashRec methods have lower computational complexity, there exists an inconsistency between optimization objective and evaluation criterion, which may lead suboptimal performance.

\begin{table*}[!t]
\setlength{\tabcolsep}{8pt}
\tiny
  \centering
  \caption{Illustration of pointwise, pairwise, listwise and heterogeneous HashRec methods, where $i$ and $j$ denote the positive and negative item of user $u$, respectively.}
    \begin{tabular}{c|c|c|c|c}
    \toprule
    \textbf{Methods} & \textbf{Training Samples} & \textbf{Advantages} & \textbf{Challenges} & \textbf{Related Work} \\
    \midrule
    \makecell[c]{Pointwise HashRec}     & ($u$, $i$, $y_{ui}$)      & Easy-to-deploy  & \makecell[c]{Ignoring the inter-dependency\\ between items} & \makecell[c]{\cite{Zhou2012BCCF},\cite{Liu14CH},\cite{Zhang2014PPH},\\\cite{Zhang2016DCF},\cite{Lian17DCMF},\cite{Liu18DFM},\\\cite{Liu18DSFCH},\cite{Zhang18DRMF},\cite{Zhang2018DDL},\\\cite{Guo2019DTMF},\cite{Liu19CCCF},\cite{Liu2019DSR},\\\cite{Hansen20NeuHash-CF},\cite{Wu2020SDMF},\cite{Xu20MFDCF},\\\cite{Zhang20CGH},\cite{Hansen2021VHPHD},\cite{Lian2021DMF},\\\cite{Luo21SDSR},\cite{Wang22HCFRec},\cite{Zhu23EDCF},\\\cite{Xu23MDCF},\cite{Xu24TSGNH},\cite{Wang19ABinCF},\\\cite{Zhang23LightFR},\cite{Hu23DLACF},\cite{Yang24DFMR}} \\
    \midrule
    \makecell[c]{Pairwise HashRec}      & ($u$, $i$, $j$)  & \makecell[c]{Capturing the relative order\\ between two items}   & \makecell[c]{Ignoring the items' \\ranking position; \\ Higher computational cost} & \makecell[c]{\cite{Zhou2012BCCF},\cite{Zhang17DPR},\cite{Kang19CIGAR},\\\cite{Lu19FHN},\cite{Wang19DGCNBinCF},\cite{Chen24BGCH+},\\\cite{Zhang22DPH},\cite{Chen22BiGeaR},\cite{Guan22BIHGH},\\\cite{Lu23FHN}, \cite{Chen25BiGeaR-SS}} \\
    \midrule
    \makecell[c]{Listwise HashRec}      & ($u$, $i_1$, $\cdots$, $i_n$)      & Matching the ultimate goal of RS   & Heavy computational cost  & \makecell[c]{\cite{Liu21DLCF},\cite{Luo22DLPR},\cite{Luo24DLFM}} \\
    \midrule
    \makecell[c]{Heterogeneous HashRec}      & ($u$, $i$, $j$)      & \makecell[c]{Taking both the preference score and \\ relative preference into account }  & \makecell[c]{Ignoring the items' \\ranking position; \\ Higher computational cost}  & \makecell[c]{\cite{Tan20HashGNN},\cite{Chen23BGCH}} \\
    \bottomrule
    \end{tabular}%
  \label{tab:divided_by_loss}%
\end{table*}%

\subsubsection{Summary}
In this section, we categorize existing HashRec methods into four types: pointwise, pairwise, listwise, and heterogeneous, based on the paradigm of LTR. A comprehensive examination of training samples, advantages, challenges, and related work is summarized in Table \ref{tab:divided_by_loss}. Although pointwise, pairwise and heterogeneous HashRec methods are easy to deploy and possess lower computational complexity, they all overlook the crucial role of items' ranking position during the training phase, which deviates from the ultimate goal of RS. To be consistent with the goal of RS, the listwise HashRec methods are proposed. Concretely, listwise HashRec methods are designed to capture the intricate relationships among items within the recommendation list to bridge the gap between learning objectives and evaluation metrics. However, this method faces challenging training efficiency problem due to the calculation of sorting operation in the training stage. Hence, there is a necessity for a highly effective and efficient learning objective.

\begin{figure}[!h]
    \centering
    \includegraphics[width=0.8\textwidth]{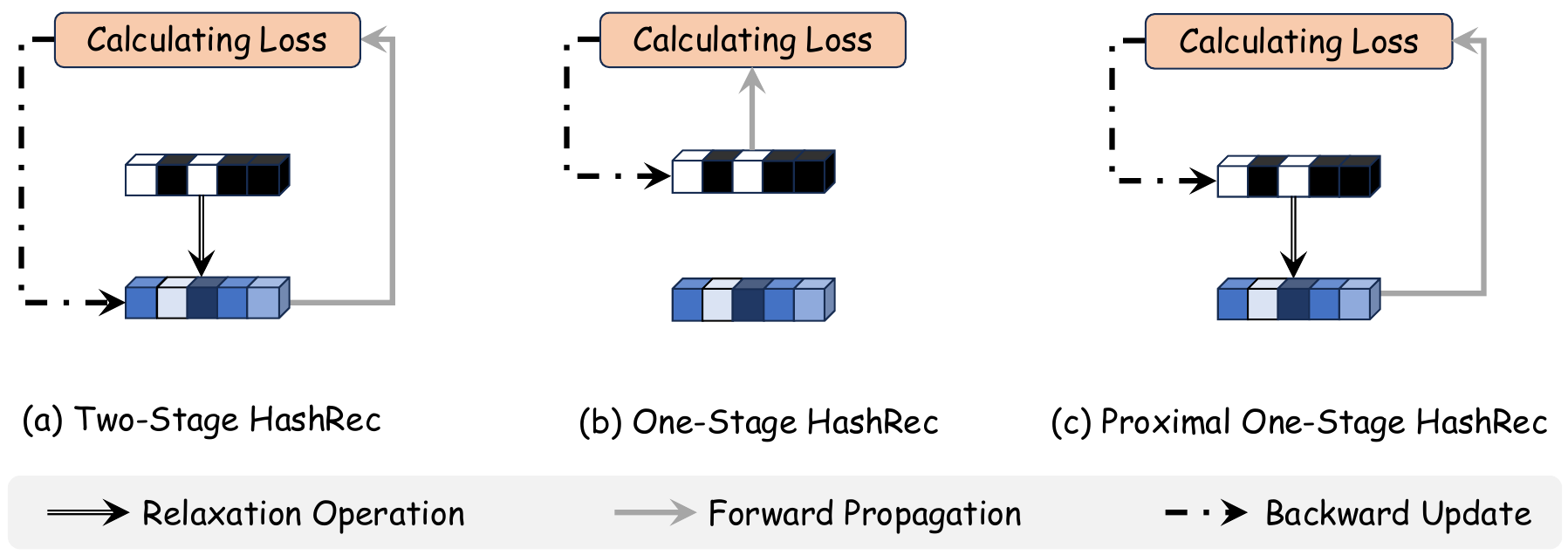}
    \caption{Illustration of three optimization strategies, where the black and white blocks represent the hash codes, while the blue blacks represent the real-valued embeddings.}
    \label{fig:optimization_strategy}
\end{figure}

\subsection{Optimization Strategy}\label{sec:opt}
According to the optimization strategy, existing methods can be divided into three types: (i) the two-stage HashRec methods, (ii) the one-stage HashRec methods, and (iii) the proximal one-stage HashRec methods. 
The comparison among three optimization strategies is visualized in Fig. \ref{fig:optimization_strategy}. 

\subsubsection{The Two-stage HashRec}
Two-stage HashRec methods consist of two key steps: relaxation and quantization. Based on the stage at which the relaxation operation is applied, existing two-stage HashRec methods can be categorized as forward approximation and backward approximation. Forward approximation method relax the non-differentiable $\rm{sign}(\cdot)$ function induced by the discrete constraints during the forward propagation process, whereas backward approximation methods achieve this relaxation during the backward propagation process. Consequently, the two-stage optimization strategy is not constrained by the learning objective, providing a flexible solution to the discrete optimization problem. 

\begin{algorithm}[!t]
\caption{Two-stage HashRec: Forward Approximation}
\label{alg:forward}
\LinesNumbered 
Relax the discrete constraints on $\mathbf{B}$ and $\mathbf{D}$ into $\mathbf{P}$ and $\mathbf{Q}$; \\
Initialize $\mathbf{P}$ and $\mathbf{Q}$; \\
\Repeat{converge}{
Optimize learning objective regarding $\mathbf{P}$ and $\mathbf{Q}$ via gradient-descend methods;
}
Conduct quantization operation on $\mathbf{P}$ and $\mathbf{Q}$; \\
$\textbf{return}\ \mathbf{B}, \mathbf{D}$
\end{algorithm}

\textbf{Forward Approximation.} The early studies typically learn real-valued latent features via gradient-descend methods \cite{Baldi95GD} for users and items by discarding the discrete constraints, and then perform quantization, such as median quantization, on real-valued users' and items' representations to obtain hash codes. Specifically, the algorithmic framework of forward approximation is given in Algorithm \ref{alg:forward}. Let's take the MSE loss in Eq. (\ref{Eq:MSE}) as an example, the relaxed optimization problem is formulated as $\widetilde{\mathcal{L}} = \frac{1}{|\Omega|}\sum_{(u, i) \in \Omega}(y_{ui} - \mathbf{p}_u^{\mathsf{T}} \mathbf{q}_i)^2$.
Then, the update rule with respect to $\mathbf{p}_u$ and $\mathbf{q}_i$ can be expressed as $\mathbf{p}_u \leftarrow \mathbf{p}_u - lr  \cdot \frac{\partial \widetilde{\mathcal{L}}}{\partial \mathbf{p}_u}$ and $\mathbf{q}_i \leftarrow \mathbf{q}_i - lr  \cdot \frac{\partial \widetilde{\mathcal{L}}}{\partial \mathbf{q}_i}$, 
where $lr$ is the learning rate. After solving the relaxed optimization problem, we can obtain real-valued representations of users and items, i.e., $\mathbf{P}=[\mathbf{p}_1, \mathbf{p}_2, \cdots, \mathbf{p}_m]^{\mathsf{T}}$ and $\mathbf{Q}=[\mathbf{q}_1, \mathbf{q}_2, \cdots, \mathbf{q}_n]^{\mathsf{T}}$. Then, we can obtain hash codes by conducting quantization on the real-valued representations. A straightforward method is median quantization, a binary thresholding operation where any feature value above the median is set to 1, and any value below it is set to -1.
While the early methods effectively address the challenging discrete optimization problem, the optimization strategy unfortunately incurs substantial quantization loss, which negatively impacts overall recommendation quality. To mitigate this issue, some researchers \cite{Kang19CIGAR,Guan22BIHGH,Xu24TSGNH} propose to map the range of real-valued representations into a tighter interval of $[-1, 1]$ via the $\rm{tanh}(\cdot)$ function. Then, the learning objective is as follows:
\begin{align}
\widetilde{\mathcal{L}} = \frac{1}{|\Omega|}\sum_{(u, i) \in \Omega}(y_{ui} - \rm{tanh}(\beta \mathbf{p}_u)^{\mathsf{T}}\rm{tanh}(\beta \mathbf{q}_i))^2,
\end{align}
where $\beta$ is a temperature coefficient, which controls the approximation between $\rm{sign}(x)$ and $\rm{tanh}(\beta x)$. And the larger $\beta$ is, the more similar they are.

\begin{algorithm}[!t]
\caption{Two-stage HashRec: Backward Approximation}
\label{alg:backward}
\LinesNumbered 
Relax the discrete constraints on $\mathbf{B}$ and $\mathbf{D}$ into $\mathbf{P}$ and $\mathbf{Q}$; \\
Initialize $\mathbf{P}$ and $\mathbf{Q}$; \\
\Repeat{converge}{
// Forward \\
Conduct quantization operation on $\mathbf{P}$ and $\mathbf{Q}$ to obtain $\mathbf{B}$ and $\mathbf{D}$;\\
Calculate loss value regarding $\mathbf{B}$ and $\mathbf{D}$;\\
// Backward \\
Optimize $\mathbf{P}$ and $\mathbf{Q}$ via gradient estimation;\\
}
Conduct quantization operation on $\mathbf{P}$ and $\mathbf{Q}$; \\
$\textbf{return}\ \mathbf{B}, \mathbf{D}$
\end{algorithm}

\begin{figure*}[!t]
\centering
\subfigure[Sign]{\label{fig:sign}
    \includegraphics[width=0.23\textwidth]{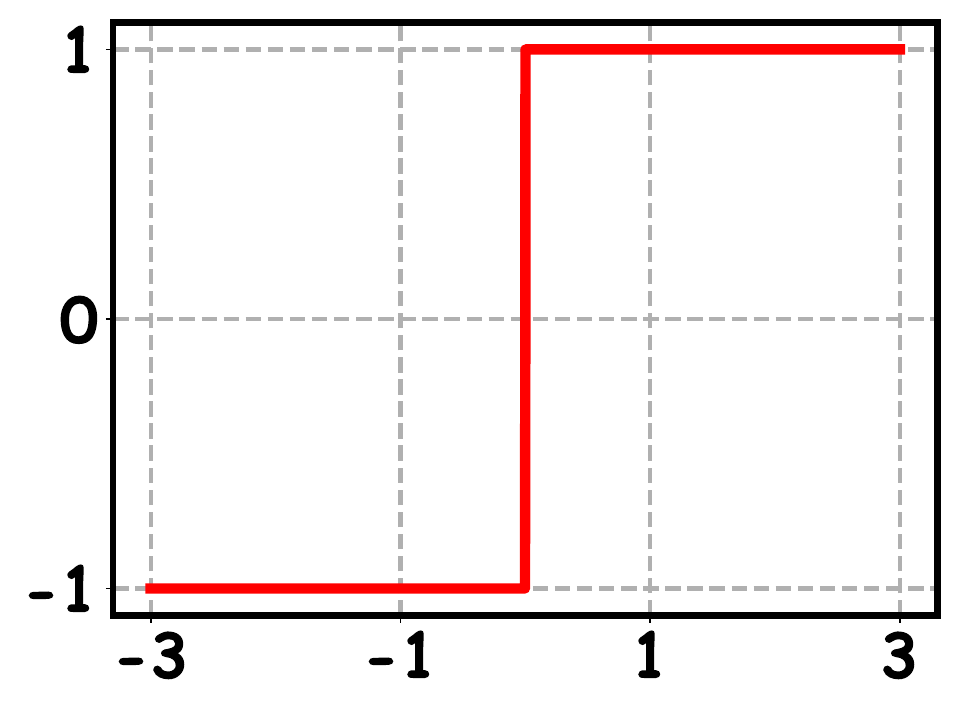}}
\subfigure[Tanh]{\label{fig:tanh}
    \includegraphics[width=0.23\textwidth]{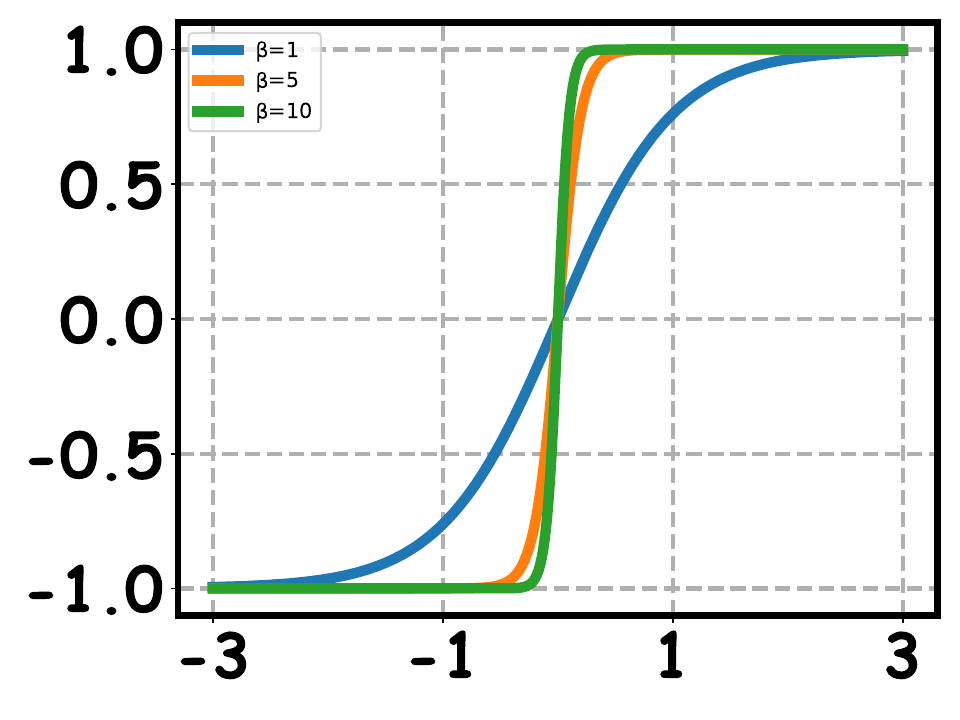}}
\subfigure[SignSwish]{\label{fig:signswish}
    \includegraphics[width=0.23\textwidth]{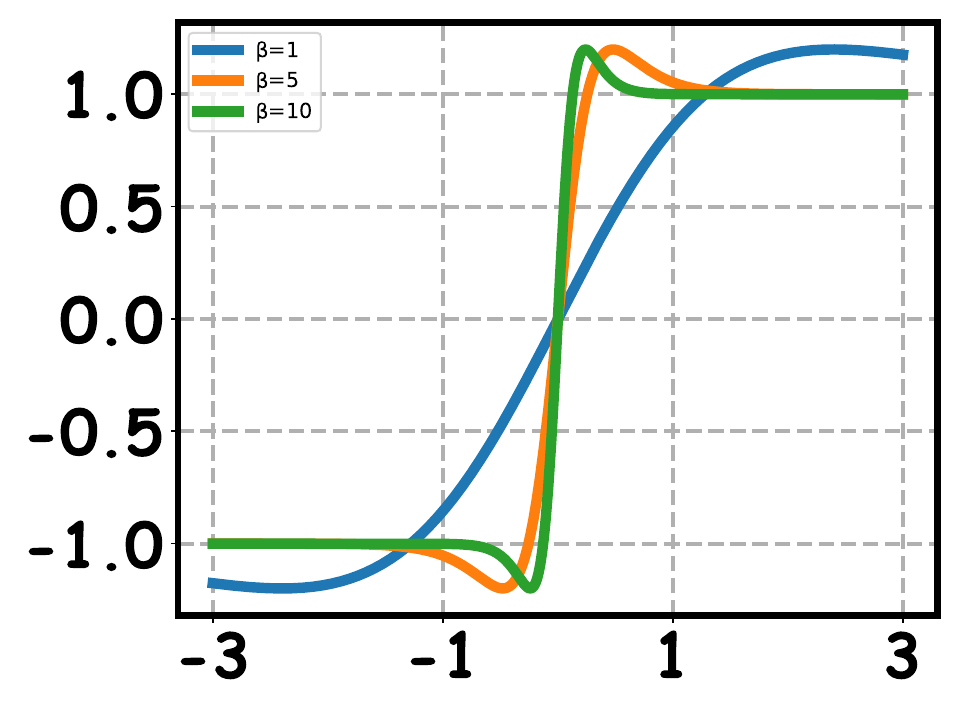}}
\subfigure[FourierSerialized]{\label{fig:fourier}
    \includegraphics[width=0.23\textwidth]{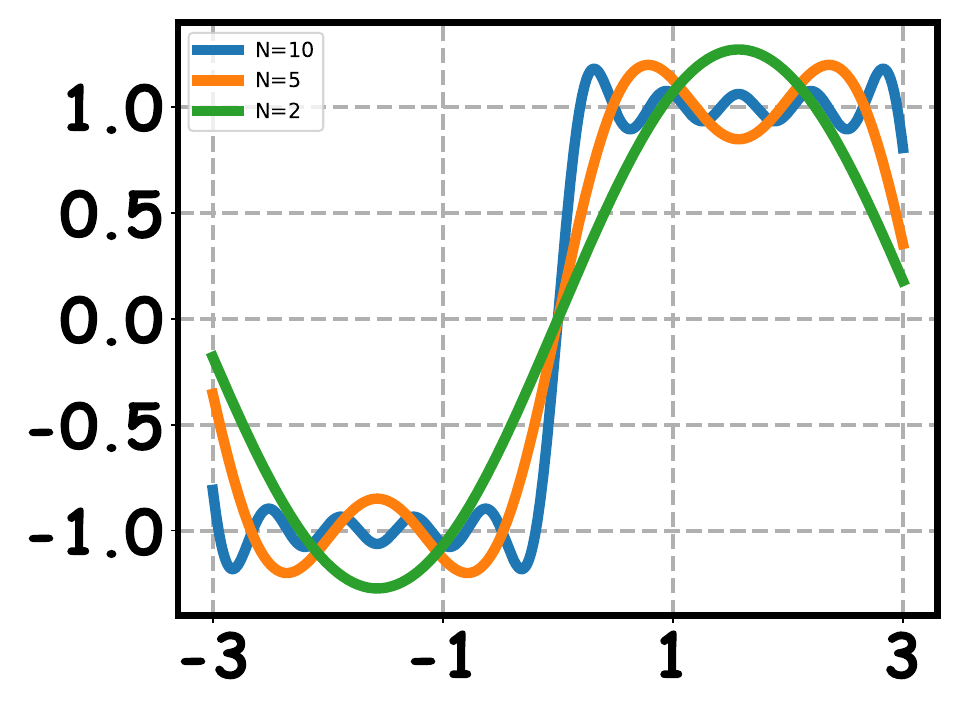}}
\subfigure[Gradient of Sign]{\label{fig:grad_sign}
    \includegraphics[width=0.23\textwidth]{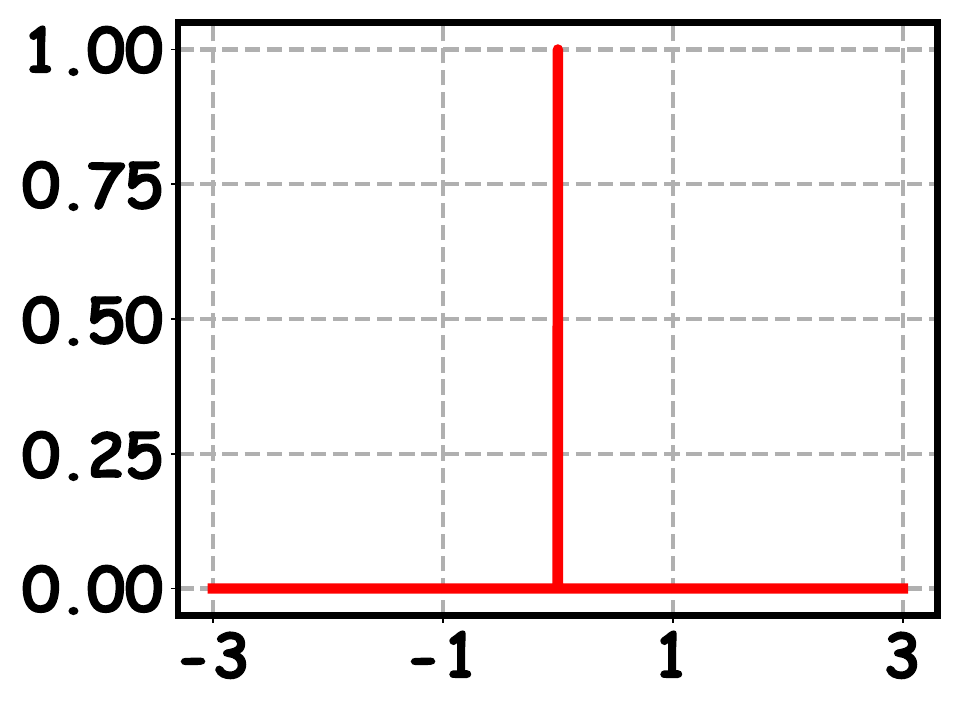}}
\subfigure[Gradient of Tanh]{\label{fig:grad_tanh}
    \includegraphics[width=0.23\textwidth]{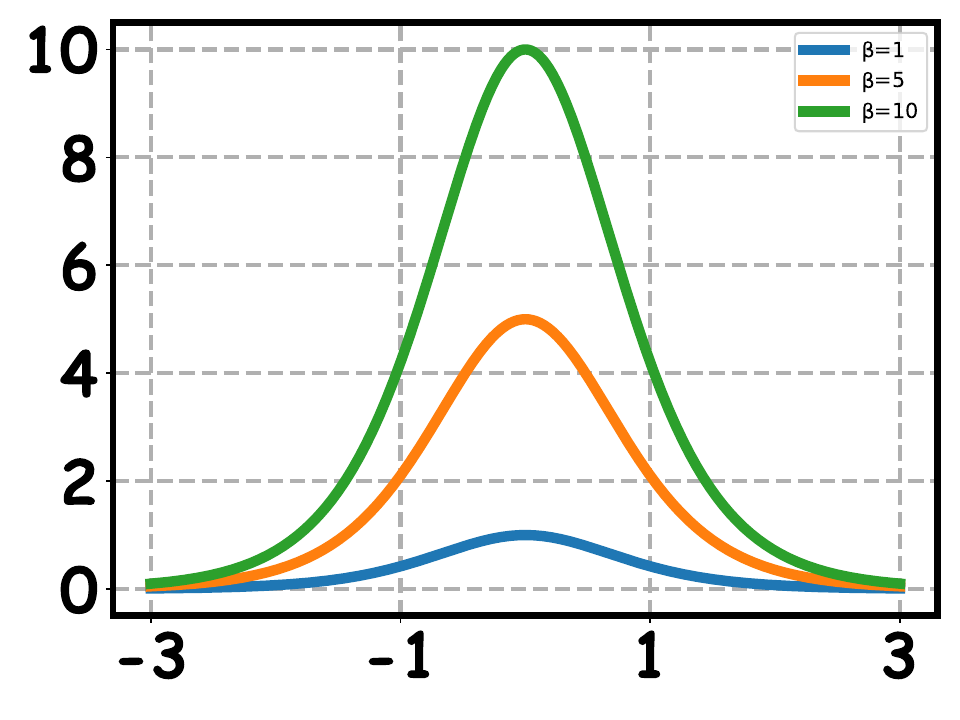}}
\subfigure[Gradient of SignSwish]{\label{fig:grad_signswish}
    \includegraphics[width=0.23\textwidth]{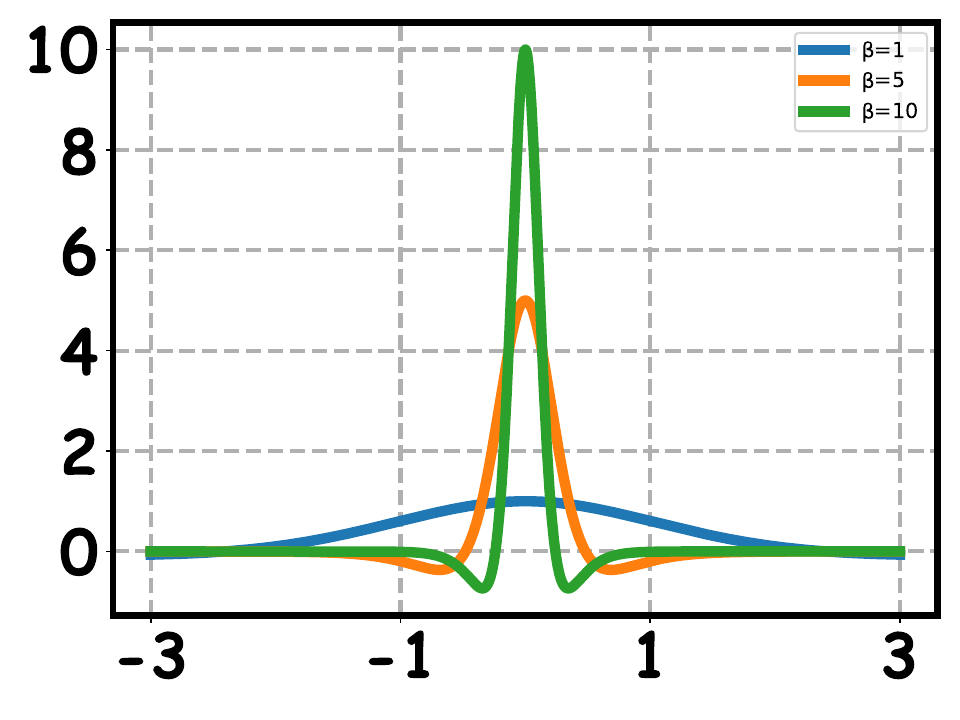}}
\subfigure[Gradient of FourierSeries]{\label{fig:grad_fourier}
    \includegraphics[width=0.23\textwidth]{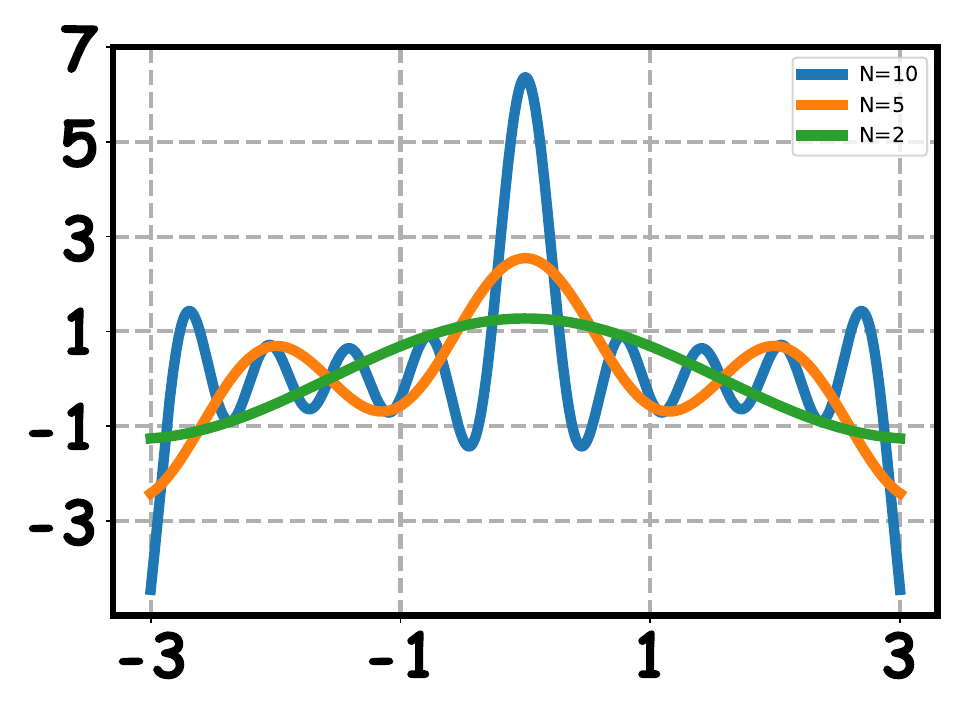}}
\caption{Illustrations of $\rm{sign}(\cdot)$ function with three approximate functions and their derivatives.}
\label{fig:sign_approximation}
\end{figure*}

\textbf{Backward Approximation.} Although the quantization loss is substantially reduced, the quantization stage in forward approximation methods remains independent of the training stage. To address this, some researchers \cite{Tan20HashGNN,Hansen2021VHPHD,Chen22BiGeaR,Wang22HCFRec,Chen23BGCH} propose an end-to-end discrete representation learning framework that optimizes over both real-valued representations and hash codes. The algorithmic framework is summarized in Algorithm \ref{alg:backward}. Specifically, it contains a hash layer ($\rm{sign}(\cdot)$ function) to generate hash codes for users and items in the forward propagation. However, the gradient of $\rm{sign}(\cdot)$ is zero for all nonzero values, which makes standard back-propagation infeasible. To address this challenging problem, a novel discrete optimization strategy based on Straight Through Estimator (STE) \cite{Oord17STE,Shen18STE} is adopted, whose core idea is to directly copy the modified gradients to the corresponding real-valued variables. The alternative gradients that are widely used are as follows:
\begin{align}
&\textbf{Tanh-alike Gradient Estimation \cite{Qin20Tanh,Gong19Tanh}:} \qquad\frac{\partial \rm{sign(x)}}{\partial x} \dot{=} \frac{\partial \rm{tanh}(\beta x)}{\partial x}, \\
&\textbf{SignSwish-alike Gradient Estimation \cite{Darabi18SignSwish}:} \qquad\frac{\partial \rm{sign(x)}}{\partial x} \dot{=} \frac{\partial 2\sigma(\beta x)(1+\beta x(1-\sigma(\beta x))) - 1}{\partial x}, \\
&\textbf{Fourier Serialized Gradient Estimation \cite{Xu21Fourier}:} \qquad\frac{\partial \rm{sign(x)}}{\partial x} \dot{=} \frac{\partial \frac{4}{\pi}\sum_{i=1,3,5}^N\frac{1}{i}\sin(\frac{\pi ix}{H})}{\partial x}, 
\end{align}
where $H$ is the length of the periodical square wave function, and $N$ is the number of terms. Intuitively, we present their original functions alongside gradient functions, as depicted in Fig. \ref{fig:sign_approximation}. It is obvious that such an approach avoids the non-differentiability of $\rm{sign}(\cdot)$ function, and combines the optimization stage and quantization stage, which achieves better performance.

\textbf{Pros and cons of the two-stage HashRec.} The two-stage optimization strategy can be seamlessly integrated into existing learning objectives with remarkable flexibility. However, despite its widespread applicability, the quantization loss between real-valued representations and the corresponding hash codes is difficult to control. 
Therefore, minimizing the quantization loss becomes a crucial aspect of implementing the two-stage optimization strategy.

\subsubsection{The One-stage HashRec}\label{subsubsec:one-stage} The one-stage optimization strategy directly tackles the challenging discrete optimization problem, thereby avoiding the large quantization loss induced by the two-stage methods. 
Existing one-stage method can be divided into DCD-style \cite{Shen2015Supervised} and ALM-style \cite{Lin10ALM,Murty07ALM}, whose algorithmic frameworks are summarized in Algorithm \ref{alg:dcd&alm}. Then, we will elaborate them in detail by taking MSE loss with balanced and de-correlated constraints as an example.

\begin{algorithm}[!h]
\caption{One-stage HashRec}
\label{alg:dcd&alm}
\LinesNumbered 
Initialize $\mathbf{B}$ and $\mathbf{D}$; \\
\Repeat{converge}{
Optimize learning objective regarding $\mathbf{B}$ and $\mathbf{D}$ via DCD/ALM methods;
}
$\textbf{return}\ \mathbf{B}, \mathbf{D}$
\end{algorithm}

\textbf{DCD-Style.} The core idea of DCD is to update hash codes bit by bit. Due to the existence of balance and decorrelation constraints, solving the optimization problem directly is infeasible. Here, we introduce two auxiliary real-valued variables $\mathbf{X}\in \mathcal{B}$ and $\mathbf{H}\in \mathcal{D}$, where $\mathcal{B} = \{\mathbf{X}\in \mathbb{R}^{f\times m}|\mathbf{X1}=0, \mathbf{X}\mathbf{X}^{\mathsf{T}}=m\mathbf{I}\}$ and $\mathcal{D} = \{\mathbf{H}\in \mathbb{R}^{f\times n}|\mathbf{H1}=0, \mathbf{H}\mathbf{H}^{\mathsf{T}}=m\mathbf{I}\}$. Then, the two constraints can be softened by $d(\mathbf{B}, \mathcal{B})=\min_{\mathbf{X}\in \mathcal{B}}||\mathbf{B}-\mathbf{X}||_F$ and $d(\mathbf{D}, \mathcal{D})=\min_{\mathbf{H}\in \mathcal{D}}||\mathbf{D}-\mathbf{H}||_F$, respectively. Therefore, the optimization problem is formulated as:
\begin{align}\label{Eq:relaxed_MSE}
&\mathcal{L} = \frac{1}{|\Omega|}\sum_{(u,i)\in \Omega} (y_{ui} - \frac{1}{2}-\frac{1}{2f}\mathbf{b}_u^{\mathsf{T}}\mathbf{d}_i)^2 + \alpha d^2(\mathbf{B}, \mathcal{B}) + \beta d^2(\mathbf{D}, \mathcal{D}), \nonumber\\
&\quad =\frac{1}{|\Omega|}\sum_{(u,i)\in \Omega} (y_{ui} - \frac{1}{2}-\frac{1}{2f}\mathbf{b}_u^{\mathsf{T}}\mathbf{d}_i)^2 - 2\alpha tr(\mathbf{B}^{\mathsf{T}}\mathbf{X}) -2\beta tr(\mathbf{D}^{\mathsf{T}}\mathbf{H}),\nonumber\\
&s.t., \mathbf{B}\in \{-1, +1\}^{f\times m}, \mathbf{D}\in \{-1, +1\}^{f\times n},\nonumber\\
&\quad\ \mathbf{X}\mathbf{1} = 0, \mathbf{H}\mathbf{1}=0, \mathbf{X}\mathbf{X}^{\mathsf{T}}=m\mathbf{I}, \mathbf{H}\mathbf{H}^{\mathsf{T}}=n\mathbf{I},
\end{align}
where $\alpha > 0$ and $\beta > 0$ are tuning parameters. Then, the $\mathbf{B}$ and $\mathbf{D}$ can be updated by parallel discrete optimization. $\mathbf{X}$ and $\mathbf{H}$ can be updated by small-scale Singular Value Decomposition (SVD) \cite{Liu14SVD}. Since the solution of non-binary variable (i.e., $\mathbf{X}$ and $\mathbf{H}$) is beyond the scope of this paper, so we omit them here. For more detail, please refer to \cite{Zhang2016DCF}. 
For the $\mathbf{B}$ subproblem, we can update $\mathbf{B}$ by updating $\mathbf{b}_u$ in parallel since Eq. (\ref{Eq:relaxed_MSE}) is based on summing over independent users:
\begin{align}\label{Eq:B-sub}
&\min\limits_{\mathbf{b}_u} \sum_{i\in \Omega_u} \frac{1}{4f^2}(\mathbf{d}_i^{\mathsf{T}} \mathbf{b}_u)^2 - \frac{1}{f}(y_{ui}-\frac{1}{2})\mathbf{d}_i^{\mathsf{T}}\mathbf{b}_u-2\alpha \mathbf{x}_u^{\mathsf{T}} \mathbf{b}_u, \nonumber\\
&s.t., \mathbf{b}_u \in \{\pm 1\}^f.
\end{align}
Due to the discrete constraints, the problem in Eq. (\ref{Eq:B-sub}) is generally NP-hard. Then, the DCD optimization algorithm is adapted to update $\mathbf{b}_u$ bit-by-bit \cite{Shen2015Supervised}. Denote $\mathbf{b}_u = [b_{uk}, \mathbf{b}_{u\bar{k}}]$ and $\mathbf{d}_i=[d_{ik}, \mathbf{d}_{i\bar{k}}]$, where $b_{uk}$ and $d_{ik}$ represent the $k$-th bit of $\mathbf{b}_u$ and $\mathbf{d}_i$, respectively. $\mathbf{b}_{u\bar{k}}$ and $\mathbf{d}_{i\bar{k}}$ are the rest hash codes excluding $b_{uk}$ and $d_{ik}$. The quadratic term in Eq. (\ref{Eq:B-sub}) w.r.t $b_{uk}$ can be written as:
\begin{align}\label{Eq:B_square}
&\sum\limits_{i\in \Omega_u}\frac{1}{4f^2}(\mathbf{d}_i^{\mathsf{T}}\mathbf{b}_u)^2 = \frac{1}{4f^2}\sum_{i\in \Omega_u}([d_{ik}, \mathbf{d}_{i\bar{k}}]^{\mathsf{T}}[b_{uk}, \mathbf{b}_{u\bar{k}}])^2=\frac{1}{4f^2}\sum\limits_{i\in \Omega_u}(d_{ik}b_{uk} + \mathbf{d}_{i\bar{k}}^{\mathsf{T}}\mathbf{b}_{u\bar{k}})^2\nonumber\\
&=\frac{1}{4f^2}\sum\limits_{i\in \Omega_u}[(d_{ik}b_{uk})^2 + (\mathbf{d}_{i\bar{k}}^{\mathsf{T}}\mathbf{b}_{u\bar{k}})^2 + 2b_{uk}\mathbf{d}_{i\bar{k}}^{\mathsf{T}}\mathbf{b}_{u\bar{k}}d_{ik}]\nonumber\\
&=\underbrace{\frac{1}{4f^2}\sum_{i\in \Omega_u}[(d_{ik}b_{uk})^2 + (\mathbf{d}_{i\bar{k}}^{\mathsf{T}}\mathbf{b}_{u\bar{k}})^2]}_{constant}+\frac{1}{2f^2}\sum\limits_{i\in \Omega_u}b_{uk}\mathbf{d}_{i\bar{k}}^{\mathsf{T}}\mathbf{b}_{u\bar{k}}d_{ik}.
\end{align}

It can be observed that the term $(d_{ik}b_{uk})^2$ is a constant (i.e., 1) since $d_{ik} \in \{-1, 1\}$ and $b_{uk} = \{-1, 1\}$. Besides, the term $(\mathbf{d}_{i\bar{k}}^{\mathsf{T}}\mathbf{b}_{u\bar{k}})^2$ can also be ignored since it does not contain the variable $b_{uk}$. And the rest terms w.r.t. $b_{uk}$ can be written as:
\begin{align}\label{Eq:B_rest}
&\sum\limits_{i\in \Omega_u}\frac{1}{f}(y_{ui} - \frac{1}{2})\mathbf{d}_i^{\mathsf{T}}\mathbf{b}_u +
2\alpha\mathbf{x}_u^{\mathsf{T}}\mathbf{b}_u=\sum\limits_{i\in \Omega_u}\frac{1}{f}(y_{ui} - \frac{1}{2})b_{uk}d_{ik} +2\alpha b_{uk}x_{uk}.
\end{align}

By combining Eq. (\ref{Eq:B_square}) and Eq. (\ref{Eq:B_rest}) and omitting the constant terms, we can get a bit-wise minimization problem:
\begin{align}\label{Eq:bit_min}
&\mathop{\min}_{b_{uk}} \hat{b}_{uk}b_{uk}, \nonumber\\
&s.t.~~ b_{uk} \in \{\pm 1\}
\end{align}
where $\hat{b}_{uk}=\sum_{i\in \Omega_u}\frac{1}{f}(y_{ui} - \frac{1}{2} - \frac{1}{2f}\mathbf{d}_{i\bar{k}}^{\mathsf{T}}\mathbf{b}_{u\bar{k}})d_{ik} + 2\alpha x_{uk}=\sum\limits_{i\in \Omega_u}\frac{1}{f}(y_{ui} - \frac{1}{2} - \frac{1}{2f}\mathbf{d}_i^{\mathsf{T}}\mathbf{b}_u)d_{ik}+\frac{1}{2f^2}b_{uk} + 2\alpha x_{uk}.$
It is obvious that the optimal of $b_{uk}$ is equal to $\mathrm{sign}(\hat{b}_{uk})$. Therefore, the update rule of $b_{uk}$ is $b_{uk} = \mathrm{sign}(\mathrm{K}(\hat{b}_{uk}, b_{uk}))$,
where $\mathrm{K}(\hat{b}_{uk}, b_{uk})$ is a function that $\mathrm{K}(\hat{b}_{uk}, b_{uk})=\hat{b}_{uk}$ if $\hat{b}_{uk} \neq 0$ and $\mathrm{K}(\hat{b}_{uk}, b_{uk}) = b_{uk}$ otherwise.
For $\mathbf{D}$ subproblem, which is similar with $\mathbf{B}$ subproblem, we can update $\mathbf{D}$ by updating $\mathbf{d}_i$ in parallel according to:
\begin{align}\label{Eq:D-sub}
&\mathop{\min}_{\mathbf{d}_i}\sum\limits_{u\in \Omega_i}\frac{1}{4f^2}(\mathbf{b}_u^{\mathsf{T}}\mathbf{d}_i)^2-\frac{1}{f}(y_{ui}-\frac{1}{2})\mathbf{b}_u^{\mathsf{T}}\mathbf{d}_i-2\beta\mathbf{h}_i^{\mathsf{T}}\mathbf{d}_i,\nonumber\\
&s.t.~~ \mathbf{d}_i\in \lbrace\pm 1\rbrace^f.
\end{align}
Denote $d_{ik}$ as the $k$-th bit of $\mathbf{d}_i$ and $\mathbf{d}_{i\bar{k}}$ as the rest hash codes excluding $d_{ik}$, the update rule for $d_{ik}$ is $d_{ik} = \mathrm{sign}(K(\hat{d}_{ik}, d_{ik}))$,
where $\hat{d}_{ik}=\sum\limits_{u\in \Omega_i}\frac{1}{f}(y_{ui}-\frac{1}{2}-\frac{1}{2f}\mathbf{b}_{u\bar{k}}^{\mathsf{T}}\mathbf{d}_{i\bar{k}})b_{uk}+2\beta h_{ik} =\sum\limits_{u\in \Omega_i}\frac{1}{f}(y_{ui}-\frac{1}{2}-\frac{1}{2f}\mathbf{b}_u^{\mathsf{T}}\mathbf{d}_i)b_{uk}+\frac{1}{2f^2}d_{ik}+2\beta h_{ik}$
The detailed derivation is similar to the method used in subproblem $\mathbf{B}$ and is therefore omitted for brevity.



\textbf{ALM-Style.} Although the DCD-style optimization algorithm alleviates the quantization loss problem caused by the two-stage optimization strategy, it learns hash codes bit-by-bit, which remains time-consuming. To address this, Xu et al. \cite{Xu20MFDCF} proposed directly learning the discrete hash codes with fast optimization with the help of ALM. According to \cite{Xu20MFDCF}, we convert the problem in Eq. (\ref{Eq:MSE}) into the matrix form and omit the balanced and decorrelation constraints:
\begin{align}\label{Eq:MSE_matrix}
&\mathcal{L} = ||\mathbf{Y}-\frac{1}{2}-\frac{1}{2f}\mathbf{B}^{\mathsf{T}}\mathbf{D}||_F^2, \nonumber\\
&s.t., \mathbf{B}\in \{-1, +1\}^{f\times m}, \mathbf{D}\in \{-1, +1\}^{f\times n}.
\end{align}
We can calculate the derivative of learning objectives with respect to $\mathbf{B}$ and $\mathbf{D}$ respectively. By setting them to zero, the closed solutions of $\mathbf{B}$ and $\mathbf{D}$ can be obtained by $\mathbf{B} = \rm{sign}\Big(\emph{f}\cdot(\mathbf{D}^{\mathsf{T}})^{-1}(2\mathbf{Y}^{\mathsf{T}}-1)\Big)$ and $\mathbf{D} = \rm{sign}\Big(\emph{f}\cdot(\mathbf{B}^{\mathsf{T}})^{-1}(2\mathbf{Y}-1)\Big)$.

\textbf{Pros and cons of the one-stage HashRec.} Compared with the two-stage optimization strategy, the one-stage optimization strategy achieves lower quantization loss by directly addressing the challenging discrete optimization problem. As a result, it produces hash codes of higher quality and improve recommendation performance. Despite its effectiveness in solving discrete optimization, the applicability of the one-stage optimization strategy is limited by the learning objective. According to \cite{Zhang2016DCF}, the DCD-style algorithm can only solve the BQP problem and it is ineffective for non-BQP problem. Regrading efficiency, the DCD-style algorithm suffers from low training efficiency due to its bit-wise update strategy. In contrast, although the ALM-style optimization algorithm can significantly accelerates training, it incurs a higher computational burden because its solutions require matrix inverse calculation. 

\begin{algorithm}[!h]
\caption{Proximal One-stage HashRec}
\label{alg:proximal}
\LinesNumbered 
Relax the discrete constraints on $\mathbf{B}$ and $\mathbf{D}$ into its convex hull; \\
\Repeat{converge}{
Optimize learning objective regarding $\mathbf{B}$ and $\mathbf{D}$ via GNCCP method;
}
$\textbf{return}\ \mathbf{B}, \mathbf{D}$
\end{algorithm}

\subsubsection{The Proximal One-stage HashRec} The proximal one-stage was first proposed in \cite{Luo22DLPR,Luo24DLFM}, which not only reduces quantization loss, but also supports a variety of learning objectives. The core idea is to first relax the problem and then gradually transform the original problem into a concave function. The algorithmic framework is given in Algorithm \ref{alg:proximal}. Solving the problem in a gradual way is motivated by the fact that the global optimal solution of a concave problem is essentially a combinatorial problem \cite{ZaslavskiyBV09}. Based on the theorem that the optimal solution to a concave problem lies in the boundary of a concave set, the proximal one-stage algorithm can directly obtain discrete solution within the continuous vector space. It is noteworthy that the starting point is important to the concave problem. Since the proximal one-stage optimization strategy can be adapted to most learning objectives, we generalize the learning objective as $\mathcal{L}$ without the detailed formulation. Specifically, we omit the balanced and decorrelated constraints since they are not our focus. Therefore, we simplify the learning objective as follows:
\begin{align}\label{Eq:general_L}
&\min_{\mathbf{B}, \mathbf{D}} \mathcal{L}(\mathbf{B}, \mathbf{D}), \nonumber\\
& s.t., \mathbf{B}\in\{-1, +1\}^{f\times m}, \mathbf{D}\in \{-1, +1\}^{f\times n}.
\end{align}
Then, the $\mathbf{B}$ and $\mathbf{D}$ subproblems can be alternatively solved. Specifically, for $\mathbf{B}$ subproblem, the optimization problem with fixed $\mathbf{D}$ is as follows:
\begin{align}\label{Eq:proximal_b_sub}
&\min_{\mathbf{B}} \mathcal{L}(\mathbf{B}), \nonumber\\
&s.t., \mathbf{B}\in \{-1, +1\}^{f\times m}. 
\end{align}
To solve the problem in Eq. (\ref{Eq:proximal_b_sub}), Graduated NonConvexity and Concavity Procedure (GNCCP) \cite{Liu14GNCCP} is introduced. Its core idea is to construct an approximation of the concave problem, beginning with a convex problem, whose optimal solution serve as the starting point for the concave problem. Concretely, the problem in Eq. (\ref{Eq:proximal_b_sub}) can be transformed as:
\begin{align}\label{Eq:b_zeta}
&\mathop{\arg\min}_{\mathbf{B}}\mathcal{L}_{\zeta}(\mathbf{B})=\left\{
             \begin{array}{ll}
             (1-\zeta)\mathcal{L}(\mathbf{B})+\zeta tr(\mathbf{B}^{\mathsf{T}}\mathbf{B}),  1 \geq \zeta \geq 0  \\
             \\
            (1+\zeta)\mathcal{L}(\mathbf{B})+\zeta tr(\mathbf{B}^{\mathsf{T}}\mathbf{B}),  0 \geq \zeta \geq -1 \nonumber\\
             \end{array}
            \right. \\
            &s.t.\, \mathbf{B} \in [-1, 1]^{f\times m},
\end{align}
where $\zeta$ gradually decreases from 1 to -1 at the descent interval $d$, implying that $\mathcal{L}_{\zeta}(\mathbf{B})$ gradually becomes from convex ($\mathcal{L}_1(\mathbf{B})=tr(\mathbf{B}^{\mathsf{T}}\mathbf{B})$) to non-convex-concave ($\mathcal{L}_0(\mathbf{B})=\mathcal{L}(\mathbf{B})$), and finally to concave ($\mathcal{L}_{-1}(\mathbf{B})=-tr(\mathbf{B}^{\mathsf{T}}\mathbf{B})$). For each currently fixed $\zeta$, $\mathcal{L}_{\zeta}(\mathbf{B})$ can be minimized by Frank-Wolfe algorithm \cite{FW1956}, using the optimal solution of $\mathcal{L}_{\zeta+d}(\mathbf{B})$ as starting point. When $\zeta = 1$, the starting point is randomly given. By repeating this iterative process, the path of solutions $\{\mathbf{B}^{*}_{\zeta}\}$ can be generated where $\mathbf{B}^{*}_{\zeta}$ is the optimal solution  of  $\mathcal{L}_{\zeta}(\mathbf{B})$ for each $\zeta \in [-1, 1]$. Finally, $\mathbf{B}^{*}_{\zeta^{*}}\in \{-1, 1\}^{f \times n}$ is the approximate solution of Eq. (\ref{Eq:proximal_b_sub}), where $\zeta^*$ makes the objective function in Eq. (\ref{Eq:proximal_b_sub}) become concave. 
Specifically, the optimal solution $\mathbf{B}^{*}_{\zeta^{*}}$ to Eq. (\ref{Eq:b_zeta}) is approximately the same as the optimal solution $\mathbf{B}^{*}$ to Eq. (\ref{Eq:general_L}), which is guaranteed by \cite{Luo22DLPR}.
The detailed optimization procedure of $\mathbf{D}$ subproblem is similar to that of $\mathbf{B}$ subproblem. For more details, please refer to \cite{Luo22DLPR,Luo24DLFM}.

\begin{table*}[!b]
\setlength{\tabcolsep}{8pt}
\tiny
  \centering
  \caption{Illustration of two-stage, one-stage and proximal one-stage HashRec methods.}
    \begin{tabular}{c|c|c|c|c}
    \toprule
    \textbf{Methods} & \textbf{Basic Idea} & \textbf{Advantages} & \textbf{Challenges} & \textbf{Related Work} \\
    \midrule
    \makecell[c]{Two-stage HashRec}      & \makecell[c]{Relaxation followed\\ by quantization}      & \makecell[c]{Easy-to-deploy; \\Suitable for most\\ learning objectives}    & High quantization loss & \makecell[c]{\cite{Zhou2012BCCF},\cite{Liu14CH},\cite{Zhang2014PPH},\\\cite{Kang19CIGAR},\cite{Lu19FHN},\cite{Wang19DGCNBinCF},\\\cite{Wang19ABinCF},\cite{Hansen20NeuHash-CF},\cite{Tan20HashGNN},\\\cite{Zhang20CGH},\cite{Hansen2021VHPHD},\cite{Chen22BiGeaR},\\\cite{Guan22BIHGH},\cite{Wang22HCFRec},\cite{Zhu23EDCF},\\\cite{Chen23BGCH},\cite{Xu24TSGNH},\cite{Chen24BGCH+},\\\cite{Zhang16NQ},\cite{Hu23DLACF},\cite{Lu23FHN},\\\cite{Chen25BiGeaR-SS}} \\
    \midrule
    \makecell[c]{One-stage HashRec}      & \makecell[c]{Directly solve the \\discrete optimization problem}   & Low quantization loss   & \makecell[c]{Only suitable for \\ BQP problems} & \makecell[c]{\cite{Zhang2016DCF},\cite{Lian17DCMF},\cite{Zhang17DPR},\\\cite{Liu18DFM},\cite{Liu18DSFCH},\cite{Zhang18DRMF},\\\cite{Zhang2018DDL},\cite{Guo2019DTMF},\cite{Liu19CCCF},\\\cite{Liu2019DSR},\cite{Wu2020SDMF},\cite{Xu20MFDCF},\\\cite{Zhang22DPH},\cite{Lian2021DMF},\cite{Liu21DLCF},\\\cite{Luo21SDSR},\cite{Xu23MDCF},\cite{Zhang23LightFR},\\\cite{Yang24DFMR}} \\
    \midrule
    \makecell[c]{Proximal one-stage HashRec}      & \makecell[c]{Relaxation and solve the\\ relaxed optimization problem\\ to obtain hash codes}     & \makecell[c]{Low quantization loss; \\Suitable for most\\ learning objectives}  & Training efficiency  & \makecell[c]{\cite{Luo22DLPR},\cite{Luo24DLFM}} \\
    \bottomrule
    \end{tabular}%
  \label{tab:divided_by_optimization_strategy}%
\end{table*}%

\textbf{Pros and cons of the proximal one-stage HashRec.} Compared with the one-stage optimization strategy, the proximal one-stage optimization strategy can be applied to most learning objectives, whereas the one-stage optimization strategy is limited to solving the BQP problems. In addition, the proximal one-stage optimization strategy can obtain hash codes directly in the continuous vector space without any extra quantization, thereby dramatically reducing the quantization loss dramatically and achieving better performance than the two-stage optimization strategy. However, training efficiency is heavily influenced by the descent interval. Concretely, the small descend interval will lead to lower training efficiency while generating subtle optimization path, and the large descend interval corresponds to higher training efficiency and coarse optimization path. Therefore, how to improve the efficiency of the proximal one-stage optimization strategy ensuring the effectiveness is an important research direction.

\subsubsection{Summary}
In this section, we divide existing HashRec methods into three types: two-stage HashRec, one-stage HashRec, and proximal one-stage HashRec. Specifically, the comprehensive examination of their basic idea, advantages, challenges, and related work is summarized in Table \ref{tab:divided_by_optimization_strategy}. In essence, the two-stage optimization strategy is easy to deploy, and is not limited by the learning objective. However, it incurs higher quantization loss due to the decoupling of optimization and quantization process. Conversely, one-stage methods substantially address this issue by directly tackling the discrete optimization problem, making them popular among researchers. Yet, this approach is constrained to solving BQP problems, limiting its applicability. In contrast, proximal one-stage methods successfully mitigate quantization loss while accommodating diverse learning objectives. Nevertheless, their training efficiency is notably affected by the descending step. Consequently, there is an urgent need for an optimization strategy that combines efficiency, effectiveness, and broad applicability.

\subsection{Recommendation Scenario}\label{sec:recommendation_task}
In addition to the classical user-item CF, the researchers have started to utilize L2H method for imporving the inference efficiency and computational burden of other recommendation scenarios, such as cold-start recommendation, social recommendation, outfit recommendation, explainable recommendation, and federated recommendation. 

\subsubsection{Cold-start HashRec}
Cold-start recommendation, a critical challenge in modern RS, addresses the problem of generating personalized suggestions for new users with sparse or no interaction data. Existing cold-start mitigation strategies often integrate content information with the help of knowledge transfer methods. In such a scenario, the computational overhead and inference efficiency become the urgent issues. 
\textbf{DCMF} \cite{Lian17DCMF}, \textbf{DMF} \cite{Lian2021DMF}, \textbf{DDL} \cite{Zhang2018DDL}, \textbf{DPH} \cite{Zhang22DPH} and \textbf{CGH} \cite{Zhang20CGH} derive compact yet informative hash codes in the presence of user/item content information. In general, the learning objective consists of three components: the preference reconstruction loss $\mathcal{L}_{Rec}$, the information align term between hash codes and real-valued representations $\mathcal{L}_{IA}$, and the content learning loss $\mathcal{L}_{content}$.
DCMF and CGH utilize off-the-shelf user and item content feature, and introduce two feature transformation matrices to construct the information alignment term between users'/items' hash codes and their real-valued representations. 
DMF only utilizes items' content features. Therefore, its learning objective only contains this term $||\mathbf{D}-\mathbf{W}_2\mathbf{Q}_c^{\mathsf{T}}||_F^2$. DDL and DPH adopt an end-to-end learning scheme by training the MF model and content learning network to obtain high-quality hash codes. 
Differently, \textbf{NeuHash-CF} \cite{Hansen20NeuHash-CF} method obtains items' hash codes directly from item content rather than item ID. \textbf{MFDCF} \cite{Xu20MFDCF} and \textbf{MDCF} \cite{Xu23MDCF} methods construct multiple information alignment terms to integrate multiple content features into hash codes, and develop an efficient discrete optimization strategy based on augmented lagrangian multiplier to directly learn the user and item hash codes with simple efficient operations.

\subsubsection{Social HashRec}
Social recommendation methods have received considerable attention recently for their abilities to capture the influence among connected users. However, the large volume of user/item latent features leads to high storage and computation cost, particularly on terminal user devices with limited resources. Consequently, when incorporating additional social information, precisely extracting the relevant items for a given user from massive candidate sets becomes even more time- and memory-intensive, posing significant challenges for efficient and accurate recommendations. 
\textbf{DTMF} \cite{Guo2019DTMF} and \textbf{DSR} \cite{Liu2019DSR} decompose the user-item rating matrix and user-user social matrix jointly, and share the user hash codes. The DCD optimization algorithm is adopted to obtain hash codes.
\textbf{SDSR} \cite{Luo21SDSR} combines discrete MF model and real-valued network embedding model to achieve social recommendation. This method considers the information propagation from each user to his/her multi-hop neighbors, and avoids extra encoding loss originating from the discrete constraints on social embedding, which is not involved in preference prediction. 
\textbf{DLACF} \cite{Hu23DLACF} incorporates limited attention into discrete optimization framework to improve the accuracy with less information loss. 
\textbf{TSGNH} \cite{Xu24TSGNH} constructs a temporal social graph network to incorporate social information, and proposes a dynamic-adaptive aggregation method to capture long-term and short-term dynamic characters of users and items. Take one user as an example, this method first obtains the neighborhood item's opinion-aware representation $\mathbf{h}_i$ by fusing item's embedding $\mathbf{e}_i$ and rating embedding $\mathbf{e}_y$: $\mathbf{h}_i=MLP_u([\mathbf{e}_i||\mathbf{e}_y])$. To obtain user $u$'s long-term habit, all his neighbor item embeddings are aggregated: $\mathbf{e}_u^L=\sum_{i\in\mathcal{I}_u}\alpha_{ui}\mathbf{h}_i$, where $\alpha_{ui}$ is the attentive score. To obtain user $u$'s short-term habit, the latest interaction item embeddings are aggregated: $\mathbf{e}_u^s=\sum_{i\in\mathcal{I}_u}\hat{\alpha}_{ui}\mathbf{h}_i$, where $\hat{\alpha}_{ui}$ is the attention coefficient. Finally, the user $u$'s embedding can be formulated as the aggregation of long-term habit embedding, short-term interest embedding, and user embedding: $\mathbf{e}_u=\tanh(\mathbf{W}\cdot [\mathbf{e}_u^L || \mathbf{e}_u^s || \mathbf{e}_u]+b)$, where $\mathbf{W}$ and $\mathbf{b}$ are weight and bias of neural networks. Similarly, the item $i$'s embedding $\mathbf{e}_i$ can be obtained in the same way. 
In addition, the balanced and de-correlated constraints are considered as the regularization terms to construct the final loss.

\subsubsection{Outfit HashRec}
Outfit recommendation aims to recommend fashion outfits for each user, consisting of multiple interacted clothing and accessories. However, the number of possible outfits grows exponentially with the number of items in each garment category. Therefore, the storage complexity and the retrieval efficiency of the outfits are essential for the deployment of a fashion recommendation system in practice. 
\textbf{DSFCH} \cite{Liu18DSFCH} learns the hash codes of clothing items from their visual content features and the matching matrix constructed based on expertise knowledge. In addition, the quality of hash codes is kept by the regularization between hash codes and real-valued visual representations. 
\textbf{FHN} \cite{Lu19FHN} captures the preferences of different users and learns the compatibility between fashion items. It consists of three components: a feature network that first extracts content features; a set of type-dependent hashing modules that convert the features and user taste representations into hash codes; and a matching block that computes the overall preference of a user for an outfit through pairwise weighted hashing. 
To solve the discrete optimization problem, FHN adopts $\tanh(\beta x)$ function to approximate the $\mathrm{sign}(x)$ function during the training stage. 
\textbf{FHN+} \cite{Lu23FHN} is an extensive version of \textbf{FHN}, which proposed a deep hashing approach by treating the hash codes as samples of underlying distributions, providing a controllable trade-off between efficiency and accuracy.
\textbf{BIHGH} \cite{Guan22BIHGH} presents a heterogeneous graph learning-based outfit recommendation scheme, where the four types of entities (i.e., users, outfits, items, and attributes) and their relations are seamlessly integrated. This scheme consists of three key components: heterogeneous graph node initialization, bi-directional sequential graph convolution, and hash code learning. In the heterogeneous graph node initialization module, it utilizes the pre-trained ResNet \cite{He16RESNET} and pre-trained BERT \cite{DevlinCLT19BERT} models to obtain the $i$-th item's visual embedding $\mathbf{e}_{v_i}\in \mathbb{R}^d$ and textual embedding $\mathbf{e}_{t_i}\in \mathbb{R}^d$. The visual and textual embeddings are then concatenated, and fed into a learnable fully-connected layer to derive the final item embedding $\mathbf{e}_{m_i}\in \mathbb{R}^d$. To maintain semantic consistency between the visual and textual embeddings of the same item, the hinge loss is adopted.
To mine the semantic content of each attribute, the pre-trained BERT model with a learnable fully-connected layer is utilized to obtain $i$-th item's attribute embedding $\mathbf{e}_{a_i}\in \mathbb{R}^d$. The embedding of each outfit $\mathbf{e}_{o_i}\in \mathbb{R}^d$  is obtained by aggregating all the embeddings of its composing items. Analogous to the outfit, the $i$-th user's embedding $\mathbf{e}_{v_i}\in \mathbb{R}^d$ is obtained by fusing all the embeddings of his/her interacted and preferred outfits. In the bi-directional sequential graph convolution module, the entire graph is decomposed into three subgraphs. Graph convolution operations are then applied in both upward and downward directions to obtain the entity embeddings. 

\subsubsection{Explainable HashRec}
Explainable recommendation methods not only provide users with recommendation results, but also generate textual explanations to clarify why certain items are recommended. Since these methods involve text generation tasks, they incur large computational burden. 
\textbf{EDCF} \cite{Zhu23EDCF} preserves the user-item interaction features and semantic text features in hash codes by adaptively exploiting the correlations between the preference prediction task and the explanation generation task. In the preference prediction task, user and item interaction features are obtained using SVD methods. Review features for users and items are extracted via the TextCNN \cite{Kim14TextCNN} method. The hash codes for user/item are then generated by concatenating the interaction and review features and feeding them into Multi-Layer Perception (MLP) with a tanh activation layer. In the explanation generation task, a Long Short-Term Memory (LSTM) \cite{LSTM1997} is employed to produce the textual explanation. 



\subsubsection{Federated HashRec}
Federated recommendation methods have received widespread attention, due to their advantages in privacy protection and strong performance. In these methods, a global model on the server is aggregated and updated from user-specific local models through collaboration between the server and clients, ensuring that users' private interaction data never leaves their devices. However, such a scheme in federated settings ignores the limited capacities in resource-constrained user devices (i.e., storage space, computational overhead). To address the efficiency concern, some researchers propose to integrate L2H techniques into federated recommendation methods.
\textbf{LightFR} \cite{Zhang23LightFR} generates high-quality hash codes by leveraging L2H techniques in a federated settings, and devised an efficient federated discrete optimization algorithm to collaboratively train model parameters between the server and clients, effectively preventing real-valued gradient attacks from malicious parties. Its learning objective is the same as that of DCF \cite{Zhang2016DCF}. Unlike DCF, LightFR proposed two global discrete aggregation methods to update the item hash codes, using gradients uploaded from a subset of clients to the central server. 
In addition to aggregating gradients from clients, this method also directly aggregates the item hash codes that have been locally updated on the clients.
\textbf{DFMR} \cite{Yang24DFMR} studies the federated multi-behavior recommendation problem under the assumption that purchase behaviors can be collected. In addition, DFMR introduces a constraint term, which lies in the summation over the missing data portion, to alleviate the data sparsity problem. Despite the effectiveness, it will bring the computational bottleneck in the update process. To this end, they design a memorization module called cache updating module, enabling some terms to be independent of the items or users, and then pre-calculating them only once in each training round. 

\begin{table}[!h]
\setlength{\tabcolsep}{11pt}
\tiny
  \centering
  \caption{Illustration of cold-start, social, outfit, explainable, and federated HashRec methods}
    \begin{tabular}{c|c|c|c}
    \toprule
    \textbf{Methods} & \textbf{Basic Idea} & \textbf{Challenges} & \textbf{Related Work} \\
    \midrule
    Cold-Start HashRec & \multirow{4}[8]{*}{\makecell[c]{By leveraging auxiliary information to \\enhance the representation capacity of hash codes, \\and simultaneously, hash codes can improve \\inference efficiency and reduce memory overhead}} & \makecell[c]{Ignoring the utilization of\\ multi-source information}      &\makecell[c]{\cite{Lian17DCMF},\cite{Lian2021DMF},\cite{Zhang2018DDL},\\\cite{Zhang22DPH},\cite{Zhang20CGH},\cite{Hansen20NeuHash-CF},\\\cite{Xu20MFDCF},\cite{Xu23MDCF}}   \\
\cmidrule{1-1}\cmidrule{3-4}    Social HashRec &       & \makecell[c]{Failing to capture dynamic character}      & \makecell[c]{\cite{Liu2019DSR},\cite{Guo2019DTMF},\cite{Luo21SDSR},\\\cite{Hu23DLACF},\cite{Xu24TSGNH}} \\
\cmidrule{1-1}\cmidrule{3-4}    Outfit HashRec &       & \makecell[c]{Lacking in-depth understanding of \\fashion trends, brand culture or\\ popular elements}      &\makecell[c]{\cite{Liu18DSFCH},\cite{Lu19FHN},\cite{Lu23FHN},\\\cite{Guan22BIHGH}}  \\
\cmidrule{1-1}\cmidrule{3-4}    Explainable HashRec &       & \makecell[c]{Ignoring the emotional consistency}      & \makecell[c]{\cite{Zhu23EDCF}} \\
    \midrule
    Federated HashRec & \makecell[c]{By utilizing L2H techniques, the recommendation\\ model can be deployed in resource-constrained\\ user devices}       & \makecell[c]{Failing to address the client drift\\ problem}      & \makecell[c]{\cite{Zhang23LightFR},\cite{Yang24DFMR}} \\
    \bottomrule
    \end{tabular}%
  \label{tab:divided_by_recommendation_task}%
\end{table}%


\subsubsection{Summary}
In this section, we divide existing HashRec methods into five types: cold-start HashRec, social HashRec, outfit HashRec, explainable HashRec, and federated HashRec. Specifically, the comprehensive examination of their points of basic idea, challenges and related work is summarized in Table \ref{tab:divided_by_recommendation_task}. Although cold-start HashRec methods address the recommendation problem without any interaction data, existing HashRec methods only utilize content features, while ignoring multi-source information. Social HashRec methods enhance the effectiveness of hash codes with the help of social connections, further improving recommendation performance. However, social networks are dynamic and existing methods fail to capture this dynamic property, leading to suboptimal performance. Outfit HashRec methods, which aim to recommend complete outfits to users, have gained popularity among consumers. Nevertheless, these methods usually make recommendations based on the attributes of the clothing items, lacking a deeper understanding of fashion trends, brand culture or popular elements. As a result, the recommended outfit lacks novelty. Explainable HashRec methods generate recommendation results along with text explanation, enhancing the credibility of the recommendations. Nevertheless, existing methods often ignore the emotional consistency between the text and results, which can lead to user dissatisfaction. Federated HashRec methods enable effective deployment. However, existing methods do not account for client drift, resulting in degraded performance on individual clients.

\section{Measurements of HashRec}\label{sec:metrics}
In this section, we introduce the effectiveness measurements and efficiency measurements, which are commonly used in existing HashRec methods.

\subsection{Effectiveness Measurements}

Performance measurements provide effective quantitative metrics for evaluating the superiority of one method over another. In this section, we introduce some widely used metrics in HashRec.
For clarity, we calculate these metrics for each user. Next, we examine each metric in detail. 



\textbf{Recall} measures the proportion of retrieved relevant items compared to the total number of relevant items. Specifically, given the positive item set $\mathcal{I}_u$ and the relevance score $\hat{y}_{ui}$, is an integer ranging from 1 to n, the ranking position $r_{ui}$ is calculated from a pairwise comparison between predicted relevance score for item $i$ and all other items:
\begin{align}
r_{ui} = 1 + \sum_{j=1\& j\neq i}^n \mathbb{I}(\hat{y}_{uj} > \hat{y}_{ui}),
\end{align}
where $\mathbb{I}(\cdot)$ denotes the indicator function. It is noteworthy that the smaller the $r_{ui}$, the higher the preference/rating score. Then, the definition of Recall at the cutoff $K$ is as follows:
\begin{align}
Recall@K = \frac{1}{|\mathcal{I}_u|}\sum_{i=1}^n y_{ui}\mathbb{I}(r_{ui} \leq K),
\end{align}
where $|\cdot|$ denotes the size of a set. It is worth noting that the value of Recall@K is in the range of [0, 1]. The closer this value is to 1, the item that the user may like appears in the top-K of the recommended list. It can be observed that Recall@K is very important in scenarios that highly focus on positive items.

\textbf{Normalized Discounted Cumulative Gain (NDCG)} is an evaluation metric that accounts for both ratings and preference, and incorporates an explicit position-based discount factor in its calculation. Formally, given the predicted relevance $\hat{y}_{ui}$ between the user $u$ and item $i$ and the ranking position $r_{ui}$ corresponding to the relevance score $\hat{y}_{ui}$, the Discounted Cumulative Gain (DCG) at the cutoff $K$ is defined as follows:
\begin{align}
DCG@K = \sum_{i=1}^n \frac{2^{y_{ui}}-1}{log_2(r_{ui}+1)}.
\end{align}
By normalizing DCG@K with its maximum possible value, the NDCG@K is proposed. That is, 
\begin{align}
NDCG@K = \frac{DCG@K}{IDCG@K} = \frac{\sum_{i=1}^n \mathbb{I}(r_{ui}<K) \frac{2^{y_{ui}}-1}{log_2(r_{ui}+1)}}{\sum_{i=1}^n \mathbb{I}(r^*_{ui}<K)\frac{2^{y_{ui}}-1}{log_2(r^*_{ui}+1)}},
\end{align}
where $r^*_{ui}$ is the ideal ranking position corresponding to the true preference score $y_{ui}$.

\textbf{Average Precision (AP)} focuses on the recommendation list quality, which is similar to NDCG. Given the positive item set $\mathcal{P}_u$ of user $u$ and the cutoff $K$, the definition of AP@K is as follows:
\begin{align}
AP@K = \frac{1}{|\mathcal{P}_u|}\sum_{i=1}^n y_{ui}\mathbb{I}(r_{ui}\leq K)P@i,
\end{align}
where $P@i$ denotes the precision of the recommendation list in terms of user $u$. $P@i$ is the proportion of positive items whose rank is higher than item $i$, and is calculated as follows:
\begin{align}
P@i = \frac{1}{r_{ui}} \sum_{j=1}^n y_{uj}\mathbb{I}(r_{uj}\leq r_{ui}).
\end{align}
The value of $AP@K$ is in the range of [0, 1]. The larger the value, the more satisfied the user is with the recommended result.

\textbf{Reciprocal Rank (RR)} is a widely used metric for evaluating ranking algorithms. Different from the NDCG and AP, the RR metric emphasizes the ranking position of the most relevant items, rather than the overall ranking quality of the recommendation list. The definition of RR is as follows:
\begin{align}\label{Eq:MRR}
RR = \sum_{i=1}^n \frac{y_{ui}}{r_{ui}} \prod_{j=1}^n (1 - y_{uj}\mathbb{I}(r_{uj} < r_{ui})).
\end{align}
From Eq. (\ref{Eq:MRR}), we can see that when the ranking position of the most relevant items is equal to 1, the RR value reaches its maximum. This indicates that the RR metric is particularly suitable for the top-1 recommendation scenario, such as sequential recommendation \cite{Boka24SeRec_Survey}. The average of RR for all users is Mean Reciprocal Rank (MRR).

\textbf{Area Under the Curve (AUC)} is a benchmark metric used to assess the performance of binary classification models. In the context of recommendation systems, it means that the probability of selecting a positive sample is higher than that of selecting a negative sample. Specifically, given the positive item set $\mathcal{P}_u$ and the negative item set $\mathcal{N}_u$ of the user $u$, the definition of AUC is as follows:
\begin{align}\label{Eq:AUC}
AUC = \frac{1}{|\mathcal{P}_u| \times |\mathcal{N}_u|} \sum_{i = 1}^{\mathcal{P}_u}\sum_{j=1}^{\mathcal{N}_{u}}\Bigg(\mathbb{I}(\hat{y}_{ui}-\hat{y}_{uj}> 0) +\frac{1}{2}\mathbb{I}(\hat{y}_{ui}-\hat{y}_{uj}= 0)\Bigg).
\end{align}
From the Eq. (\ref{Eq:AUC}), we can find that AUC metric is independent of the model's predicted absolute scores, thereby eliminating the influence of manually set thresholds on the recommendation results. In addition, AUC metric accounts for the classification performance of both positive and negative items, allowing it to provide a reliable evaluation even with imbalanced sample distributions \cite{Yang23AUC_Survey}.

\textbf{Hit Ratio (HR)} is a key metric for evaluating the effectiveness of predicting or recommending relevant items to users. It measures how often the system's recommendations include the items that users actually interact with or choose, reflecting how well the system captures user preferences. The definition of HR is as follows:
\begin{align}
HR = \frac{1}{m}\sum_{u=1}^m hits(u) \times 100\%,
\end{align}
where $hits(u)$ indicates whether the item liked by the $u$-th user is in the recommended list, with 1 representing yes and 0 representing no. Suppose that there are 100 users interacting with the system, and the items they eventually interact if found in the top-5 list for 60 users, then the HR value is: $HR@5 = \frac{60}{100}\times 100\% = 60\%$. It means the system correctly recommends at least one relevant item to 60\% of users.

\textbf{Accuracy} is a metric used to evaluate how closely the recommendations align with the actual preferences or behaviors of users. Since the model's output is typically a floating-point number, a threshold must be set to determine whether the current item is predicted to a positive or negative sample. Formally, it is defined as the percentage of all items that are predicted accurately:
\begin{align}
Accuracy = \frac{1}{|\Omega_u|}\sum_{i\in \Omega_u} \mathbb{I}(y_{ui} = \mathcal{T}_{\theta}(\hat{y}_{ui})),
\end{align}
where $\mathcal{T}_{\theta}(\hat{y}_{ui}) = \mathbb{I}(\hat{y}_{ui}-\theta)$ represents the threshold function. When the predicted value of an item exceeds the threshold, it is classified as a positive item; otherwise, it is negative.

\subsection{Efficiency Measurements}
Efficiency measurements focus on evaluating how efficiently a method can provide a recommendation list. Here, we delve into two efficiency measurements: time cost and storage cost, which are widely employed in the evaluation of HashRec techniques.

\textbf{Time Cost} is a key metric for evaluating the latency of generating recommendation results. The promptness of recommendations directly influences user satisfaction, as timely suggestions are essential for a seamless user experience. Formally, it is defined as the total time required for model inference in a production environment.

\textbf{Storage Cost} is a critical metric that determines the practical deployability of a method on resource-constrained devices. It reflects the model's efficiency in terms of memory footprint and is often measured as the total amount of memory consumed by the model parameters and necessary runtime data.


\section{Future Research Directions}\label{sec:discussion}
In this section, we outline several promising directions for future research.

\subsection{Designing a General HashRec Framework}

Based on a comprehensive review of previous research, it is evident that existing HashRec methods are generally tailored with specific optimization strategies for particular learning objectives. However, the landscape of recommendation task is diverse, encompassing a wide array of scenarios and objectives. The diversity necessitates the adoption of different learning objectives and optimization strategies to effectively address various recommendation tasks. For instance, the BCE loss is commonly used in Click-Through Rate (CTR) \cite{Zhu21CTR,Luo25RgsAUC} prediction task, where the model predicts whether a user will click on a given item or not. In contrast, the BPR loss serves as the benchmark for Top-K recommendation tasks, which aim to provide a personalized item list that users may be interested in. These examples highlight that different optimization strategies are required to meet different learning objectives, often making recommendation models highly task-specific. Therefore, there is a pressing need for a general framework for HashRec methods that can handle the diverse learning objectives. It is a promising but largely under-explored area where more studies are expected.

\subsection{Achieving Efficiency-Effectiveness Trade-off}
The trade-off between efficiency and effectiveness is particularly crucial in the recall phase of RS \cite{DNN2016YouTube,Yi19sampleingbias}. Efficiency, in this context, ensures that the recall model can swiftly identify and retrieve candidate items from a vast item corpus. Rapid processing is important for maintaining user experience and system responsiveness, especially in scenarios where real-time interactions are expected. Effectiveness, on the other hand, reflects the model's ability to select relevant and accurate items from the multitude of possibilities. It measures how well the recall model aligns with user intent and the precision of the retrieved results. High effectiveness indicates that the model retrieves items with a high degree of accuracy, thereby enhancing the overall quality and relevance of the search outcomes. Balancing these two aspects is a delicate task, as optimizing for one often comes at the expense of the other \cite{Zhang19RS_Survey}. As indicated in this survey, hash codes, which can speed up inference efficiency, contain less information compared with their real-valued counterpart, leading to suboptimal recommendation performance. Achieving an optimal equilibrium ensures that the recall model not only operates swiftly but also delivers reliable and pertinent results, ultimately satisfying user needs and expectations.


\subsection{Making Large Language Model Lightweight}
With the rapid development of artificial intelligence technology, Large Language Models (LLMs) have been integrated into RS, serving both as knowledge extractors \cite{Hou23VQSR,Hou22UniSR} and as recommenders \cite{Dai23LLMRS,Hou24LLMRS}. They improve the accuracy and relevance of recommendations, while further enhancing user satisfaction. Specifically, LLMs offer new possibilities to RS through zero/few-shot recommendation capabilities, addressing common data sparsity issues caused by limited historical interactions \cite{Sileo22ZeroShotRS}. Although LLMs consistently exhibit remarkable performance across various tasks, their exceptional capabilities come with significant challenges stemming from their extensive size and computational requirements \cite{Zhu23LLM_Compression}. For instance, the GPT-175B model with an impressive 175 billion parameters, demands a minimum of 350GB of memory in FP16 format \cite{Brown20LLM}. To this end, L2H techniques can serve as a bridge between LLMs and RS, facilitating efficient and lightweight LLMs-based RS. 

\subsection{Learning with Multi-Objective}
Traditional recommendation models typically focus on addressing one single objective, such as minimizing the prediction errors or maximizing the ranking quality. However, there are several key considerations that researchers or engineers face in real-world scenarios. For example, e-commerce platforms need to balance clicks, add-to-carts and conversions with potential advertising revenue \cite{Gu20MultiObjective,Lin19MultiObjective}. Similarly, streaming platforms encounter their own unique set of challenges, aiming to optimize both short-term and long-term engagement, with the ultimate goal of enhancing retention rates and maximizing lifetime customer value. This requires understanding user behavior over extended periods, predicting future preferences, and ensuring a seamless, personalized experience \cite{Bugliarello22MultiObjective,Mehrotra20MultiObjective}. In the recall phase where model retrieves a larger set of potentially relevant items, additional metrics beyond recall also deserve attention. For example, item diversity is paramount to prevent the user from being trapped in a filter bubble, ensuring they are exposed to a wide range of content. Ensuring diverse recommendations not only enhances user satisfaction but also aligns with broader societal values. Thus, the development of recommendation models must increasingly incorporate these multi-objective frameworks to reflect the complexities of real-world applications.

\subsection{Alleviating Bias in HashRec}
User behavioral data is observational rather than experimental \cite{Chen23Bias_Survey}, which introduces many biases, such as selection bias, exposure bias, popularity bias, and fairness bias. Blindly fitting user behavioral data without accounting for biases can be problematic. For instance, regarding popularity bias, popular items tend to be recommended even more frequently than actual popularity warrants, if the bias is not addressed. Over time, the feedback loop in RS not only introduces additional inherent biases, but also amplifies existing ones, resulting in ``the rich get richer" Matthew effect. To mitigate this issue, extensive researches are improvements \cite{Tobias16IPS,Wang19DR,Chen21AutoDebias,Ding22InterD,zhang2024uncovering,Wu24SSM}. Although these methods achieve significant performance improvements, they are primarily designed for ranking models rather than recall models. Furthermore, it is unclear whether the biases inherent in recall phase are the same as those in ranking stage. Consequently, it is crucial and meaningful to exploit the specific biases in the recall stage for accurate user preference modeling.

\section{Conclusion}\label{sec:conclusion}
In this survey, we provide a comprehensive and systematic overview of research on learning to hash for recommendation from 2012 to 2025. Our primary objective is to consolidate the vast array of existing work into a coherent and structured framework. We first propose a taxonomy based on key components, including the type of learning objective and optimization strategy, to categorize existing works. For each component, we provide detailed descriptions and discussions. Next, we discuss the evaluation metrics used in the literature to assess the performance. Finally, we outline promising research directions and open issues to inspire some future studies in this area. We hope this survey provides researchers in academia and industry with a comprehensive understanding of this promising yet often overlooked field and offer insights for potential future research.

\begin{acks}
The authors would like to thank the anonymous reviewers for their constructive comments.
\end{acks}

\bibliographystyle{ACM-Reference-Format}
\bibliography{sample-base-simplified}


\begin{thebibliography}{144}


\ifx \showCODEN    \undefined \def \showCODEN     #1{\unskip}     \fi
\ifx \showISBNx    \undefined \def \showISBNx     #1{\unskip}     \fi
\ifx \showISBNxiii \undefined \def \showISBNxiii  #1{\unskip}     \fi
\ifx \showISSN     \undefined \def \showISSN      #1{\unskip}     \fi
\ifx \showLCCN     \undefined \def \showLCCN      #1{\unskip}     \fi
\ifx \shownote     \undefined \def \shownote      #1{#1}          \fi
\ifx \showarticletitle \undefined \def \showarticletitle #1{#1}   \fi
\ifx \showURL      \undefined \def \showURL       {\relax}        \fi
\providecommand\bibfield[2]{#2}
\providecommand\bibinfo[2]{#2}
\providecommand\natexlab[1]{#1}
\providecommand\showeprint[2][]{arXiv:#2}

\bibitem[Adomavicius and Tuzhilin(2005)]%
        {Adomavicius05CF}
\bibfield{author}{\bibinfo{person}{Gediminas Adomavicius} {and}
  \bibinfo{person}{Alexander Tuzhilin}.} \bibinfo{year}{2005}\natexlab{}.
\newblock \showarticletitle{Toward the Next Generation of Recommender Systems:
  {A} Survey of the State-of-the-Art and Possible Extensions}.
\newblock \bibinfo{journal}{\emph{IEEE Transactions on Knowledge and Data
  Engineering}} \bibinfo{volume}{17}, \bibinfo{number}{6}
  (\bibinfo{year}{2005}), \bibinfo{pages}{734--749}.
\newblock


\bibitem[Alhijawi and Kilani(2020)]%
        {Alhijawi20RS_Survey}
\bibfield{author}{\bibinfo{person}{Bushra Alhijawi} {and}
  \bibinfo{person}{Yousef Kilani}.} \bibinfo{year}{2020}\natexlab{}.
\newblock \showarticletitle{The recommender system: a survey}.
\newblock \bibinfo{journal}{\emph{International Journal of Advanced
  Intelligence Paradigms}} \bibinfo{volume}{15}, \bibinfo{number}{3}
  (\bibinfo{year}{2020}), \bibinfo{pages}{229--251}.
\newblock


\bibitem[Alsini et~al\mbox{.}(2021)]%
        {Alsini21Weibo}
\bibfield{author}{\bibinfo{person}{Areej Alsini}, \bibinfo{person}{Du~Q.
  Huynh}, {and} \bibinfo{person}{Amitava Datta}.}
  \bibinfo{year}{2021}\natexlab{}.
\newblock \showarticletitle{Hashtag Recommendation Methods for Twitter and Sina
  Weibo: {A} Review}.
\newblock \bibinfo{journal}{\emph{Future Internet}} \bibinfo{volume}{13},
  \bibinfo{number}{5} (\bibinfo{year}{2021}), \bibinfo{pages}{129}.
\newblock


\bibitem[Baldi(1995)]%
        {Baldi95GD}
\bibfield{author}{\bibinfo{person}{Pierre Baldi}.}
  \bibinfo{year}{1995}\natexlab{}.
\newblock \showarticletitle{Gradient descent learning algorithm overview: a
  general dynamical systems perspective}.
\newblock \bibinfo{journal}{\emph{IEEE Transactions on Neural Networks}}
  \bibinfo{volume}{6}, \bibinfo{number}{1} (\bibinfo{year}{1995}),
  \bibinfo{pages}{182--195}.
\newblock


\bibitem[Boka et~al\mbox{.}(2024)]%
        {Boka24SeRec_Survey}
\bibfield{author}{\bibinfo{person}{Tesfaye~Fenta Boka},
  \bibinfo{person}{Zhendong Niu}, {and} \bibinfo{person}{Rama~Bastola
  Neupane}.} \bibinfo{year}{2024}\natexlab{}.
\newblock \showarticletitle{A survey of sequential recommendation systems:
  Techniques, evaluation, and future directions}.
\newblock \bibinfo{journal}{\emph{Information Systems}}  \bibinfo{volume}{125}
  (\bibinfo{year}{2024}), \bibinfo{pages}{102427}.
\newblock


\bibitem[Broder(1997)]%
        {Broder97LSH1}
\bibfield{author}{\bibinfo{person}{Andrei~Z. Broder}.}
  \bibinfo{year}{1997}\natexlab{}.
\newblock \showarticletitle{On the resemblance and containment of documents}.
  In \bibinfo{booktitle}{\emph{Proceedings of the SEQUENCES}},
  \bibfield{editor}{\bibinfo{person}{Bruno Carpentieri},
  \bibinfo{person}{Alfredo~De Santis}, \bibinfo{person}{Ugo Vaccaro}, {and}
  \bibinfo{person}{James~A. Storer}} (Eds.). \bibinfo{pages}{21--29}.
\newblock


\bibitem[Broder et~al\mbox{.}(1997)]%
        {Broder97LSH2}
\bibfield{author}{\bibinfo{person}{Andrei~Z. Broder},
  \bibinfo{person}{Steven~C. Glassman}, \bibinfo{person}{Mark~S. Manasse},
  {and} \bibinfo{person}{Geoffrey Zweig}.} \bibinfo{year}{1997}\natexlab{}.
\newblock \showarticletitle{Syntactic Clustering of the Web}.
\newblock \bibinfo{journal}{\emph{Computer Networks}} \bibinfo{volume}{29},
  \bibinfo{number}{8-13} (\bibinfo{year}{1997}), \bibinfo{pages}{1157--1166}.
\newblock


\bibitem[Brown et~al\mbox{.}(2020)]%
        {Brown20LLM}
\bibfield{author}{\bibinfo{person}{Tom~B. Brown}, \bibinfo{person}{Benjamin
  Mann}, \bibinfo{person}{Nick Ryder}, \bibinfo{person}{Melanie Subbiah},
  \bibinfo{person}{Jared Kaplan}, \bibinfo{person}{Prafulla Dhariwal},
  \bibinfo{person}{Arvind Neelakantan}, \bibinfo{person}{Pranav Shyam},
  \bibinfo{person}{Girish Sastry}, \bibinfo{person}{Amanda Askell},
  \bibinfo{person}{Sandhini Agarwal}, \bibinfo{person}{Ariel Herbert{-}Voss},
  \bibinfo{person}{Gretchen Krueger}, \bibinfo{person}{Tom Henighan},
  \bibinfo{person}{Rewon Child}, \bibinfo{person}{Aditya Ramesh},
  \bibinfo{person}{Daniel~M. Ziegler}, \bibinfo{person}{Jeffrey Wu},
  \bibinfo{person}{Clemens Winter}, \bibinfo{person}{Christopher Hesse},
  \bibinfo{person}{Mark Chen}, \bibinfo{person}{Eric Sigler},
  \bibinfo{person}{Mateusz Litwin}, \bibinfo{person}{Scott Gray},
  \bibinfo{person}{Benjamin Chess}, \bibinfo{person}{Jack Clark},
  \bibinfo{person}{Christopher Berner}, \bibinfo{person}{Sam McCandlish},
  \bibinfo{person}{Alec Radford}, \bibinfo{person}{Ilya Sutskever}, {and}
  \bibinfo{person}{Dario Amodei}.} \bibinfo{year}{2020}\natexlab{}.
\newblock \showarticletitle{Language Models are Few-Shot Learners}. In
  \bibinfo{booktitle}{\emph{Proceedings of the NIPS}}.
\newblock


\bibitem[Bugliarello et~al\mbox{.}(2022)]%
        {Bugliarello22MultiObjective}
\bibfield{author}{\bibinfo{person}{Emanuele Bugliarello},
  \bibinfo{person}{Rishabh Mehrotra}, \bibinfo{person}{James Kirk}, {and}
  \bibinfo{person}{Mounia Lalmas}.} \bibinfo{year}{2022}\natexlab{}.
\newblock \showarticletitle{Mostra: {A} Flexible Balancing Framework to
  Trade-off User, Artist and Platform Objectives for Music Sequencing}. In
  \bibinfo{booktitle}{\emph{Proceedings of the WWW}}.
  \bibinfo{pages}{2936--2945}.
\newblock


\bibitem[Cao et~al\mbox{.}(2007)]%
        {Cao07ListNET}
\bibfield{author}{\bibinfo{person}{Zhe Cao}, \bibinfo{person}{Tao Qin},
  \bibinfo{person}{Tie{-}Yan Liu}, \bibinfo{person}{Ming{-}Feng Tsai}, {and}
  \bibinfo{person}{Hang Li}.} \bibinfo{year}{2007}\natexlab{}.
\newblock \showarticletitle{Learning to Rank: From Pairwise Approach to
  Listwise Approach}. In \bibinfo{booktitle}{\emph{Proceedings of the ICML}},
  Vol.~\bibinfo{volume}{227}. \bibinfo{pages}{129--136}.
\newblock


\bibitem[Cayton and Dasgupta(2007)]%
        {Cayton07DD1}
\bibfield{author}{\bibinfo{person}{Lawrence Cayton} {and}
  \bibinfo{person}{Sanjoy Dasgupta}.} \bibinfo{year}{2007}\natexlab{}.
\newblock \showarticletitle{A learning framework for nearest neighbor search}.
  In \bibinfo{booktitle}{\emph{Proceedings of the NIPS}}.
  \bibinfo{pages}{233--240}.
\newblock


\bibitem[Chang et~al\mbox{.}(2023)]%
        {Chang23Kuaishou}
\bibfield{author}{\bibinfo{person}{Jianxin Chang}, \bibinfo{person}{Chenbin
  Zhang}, \bibinfo{person}{Zhiyi Fu}, \bibinfo{person}{Xiaoxue Zang},
  \bibinfo{person}{Lin Guan}, \bibinfo{person}{Jing Lu}, \bibinfo{person}{Yiqun
  Hui}, \bibinfo{person}{Dewei Leng}, \bibinfo{person}{Yanan Niu},
  \bibinfo{person}{Yang Song}, {and} \bibinfo{person}{Kun Gai}.}
  \bibinfo{year}{2023}\natexlab{}.
\newblock \showarticletitle{{TWIN:} TWo-stage Interest Network for Lifelong
  User Behavior Modeling in {CTR} Prediction at Kuaishou}. In
  \bibinfo{booktitle}{\emph{Proceedings of the SIGKDD}}.
  \bibinfo{pages}{3785--3794}.
\newblock


\bibitem[Charikar(2002)]%
        {Charikar02LSH3}
\bibfield{author}{\bibinfo{person}{Moses Charikar}.}
  \bibinfo{year}{2002}\natexlab{}.
\newblock \showarticletitle{Similarity estimation techniques from rounding
  algorithms}. In \bibinfo{booktitle}{\emph{Proceedings of the STOC}}.
  \bibinfo{pages}{380--388}.
\newblock


\bibitem[Chen et~al\mbox{.}(2021)]%
        {Chen21AutoDebias}
\bibfield{author}{\bibinfo{person}{Jiawei Chen}, \bibinfo{person}{Hande Dong},
  \bibinfo{person}{Yang Qiu}, \bibinfo{person}{Xiangnan He},
  \bibinfo{person}{Xin Xin}, \bibinfo{person}{Liang Chen},
  \bibinfo{person}{Guli Lin}, {and} \bibinfo{person}{Keping Yang}.}
  \bibinfo{year}{2021}\natexlab{}.
\newblock \showarticletitle{AutoDebias: Learning to Debias for Recommendation}.
  In \bibinfo{booktitle}{\emph{Proceedings of the SIGIR}}.
  \bibinfo{pages}{21--30}.
\newblock


\bibitem[Chen et~al\mbox{.}(2023a)]%
        {Chen23Bias_Survey}
\bibfield{author}{\bibinfo{person}{Jiawei Chen}, \bibinfo{person}{Hande Dong},
  \bibinfo{person}{Xiang Wang}, \bibinfo{person}{Fuli Feng},
  \bibinfo{person}{Meng Wang}, {and} \bibinfo{person}{Xiangnan He}.}
  \bibinfo{year}{2023}\natexlab{a}.
\newblock \showarticletitle{Bias and Debias in Recommender System: {A} Survey
  and Future Directions}.
\newblock \bibinfo{journal}{\emph{ACM Transactions on Information Systems}}
  \bibinfo{volume}{41}, \bibinfo{number}{3} (\bibinfo{year}{2023}),
  \bibinfo{pages}{67:1--67:39}.
\newblock


\bibitem[Chen et~al\mbox{.}(2023b)]%
        {Chen23BGCH}
\bibfield{author}{\bibinfo{person}{Yankai Chen}, \bibinfo{person}{Yixiang
  Fang}, \bibinfo{person}{Yifei Zhang}, {and} \bibinfo{person}{Irwin King}.}
  \bibinfo{year}{2023}\natexlab{b}.
\newblock \showarticletitle{Bipartite Graph Convolutional Hashing for Effective
  and Efficient Top-N Search in Hamming Space}. In
  \bibinfo{booktitle}{\emph{Proceedings of the WWW}}.
  \bibinfo{pages}{3164--3172}.
\newblock


\bibitem[Chen et~al\mbox{.}(2024)]%
        {Chen24BGCH+}
\bibfield{author}{\bibinfo{person}{Yankai Chen}, \bibinfo{person}{Yixiang
  Fang}, \bibinfo{person}{Yifei Zhang}, \bibinfo{person}{Chenhao Ma},
  \bibinfo{person}{Yang Hong}, {and} \bibinfo{person}{Irwin King}.}
  \bibinfo{year}{2024}\natexlab{}.
\newblock \showarticletitle{Towards Effective Top-N Hamming Search via
  Bipartite Graph Contrastive Hashing}.
\newblock \bibinfo{journal}{\emph{IEEE Transactions on Knowledge and Data
  Engineering}} (\bibinfo{year}{2024}), \bibinfo{pages}{1--14}.
\newblock


\bibitem[Chen et~al\mbox{.}(2022)]%
        {Chen22BiGeaR}
\bibfield{author}{\bibinfo{person}{Yankai Chen}, \bibinfo{person}{Huifeng Guo},
  \bibinfo{person}{Yingxue Zhang}, \bibinfo{person}{Chen Ma},
  \bibinfo{person}{Ruiming Tang}, \bibinfo{person}{Jingjie Li}, {and}
  \bibinfo{person}{Irwin King}.} \bibinfo{year}{2022}\natexlab{}.
\newblock \showarticletitle{Learning Binarized Graph Representations with
  Multi-faceted Quantization Reinforcement for Top-K Recommendation}. In
  \bibinfo{booktitle}{\emph{Proceedings of the SIGKDD}}.
  \bibinfo{publisher}{{ACM}}, \bibinfo{pages}{168--178}.
\newblock


\bibitem[Chen et~al\mbox{.}(2025)]%
        {Chen25BiGeaR-SS}
\bibfield{author}{\bibinfo{person}{Yankai Chen}, \bibinfo{person}{Yue Que},
  \bibinfo{person}{Xinni Zhang}, \bibinfo{person}{Chen Ma}, {and}
  \bibinfo{person}{Irwin King}.} \bibinfo{year}{2025}\natexlab{}.
\newblock \showarticletitle{Learning Binarized Representations with
  Pseudo-positive Sample Enhancement for Efficient Graph Collaborative
  Filtering}.
\newblock \bibinfo{journal}{\emph{ACM Transactions on Information Systems}}
  \bibinfo{volume}{43}, \bibinfo{number}{5} (\bibinfo{year}{2025}),
  \bibinfo{pages}{131:1--131:28}.
\newblock


\bibitem[Covington et~al\mbox{.}(2016)]%
        {DNN2016YouTube}
\bibfield{author}{\bibinfo{person}{Paul Covington}, \bibinfo{person}{Jay
  Adams}, {and} \bibinfo{person}{Emre Sargin}.}
  \bibinfo{year}{2016}\natexlab{}.
\newblock \showarticletitle{Deep Neural Networks for YouTube Recommendations}.
  In \bibinfo{booktitle}{\emph{Proceedings of the RecSys}}.
  \bibinfo{pages}{191--198}.
\newblock


\bibitem[Dai et~al\mbox{.}(2023)]%
        {Dai23LLMRS}
\bibfield{author}{\bibinfo{person}{Sunhao Dai}, \bibinfo{person}{Ninglu Shao},
  \bibinfo{person}{Haiyuan Zhao}, \bibinfo{person}{Weijie Yu},
  \bibinfo{person}{Zihua Si}, \bibinfo{person}{Chen Xu},
  \bibinfo{person}{Zhongxiang Sun}, \bibinfo{person}{Xiao Zhang}, {and}
  \bibinfo{person}{Jun Xu}.} \bibinfo{year}{2023}\natexlab{}.
\newblock \showarticletitle{Uncovering ChatGPT's Capabilities in Recommender
  Systems}. In \bibinfo{booktitle}{\emph{Proceedings of the RecSys}}.
  \bibinfo{pages}{1126--1132}.
\newblock


\bibitem[Darabi et~al\mbox{.}(2018)]%
        {Darabi18SignSwish}
\bibfield{author}{\bibinfo{person}{Sajad Darabi}, \bibinfo{person}{Mouloud
  Belbahri}, \bibinfo{person}{Matthieu Courbariaux}, {and}
  \bibinfo{person}{Vahid~Partovi Nia}.} \bibinfo{year}{2018}\natexlab{}.
\newblock \showarticletitle{{BNN+:} Improved Binary Network Training}.
\newblock \bibinfo{journal}{\emph{CoRR}}  \bibinfo{volume}{abs/1812.11800}
  (\bibinfo{year}{2018}).
\newblock


\bibitem[Dasgupta et~al\mbox{.}(2011)]%
        {Dasgupta11FLSH}
\bibfield{author}{\bibinfo{person}{Anirban Dasgupta}, \bibinfo{person}{Ravi
  Kumar}, {and} \bibinfo{person}{Tam{\'{a}}s Sarl{\'{o}}s}.}
  \bibinfo{year}{2011}\natexlab{}.
\newblock \showarticletitle{Fast locality-sensitive hashing}. In
  \bibinfo{booktitle}{\emph{Proceedings of the SIGKDD}}.
  \bibinfo{pages}{1073--1081}.
\newblock


\bibitem[Datar et~al\mbox{.}(2004)]%
        {Datar04LSH4}
\bibfield{author}{\bibinfo{person}{Mayur Datar}, \bibinfo{person}{Nicole
  Immorlica}, \bibinfo{person}{Piotr Indyk}, {and} \bibinfo{person}{Vahab~S.
  Mirrokni}.} \bibinfo{year}{2004}\natexlab{}.
\newblock \showarticletitle{Locality-sensitive hashing scheme based on p-stable
  distributions}. In \bibinfo{booktitle}{\emph{Proceedings of the SCG}}.
  \bibinfo{pages}{253--262}.
\newblock


\bibitem[Devlin et~al\mbox{.}(2019)]%
        {DevlinCLT19BERT}
\bibfield{author}{\bibinfo{person}{Jacob Devlin}, \bibinfo{person}{Ming{-}Wei
  Chang}, \bibinfo{person}{Kenton Lee}, {and} \bibinfo{person}{Kristina
  Toutanova}.} \bibinfo{year}{2019}\natexlab{}.
\newblock \showarticletitle{{BERT:} Pre-training of Deep Bidirectional
  Transformers for Language Understanding}. In
  \bibinfo{booktitle}{\emph{Proceedings of the NAACL-HLT}}.
  \bibinfo{pages}{4171--4186}.
\newblock


\bibitem[Ding et~al\mbox{.}(2022)]%
        {Ding22InterD}
\bibfield{author}{\bibinfo{person}{Sihao Ding}, \bibinfo{person}{Fuli Feng},
  \bibinfo{person}{Xiangnan He}, \bibinfo{person}{Jinqiu Jin},
  \bibinfo{person}{Wenjie Wang}, \bibinfo{person}{Yong Liao}, {and}
  \bibinfo{person}{Yongdong Zhang}.} \bibinfo{year}{2022}\natexlab{}.
\newblock \showarticletitle{Interpolative Distillation for Unifying Biased and
  Debiased Recommendation}. In \bibinfo{booktitle}{\emph{Proceedings of the
  SIGIR}}. \bibinfo{pages}{40--49}.
\newblock


\bibitem[Donmez et~al\mbox{.}(2009)]%
        {Donmez09}
\bibfield{author}{\bibinfo{person}{Pinar Donmez}, \bibinfo{person}{Krysta~M.
  Svore}, {and} \bibinfo{person}{Christopher J.~C. Burges}.}
  \bibinfo{year}{2009}\natexlab{}.
\newblock \showarticletitle{On the local optimality of LambdaRank}. In
  \bibinfo{booktitle}{\emph{Proceedings of the SIGIR}},
  \bibfield{editor}{\bibinfo{person}{James Allan}, \bibinfo{person}{Javed~A.
  Aslam}, \bibinfo{person}{Mark Sanderson}, \bibinfo{person}{ChengXiang Zhai},
  {and} \bibinfo{person}{Justin Zobel}} (Eds.). \bibinfo{pages}{460--467}.
\newblock


\bibitem[Gan et~al\mbox{.}(2012)]%
        {Gan12LSH7}
\bibfield{author}{\bibinfo{person}{Junhao Gan}, \bibinfo{person}{Jianlin Feng},
  \bibinfo{person}{Qiong Fang}, {and} \bibinfo{person}{Wilfred Ng}.}
  \bibinfo{year}{2012}\natexlab{}.
\newblock \showarticletitle{Locality-sensitive hashing scheme based on dynamic
  collision counting}. In \bibinfo{booktitle}{\emph{Proceedings of the
  SIGMOD}}. \bibinfo{pages}{541--552}.
\newblock


\bibitem[Gionis et~al\mbox{.}(1999)]%
        {Gionis99LSH}
\bibfield{author}{\bibinfo{person}{Aristides Gionis}, \bibinfo{person}{Piotr
  Indyk}, {and} \bibinfo{person}{Rajeev Motwani}.}
  \bibinfo{year}{1999}\natexlab{}.
\newblock \showarticletitle{Similarity Search in High Dimensions via Hashing}.
  In \bibinfo{booktitle}{\emph{Proceedings of the VLDB}}.
  \bibinfo{pages}{518--529}.
\newblock


\bibitem[Gong et~al\mbox{.}(2019)]%
        {Gong19Tanh}
\bibfield{author}{\bibinfo{person}{Ruihao Gong}, \bibinfo{person}{Xianglong
  Liu}, \bibinfo{person}{Shenghu Jiang}, \bibinfo{person}{Tianxiang Li},
  \bibinfo{person}{Peng Hu}, \bibinfo{person}{Jiazhen Lin},
  \bibinfo{person}{Fengwei Yu}, {and} \bibinfo{person}{Junjie Yan}.}
  \bibinfo{year}{2019}\natexlab{}.
\newblock \showarticletitle{Differentiable Soft Quantization: Bridging
  Full-Precision and Low-Bit Neural Networks}. In
  \bibinfo{booktitle}{\emph{Proceedings of the ICCV}}.
  \bibinfo{pages}{4851--4860}.
\newblock


\bibitem[Gong et~al\mbox{.}(2020)]%
        {Gong20Taobao}
\bibfield{author}{\bibinfo{person}{Yu Gong}, \bibinfo{person}{Ziwen Jiang},
  \bibinfo{person}{Yufei Feng}, \bibinfo{person}{Binbin Hu},
  \bibinfo{person}{Kaiqi Zhao}, \bibinfo{person}{Qingwen Liu}, {and}
  \bibinfo{person}{Wenwu Ou}.} \bibinfo{year}{2020}\natexlab{}.
\newblock \showarticletitle{EdgeRec: Recommender System on Edge in Mobile
  Taobao}. In \bibinfo{booktitle}{\emph{Proceedings of the CIKM}}.
  \bibinfo{pages}{2477--2484}.
\newblock


\bibitem[Grover et~al\mbox{.}(2019)]%
        {Grover19NeuralSort}
\bibfield{author}{\bibinfo{person}{Aditya Grover}, \bibinfo{person}{Eric Wang},
  \bibinfo{person}{Aaron Zweig}, {and} \bibinfo{person}{Stefano Ermon}.}
  \bibinfo{year}{2019}\natexlab{}.
\newblock \showarticletitle{Stochastic Optimization of Sorting Networks via
  Continuous Relaxations}. In \bibinfo{booktitle}{\emph{Proceedings of the
  ICLR}}.
\newblock


\bibitem[Gu et~al\mbox{.}(2020)]%
        {Gu20MultiObjective}
\bibfield{author}{\bibinfo{person}{Yulong Gu}, \bibinfo{person}{Zhuoye Ding},
  \bibinfo{person}{Shuaiqiang Wang}, \bibinfo{person}{Lixin Zou},
  \bibinfo{person}{Yiding Liu}, {and} \bibinfo{person}{Dawei Yin}.}
  \bibinfo{year}{2020}\natexlab{}.
\newblock \showarticletitle{Deep Multifaceted Transformers for Multi-objective
  Ranking in Large-Scale E-commerce Recommender Systems}. In
  \bibinfo{booktitle}{\emph{Proceedings of the CIKM}}.
  \bibinfo{pages}{2493--2500}.
\newblock


\bibitem[Guan et~al\mbox{.}(2022)]%
        {Guan22BIHGH}
\bibfield{author}{\bibinfo{person}{Weili Guan}, \bibinfo{person}{Xuemeng Song},
  \bibinfo{person}{Haoyu Zhang}, \bibinfo{person}{Meng Liu},
  \bibinfo{person}{Chung{-}Hsing Yeh}, {and} \bibinfo{person}{Xiaojun Chang}.}
  \bibinfo{year}{2022}\natexlab{}.
\newblock \showarticletitle{Bi-directional Heterogeneous Graph Hashing towards
  Efficient Outfit Recommendation}. In \bibinfo{booktitle}{\emph{Proceedings of
  the ACM MM}}. \bibinfo{pages}{268--276}.
\newblock


\bibitem[Guo et~al\mbox{.}(2019)]%
        {Guo2019DTMF}
\bibfield{author}{\bibinfo{person}{Guibing Guo}, \bibinfo{person}{Enneng Yang},
  \bibinfo{person}{Li Shen}, \bibinfo{person}{Xiaochun Yang}, {and}
  \bibinfo{person}{Xiaodong He}.} \bibinfo{year}{2019}\natexlab{}.
\newblock \showarticletitle{Discrete Trust-aware Matrix Factorization for Fast
  Social Recommendation}. In \bibinfo{booktitle}{\emph{Proceedings of the
  IJCAI}}. \bibinfo{pages}{1380–1386}.
\newblock


\bibitem[Hansen et~al\mbox{.}(2020)]%
        {Hansen20NeuHash-CF}
\bibfield{author}{\bibinfo{person}{Casper Hansen}, \bibinfo{person}{Christian
  Hansen}, \bibinfo{person}{Jakob~Grue Simonsen}, \bibinfo{person}{Stephen
  Alstrup}, {and} \bibinfo{person}{Christina Lioma}.}
  \bibinfo{year}{2020}\natexlab{}.
\newblock \showarticletitle{Content-aware Neural Hashing for Cold-start
  Recommendation}. In \bibinfo{booktitle}{\emph{Proceedings of the SIGIR}}.
  \bibinfo{pages}{971--980}.
\newblock


\bibitem[Hansen et~al\mbox{.}(2021)]%
        {Hansen2021VHPHD}
\bibfield{author}{\bibinfo{person}{Christian Hansen}, \bibinfo{person}{Casper
  Hansen}, \bibinfo{person}{Jakob~Grue Simonsen}, {and}
  \bibinfo{person}{Christina Lioma}.} \bibinfo{year}{2021}\natexlab{}.
\newblock \showarticletitle{Projected Hamming Dissimilarity for Bit-Level
  Importance Coding in Collaborative Filtering}. In
  \bibinfo{booktitle}{\emph{Proceedings of the WWW}}.
  \bibinfo{pages}{261--269}.
\newblock


\bibitem[He et~al\mbox{.}(2011)]%
        {He11DD2}
\bibfield{author}{\bibinfo{person}{Junfeng He}, \bibinfo{person}{Shih{-}Fu
  Chang}, \bibinfo{person}{Regunathan Radhakrishnan}, {and}
  \bibinfo{person}{Claus Bauer}.} \bibinfo{year}{2011}\natexlab{}.
\newblock \showarticletitle{Compact hashing with joint optimization of search
  accuracy and time}. In \bibinfo{booktitle}{\emph{Proceedings of the CVPR}}.
  \bibinfo{pages}{753--760}.
\newblock


\bibitem[He et~al\mbox{.}(2016)]%
        {He16RESNET}
\bibfield{author}{\bibinfo{person}{Kaiming He}, \bibinfo{person}{Xiangyu
  Zhang}, \bibinfo{person}{Shaoqing Ren}, {and} \bibinfo{person}{Jian Sun}.}
  \bibinfo{year}{2016}\natexlab{}.
\newblock \showarticletitle{Deep Residual Learning for Image Recognition}. In
  \bibinfo{booktitle}{\emph{Proceedings of the CVPR}}.
  \bibinfo{pages}{770--778}.
\newblock


\bibitem[He et~al\mbox{.}(2020)]%
        {He2020LightGCN}
\bibfield{author}{\bibinfo{person}{Xiangnan He}, \bibinfo{person}{Kuan Deng},
  \bibinfo{person}{Xiang Wang}, \bibinfo{person}{Yan Li},
  \bibinfo{person}{Yong{-}Dong Zhang}, {and} \bibinfo{person}{Meng Wang}.}
  \bibinfo{year}{2020}\natexlab{}.
\newblock \showarticletitle{LightGCN: Simplifying and Powering Graph
  Convolution Network for Recommendation}. In
  \bibinfo{booktitle}{\emph{Proceedings of the SIGIR}}.
  \bibinfo{pages}{639--648}.
\newblock


\bibitem[Hochreiter and Schmidhuber(1997)]%
        {LSTM1997}
\bibfield{author}{\bibinfo{person}{Sepp Hochreiter} {and}
  \bibinfo{person}{Jürgen Schmidhuber}.} \bibinfo{year}{1997}\natexlab{}.
\newblock \showarticletitle{Long Short-Term Memory}.
\newblock \bibinfo{journal}{\emph{Neural Computation}} \bibinfo{volume}{9},
  \bibinfo{number}{8} (\bibinfo{year}{1997}), \bibinfo{pages}{1735--1780}.
\newblock


\bibitem[Hou et~al\mbox{.}(2023)]%
        {Hou23VQSR}
\bibfield{author}{\bibinfo{person}{Yupeng Hou}, \bibinfo{person}{Zhankui He},
  \bibinfo{person}{Julian~J. McAuley}, {and} \bibinfo{person}{Wayne~Xin Zhao}.}
  \bibinfo{year}{2023}\natexlab{}.
\newblock \showarticletitle{Learning Vector-Quantized Item Representation for
  Transferable Sequential Recommenders}. In
  \bibinfo{booktitle}{\emph{Proceedings of the WWW}}.
  \bibinfo{pages}{1162--1171}.
\newblock


\bibitem[Hou et~al\mbox{.}(2022)]%
        {Hou22UniSR}
\bibfield{author}{\bibinfo{person}{Yupeng Hou}, \bibinfo{person}{Shanlei Mu},
  \bibinfo{person}{Wayne~Xin Zhao}, \bibinfo{person}{Yaliang Li},
  \bibinfo{person}{Bolin Ding}, {and} \bibinfo{person}{Ji{-}Rong Wen}.}
  \bibinfo{year}{2022}\natexlab{}.
\newblock \showarticletitle{Towards Universal Sequence Representation Learning
  for Recommender Systems}. In \bibinfo{booktitle}{\emph{Proceedings of the
  SIGKDD}}. \bibinfo{pages}{585--593}.
\newblock


\bibitem[Hou et~al\mbox{.}(2024)]%
        {Hou24LLMRS}
\bibfield{author}{\bibinfo{person}{Yupeng Hou}, \bibinfo{person}{Junjie Zhang},
  \bibinfo{person}{Zihan Lin}, \bibinfo{person}{Hongyu Lu},
  \bibinfo{person}{Ruobing Xie}, \bibinfo{person}{Julian~J. McAuley}, {and}
  \bibinfo{person}{Wayne~Xin Zhao}.} \bibinfo{year}{2024}\natexlab{}.
\newblock \showarticletitle{Large Language Models are Zero-Shot Rankers for
  Recommender Systems}. In \bibinfo{booktitle}{\emph{Proceedings of the ECIR}},
  Vol.~\bibinfo{volume}{14609}. \bibinfo{pages}{364--381}.
\newblock


\bibitem[Hu et~al\mbox{.}(2021)]%
        {Hu21Deep_Survey}
\bibfield{author}{\bibinfo{person}{Xia Hu}, \bibinfo{person}{Lingyang Chu},
  \bibinfo{person}{Jian Pei}, \bibinfo{person}{Weiqing Liu}, {and}
  \bibinfo{person}{Jiang Bian}.} \bibinfo{year}{2021}\natexlab{}.
\newblock \showarticletitle{Model complexity of deep learning: a survey}.
\newblock \bibinfo{journal}{\emph{Knowledge And Information Systems}}
  \bibinfo{volume}{63}, \bibinfo{number}{10} (\bibinfo{year}{2021}),
  \bibinfo{pages}{2585--2619}.
\newblock


\bibitem[Hu et~al\mbox{.}(2023)]%
        {Hu23DLACF}
\bibfield{author}{\bibinfo{person}{Zhibin Hu}, \bibinfo{person}{Xuebin Zhou},
  \bibinfo{person}{Zhiwei He}, \bibinfo{person}{Zehang Yang},
  \bibinfo{person}{Jian Chen}, {and} \bibinfo{person}{Jin Huang}.}
  \bibinfo{year}{2023}\natexlab{}.
\newblock \showarticletitle{Discrete limited attentional collaborative
  filtering for fast social recommendation}.
\newblock \bibinfo{journal}{\emph{Engineering Applications of Artificial
  Intelligence}} \bibinfo{volume}{123}, \bibinfo{number}{Part {C}}
  (\bibinfo{year}{2023}), \bibinfo{pages}{106437}.
\newblock


\bibitem[Huang et~al\mbox{.}(2013)]%
        {DSSM13}
\bibfield{author}{\bibinfo{person}{Po-Sen Huang}, \bibinfo{person}{Xiaodong
  He}, \bibinfo{person}{Jianfeng Gao}, \bibinfo{person}{Li Deng},
  \bibinfo{person}{Alex Acero}, {and} \bibinfo{person}{Larry Heck}.}
  \bibinfo{year}{2013}\natexlab{}.
\newblock \showarticletitle{Learning deep structured semantic models for web
  search using clickthrough data}. In \bibinfo{booktitle}{\emph{Proceedings of
  the CIKM}}. \bibinfo{pages}{2333–2338}.
\newblock


\bibitem[Ji et~al\mbox{.}(2013)]%
        {Ji13LSH12}
\bibfield{author}{\bibinfo{person}{Jianqiu Ji}, \bibinfo{person}{Jianmin Li},
  \bibinfo{person}{Shuicheng Yan}, \bibinfo{person}{Qi Tian}, {and}
  \bibinfo{person}{Bo Zhang}.} \bibinfo{year}{2013}\natexlab{}.
\newblock \showarticletitle{Min-Max Hash for Jaccard Similarity}. In
  \bibinfo{booktitle}{\emph{Proceedings of the ICDM}}.
  \bibinfo{pages}{301--309}.
\newblock


\bibitem[Ji et~al\mbox{.}(2012)]%
        {Ji12LSH10}
\bibfield{author}{\bibinfo{person}{Jianqiu Ji}, \bibinfo{person}{Jianmin Li},
  \bibinfo{person}{Shuicheng Yan}, \bibinfo{person}{Bo Zhang}, {and}
  \bibinfo{person}{Qi Tian}.} \bibinfo{year}{2012}\natexlab{}.
\newblock \showarticletitle{Super-Bit Locality-Sensitive Hashing}. In
  \bibinfo{booktitle}{\emph{Proceedings of the NIPS}}.
  \bibinfo{pages}{108--116}.
\newblock


\bibitem[Kang and McAuley(2019)]%
        {Kang19CIGAR}
\bibfield{author}{\bibinfo{person}{Wang{-}Cheng Kang} {and}
  \bibinfo{person}{Julian~John McAuley}.} \bibinfo{year}{2019}\natexlab{}.
\newblock \showarticletitle{Candidate Generation with Binary Codes for
  Large-Scale Top-N Recommendation}. In \bibinfo{booktitle}{\emph{Proceedings
  of the CIKM}}. \bibinfo{pages}{1523--1532}.
\newblock


\bibitem[Kim(2014)]%
        {Kim14TextCNN}
\bibfield{author}{\bibinfo{person}{Yoon Kim}.} \bibinfo{year}{2014}\natexlab{}.
\newblock \showarticletitle{Convolutional Neural Networks for Sentence
  Classification}. In \bibinfo{booktitle}{\emph{Proceedings of the EMNLP}}.
  \bibinfo{pages}{1746--1751}.
\newblock


\bibitem[Kipf and Welling(2017)]%
        {Kipf17GCN}
\bibfield{author}{\bibinfo{person}{Thomas~N. Kipf} {and} \bibinfo{person}{Max
  Welling}.} \bibinfo{year}{2017}\natexlab{}.
\newblock \showarticletitle{Semi-Supervised Classification with Graph
  Convolutional Networks}. In \bibinfo{booktitle}{\emph{Proceedings of the
  ICLR}}.
\newblock


\bibitem[Knuth(1998)]%
        {Knuth98L2H}
\bibfield{author}{\bibinfo{person}{Donald~E. Knuth}.}
  \bibinfo{year}{1998}\natexlab{}.
\newblock \bibinfo{booktitle}{\emph{Sorting and Searching}}.
\newblock \bibinfo{publisher}{Addison Wesley Longman Publishing Co., Inc.}
\newblock


\bibitem[Koren et~al\mbox{.}(2009)]%
        {Koren09MF}
\bibfield{author}{\bibinfo{person}{Yehuda Koren}, \bibinfo{person}{Robert
  Bell}, {and} \bibinfo{person}{Chris Volinsky}.}
  \bibinfo{year}{2009}\natexlab{}.
\newblock \showarticletitle{Matrix Factorization Techniques for Recommender
  Systems}.
\newblock \bibinfo{journal}{\emph{Computer}} \bibinfo{volume}{42},
  \bibinfo{number}{8} (\bibinfo{year}{2009}), \bibinfo{pages}{30--37}.
\newblock


\bibitem[Kulis and Darrell(2009)]%
        {Kulis09DD4}
\bibfield{author}{\bibinfo{person}{Brian Kulis} {and} \bibinfo{person}{Trevor
  Darrell}.} \bibinfo{year}{2009}\natexlab{}.
\newblock \showarticletitle{Learning to Hash with Binary Reconstructive
  Embeddings}. In \bibinfo{booktitle}{\emph{Proceedings of the NIPS}}.
  \bibinfo{pages}{1042--1050}.
\newblock


\bibitem[Kulis and Grauman(2012)]%
        {Kulis12KLSH}
\bibfield{author}{\bibinfo{person}{Brian Kulis} {and} \bibinfo{person}{Kristen
  Grauman}.} \bibinfo{year}{2012}\natexlab{}.
\newblock \showarticletitle{Kernelized Locality-Sensitive Hashing}.
\newblock \bibinfo{journal}{\emph{IEEE Transactions on Pattern Analysis and
  Machine Intelligence}} \bibinfo{volume}{34}, \bibinfo{number}{6}
  (\bibinfo{year}{2012}), \bibinfo{pages}{1092--1104}.
\newblock


\bibitem[Li(2011)]%
        {Li11LTR}
\bibfield{author}{\bibinfo{person}{Hang Li}.} \bibinfo{year}{2011}\natexlab{}.
\newblock \showarticletitle{A Short Introduction to Learning to Rank}.
\newblock \bibinfo{journal}{\emph{IEICE Transactions on Information and
  Systems}} \bibinfo{volume}{94-D}, \bibinfo{number}{10}
  (\bibinfo{year}{2011}), \bibinfo{pages}{1854--1862}.
\newblock


\bibitem[Li et~al\mbox{.}(2006a)]%
        {Li06LSH9}
\bibfield{author}{\bibinfo{person}{Ping Li}, \bibinfo{person}{Kenneth~Ward
  Church}, {and} \bibinfo{person}{Trevor Hastie}.}
  \bibinfo{year}{2006}\natexlab{a}.
\newblock \showarticletitle{Conditional Random Sampling: {A} Sketch-based
  Sampling Technique for Sparse Data}. In \bibinfo{booktitle}{\emph{Proceedings
  of the NIPS}}. \bibinfo{pages}{873--880}.
\newblock


\bibitem[Li et~al\mbox{.}(2006b)]%
        {Li06LSH13}
\bibfield{author}{\bibinfo{person}{Ping Li}, \bibinfo{person}{Trevor Hastie},
  {and} \bibinfo{person}{Kenneth~Ward Church}.}
  \bibinfo{year}{2006}\natexlab{b}.
\newblock \showarticletitle{Very sparse random projections}. In
  \bibinfo{booktitle}{\emph{Proceedings of the SIGKDD}}.
  \bibinfo{pages}{287--296}.
\newblock


\bibitem[Li et~al\mbox{.}(2012)]%
        {Li12LSH11}
\bibfield{author}{\bibinfo{person}{Ping Li}, \bibinfo{person}{Art~B. Owen},
  {and} \bibinfo{person}{Cun{-}Hui Zhang}.} \bibinfo{year}{2012}\natexlab{}.
\newblock \showarticletitle{One Permutation Hashing}. In
  \bibinfo{booktitle}{\emph{Proceedings of the NIPS}}.
  \bibinfo{pages}{3122--3130}.
\newblock


\bibitem[Li et~al\mbox{.}(2021)]%
        {Li21MetricOpt}
\bibfield{author}{\bibinfo{person}{Roger~Zhe Li}, \bibinfo{person}{Juli{\'{a}}n
  Urbano}, {and} \bibinfo{person}{Alan Hanjalic}.}
  \bibinfo{year}{2021}\natexlab{}.
\newblock \showarticletitle{New Insights into Metric Optimization for
  Ranking-based Recommendation}. In \bibinfo{booktitle}{\emph{Proceedings of
  the SIGIR}}. \bibinfo{pages}{932--941}.
\newblock


\bibitem[Lian et~al\mbox{.}(2017)]%
        {Lian17DCMF}
\bibfield{author}{\bibinfo{person}{Defu Lian}, \bibinfo{person}{Rui Liu},
  \bibinfo{person}{Yong Ge}, \bibinfo{person}{Kai Zheng}, \bibinfo{person}{Xing
  Xie}, {and} \bibinfo{person}{Longbing Cao}.} \bibinfo{year}{2017}\natexlab{}.
\newblock \showarticletitle{Discrete Content-aware Matrix Factorization}. In
  \bibinfo{booktitle}{\emph{Proceedings of the SIGKDD}}.
  \bibinfo{pages}{325--334}.
\newblock


\bibitem[Lian et~al\mbox{.}(2021)]%
        {Lian2021DMF}
\bibfield{author}{\bibinfo{person}{Defu Lian}, \bibinfo{person}{Xing Xie},
  {and} \bibinfo{person}{Enhong Chen}.} \bibinfo{year}{2021}\natexlab{}.
\newblock \showarticletitle{Discrete Matrix Factorization and Extension for
  Fast Item Recommendation}.
\newblock \bibinfo{journal}{\emph{IEEE Transactions on Knowledge and Data
  Engineering}} \bibinfo{volume}{33}, \bibinfo{number}{5}
  (\bibinfo{year}{2021}), \bibinfo{pages}{1919--1933}.
\newblock


\bibitem[Lin et~al\mbox{.}(2019)]%
        {Lin19MultiObjective}
\bibfield{author}{\bibinfo{person}{Xiao Lin}, \bibinfo{person}{Hongjie Chen},
  \bibinfo{person}{Changhua Pei}, \bibinfo{person}{Fei Sun},
  \bibinfo{person}{Xuanji Xiao}, \bibinfo{person}{Hanxiao Sun},
  \bibinfo{person}{Yongfeng Zhang}, \bibinfo{person}{Wenwu Ou}, {and}
  \bibinfo{person}{Peng Jiang}.} \bibinfo{year}{2019}\natexlab{}.
\newblock \showarticletitle{A pareto-efficient algorithm for multiple objective
  optimization in e-commerce recommendation}. In
  \bibinfo{booktitle}{\emph{Proceedings of the RecSys}}.
  \bibinfo{pages}{20--28}.
\newblock


\bibitem[Lin et~al\mbox{.}(2010)]%
        {Lin10ALM}
\bibfield{author}{\bibinfo{person}{Zhouchen Lin}, \bibinfo{person}{Minming
  Chen}, {and} \bibinfo{person}{Yi Ma}.} \bibinfo{year}{2010}\natexlab{}.
\newblock \showarticletitle{The Augmented Lagrange Multiplier Method for Exact
  Recovery of Corrupted Low-Rank Matrices}.
\newblock \bibinfo{journal}{\emph{CoRR}}  \bibinfo{volume}{abs/1009.5055}
  (\bibinfo{year}{2010}).
\newblock


\bibitem[Linden et~al\mbox{.}(2003)]%
        {Linden03Amazon}
\bibfield{author}{\bibinfo{person}{Greg Linden}, \bibinfo{person}{Brent Smith},
  {and} \bibinfo{person}{Jeremy York}.} \bibinfo{year}{2003}\natexlab{}.
\newblock \showarticletitle{Amazon.com Recommendations: Item-to-Item
  Collaborative Filtering}.
\newblock \bibinfo{journal}{\emph{IEEE Internet Computing}}
  \bibinfo{volume}{7}, \bibinfo{number}{1} (\bibinfo{year}{2003}),
  \bibinfo{pages}{76--80}.
\newblock


\bibitem[Liu et~al\mbox{.}(2021)]%
        {Liu21DLCF}
\bibfield{author}{\bibinfo{person}{Chenghao Liu}, \bibinfo{person}{Tao Lu},
  \bibinfo{person}{Zhiyong Cheng}, \bibinfo{person}{Xin Wang},
  \bibinfo{person}{Jianling Sun}, {and} \bibinfo{person}{Steven C.~H. Hoi}.}
  \bibinfo{year}{2021}\natexlab{}.
\newblock \showarticletitle{Discrete Listwise Collaborative Filtering for Fast
  Recommendation}. In \bibinfo{booktitle}{\emph{Proceedings of the SDM}}.
  \bibinfo{pages}{46--54}.
\newblock


\bibitem[Liu et~al\mbox{.}(2019a)]%
        {Liu19CCCF}
\bibfield{author}{\bibinfo{person}{Chenghao Liu}, \bibinfo{person}{Tao Lu},
  \bibinfo{person}{Xin Wang}, \bibinfo{person}{Zhiyong Cheng},
  \bibinfo{person}{Jianling Sun}, {and} \bibinfo{person}{Steven C.~H. Hoi}.}
  \bibinfo{year}{2019}\natexlab{a}.
\newblock \showarticletitle{Compositional Coding for Collaborative Filtering}.
  In \bibinfo{booktitle}{\emph{Proceedings of the SIGIR}}.
  \bibinfo{pages}{145--154}.
\newblock


\bibitem[Liu et~al\mbox{.}(2019b)]%
        {Liu2019DSR}
\bibfield{author}{\bibinfo{person}{Chenghao Liu}, \bibinfo{person}{Xin Wang},
  \bibinfo{person}{Tao Lu}, \bibinfo{person}{Wenwu Zhu},
  \bibinfo{person}{Jianling Sun}, {and} \bibinfo{person}{Steven Hoi}.}
  \bibinfo{year}{2019}\natexlab{b}.
\newblock \showarticletitle{Discrete Social Recommendation}. In
  \bibinfo{booktitle}{\emph{Proceedings of the AAAI}}.
  \bibinfo{pages}{208--215}.
\newblock


\bibitem[Liu et~al\mbox{.}(2018b)]%
        {Liu18DFM}
\bibfield{author}{\bibinfo{person}{Han Liu}, \bibinfo{person}{Xiangnan He},
  \bibinfo{person}{Fuli Feng}, \bibinfo{person}{Liqiang Nie},
  \bibinfo{person}{Rui Liu}, {and} \bibinfo{person}{Hanwang Zhang}.}
  \bibinfo{year}{2018}\natexlab{b}.
\newblock \showarticletitle{Discrete Factorization Machines for Fast
  Feature-based Recommendation}. In \bibinfo{booktitle}{\emph{Proceedings of
  the IJCAI}}. \bibinfo{pages}{3449--3455}.
\newblock


\bibitem[Liu et~al\mbox{.}(2018a)]%
        {Liu18DSFCH}
\bibfield{author}{\bibinfo{person}{Luyao Liu}, \bibinfo{person}{Xingzhong Du},
  \bibinfo{person}{Lei Zhu}, \bibinfo{person}{Fumin Shen}, {and}
  \bibinfo{person}{Zi Huang}.} \bibinfo{year}{2018}\natexlab{a}.
\newblock \showarticletitle{Learning Discrete Hashing Towards Efficient Fashion
  Recommendation}.
\newblock \bibinfo{journal}{\emph{Data Science and Engineering}}
  \bibinfo{volume}{3}, \bibinfo{number}{4} (\bibinfo{year}{2018}),
  \bibinfo{pages}{307--322}.
\newblock


\bibitem[Liu(2011)]%
        {Liu11LTR}
\bibfield{author}{\bibinfo{person}{Tie{-}Yan Liu}.}
  \bibinfo{year}{2011}\natexlab{}.
\newblock \bibinfo{booktitle}{\emph{Learning to Rank for Information
  Retrieval}}.
\newblock \bibinfo{publisher}{Springer}.
\newblock
\href{https://doi.org/10.1007/978-3-642-14267-3}{doi:\nolinkurl{10.1007/978-3-642-14267-3}}


\bibitem[Liu et~al\mbox{.}(2014b)]%
        {Liu14SVD}
\bibfield{author}{\bibinfo{person}{Wei Liu}, \bibinfo{person}{Cun Mu},
  \bibinfo{person}{Sanjiv Kumar}, {and} \bibinfo{person}{Shih{-}Fu Chang}.}
  \bibinfo{year}{2014}\natexlab{b}.
\newblock \showarticletitle{Discrete Graph Hashing}. In
  \bibinfo{booktitle}{\emph{Proceedings of the NIPS}}.
  \bibinfo{pages}{3419--3427}.
\newblock


\bibitem[Liu et~al\mbox{.}(2012)]%
        {Liu12DD3}
\bibfield{author}{\bibinfo{person}{Wei Liu}, \bibinfo{person}{Jun Wang},
  \bibinfo{person}{Rongrong Ji}, \bibinfo{person}{Yu{-}Gang Jiang}, {and}
  \bibinfo{person}{Shih{-}Fu Chang}.} \bibinfo{year}{2012}\natexlab{}.
\newblock \showarticletitle{Supervised hashing with kernels}. In
  \bibinfo{booktitle}{\emph{Proceedings of the CVPR}}.
  \bibinfo{pages}{2074--2081}.
\newblock


\bibitem[Liu et~al\mbox{.}(2014a)]%
        {Liu14CH}
\bibfield{author}{\bibinfo{person}{Xianglong Liu}, \bibinfo{person}{Junfeng
  He}, \bibinfo{person}{Cheng Deng}, {and} \bibinfo{person}{Bo Lang}.}
  \bibinfo{year}{2014}\natexlab{a}.
\newblock \showarticletitle{Collaborative Hashing}. In
  \bibinfo{booktitle}{\emph{Proceedings of the CVPR}}.
  \bibinfo{pages}{2147--2154}.
\newblock


\bibitem[Liu et~al\mbox{.}(2024)]%
        {liu24RED}
\bibfield{author}{\bibinfo{person}{Yifan Liu}, \bibinfo{person}{Kangning
  Zhang}, \bibinfo{person}{Xiangyuan Ren}, \bibinfo{person}{Yanhua Huang},
  \bibinfo{person}{Jiarui Jin}, \bibinfo{person}{Yingjie Qin},
  \bibinfo{person}{Ruilong Su}, \bibinfo{person}{Ruiwen Xu},
  \bibinfo{person}{Yong Yu}, \bibinfo{person}{}, {and} \bibinfo{person}{Weinan
  Zhang}.} \bibinfo{year}{2024}\natexlab{}.
\newblock \showarticletitle{AlignRec: Aligning and Training in Multimodal
  Recommendations}. In \bibinfo{booktitle}{\emph{Proceedings of the CIKM}}.
\newblock


\bibitem[Liu and Qiao(2014)]%
        {Liu14GNCCP}
\bibfield{author}{\bibinfo{person}{Zhiyong Liu} {and} \bibinfo{person}{Hong
  Qiao}.} \bibinfo{year}{2014}\natexlab{}.
\newblock \showarticletitle{{GNCCP} - Graduated NonConvexity and Graduated
  Concavity Procedure}.
\newblock \bibinfo{journal}{\emph{IEEE Transactions on Pattern Analysis and
  Machine Intelligence}} \bibinfo{volume}{36}, \bibinfo{number}{6}
  (\bibinfo{year}{2014}), \bibinfo{pages}{1258--1267}.
\newblock


\bibitem[Lu et~al\mbox{.}(2019)]%
        {Lu19FHN}
\bibfield{author}{\bibinfo{person}{Zhi Lu}, \bibinfo{person}{Yang Hu},
  \bibinfo{person}{Yunchao Jiang}, \bibinfo{person}{Yan Chen}, {and}
  \bibinfo{person}{Bing Zeng}.} \bibinfo{year}{2019}\natexlab{}.
\newblock \showarticletitle{Learning Binary Code for Personalized Fashion
  Recommendation}. In \bibinfo{booktitle}{\emph{Proceedings of the CVPR}}.
  \bibinfo{pages}{10562--10570}.
\newblock


\bibitem[Lu et~al\mbox{.}(2023)]%
        {Lu23FHN}
\bibfield{author}{\bibinfo{person}{Zhi Lu}, \bibinfo{person}{Yang Hu},
  \bibinfo{person}{Cong Yu}, \bibinfo{person}{Yunchao Jiang},
  \bibinfo{person}{Yan Chen}, {and} \bibinfo{person}{Bing Zeng}.}
  \bibinfo{year}{2023}\natexlab{}.
\newblock \showarticletitle{Personalized Fashion Recommendation With Discrete
  Content-Based Tensor Factorization}.
\newblock \bibinfo{journal}{\emph{IEEE Transactions on Multimedia}}
  \bibinfo{volume}{25} (\bibinfo{year}{2023}), \bibinfo{pages}{5053--5064}.
\newblock


\bibitem[Luo et~al\mbox{.}(2025)]%
        {Luo25RgsAUC}
\bibfield{author}{\bibinfo{person}{Fangyuan Luo}, \bibinfo{person}{Yankai
  Chen}, \bibinfo{person}{Jun Wu}, {and} \bibinfo{person}{Yidong Li}.}
  \bibinfo{year}{2025}\natexlab{}.
\newblock \showarticletitle{Rank Gap Sensitive Deep {AUC} maximization for
  {CTR} prediction}.
\newblock \bibinfo{journal}{\emph{Pattern Recognition}}  \bibinfo{volume}{164}
  (\bibinfo{year}{2025}), \bibinfo{pages}{111496}.
\newblock


\bibitem[Luo et~al\mbox{.}(2021)]%
        {Luo21SDSR}
\bibfield{author}{\bibinfo{person}{Fangyuan Luo}, \bibinfo{person}{Jun Wu},
  {and} \bibinfo{person}{Haishuai Wang}.} \bibinfo{year}{2021}\natexlab{}.
\newblock \showarticletitle{Semi-Discrete Social Recommendation (Student
  Abstract)}. In \bibinfo{booktitle}{\emph{Proceedings of the AAAI}}.
  \bibinfo{pages}{15835--15836}.
\newblock


\bibitem[Luo et~al\mbox{.}(2022)]%
        {Luo22DLPR}
\bibfield{author}{\bibinfo{person}{Fangyuan Luo}, \bibinfo{person}{Jun Wu},
  {and} \bibinfo{person}{Tao Wang}.} \bibinfo{year}{2022}\natexlab{}.
\newblock \showarticletitle{Discrete Listwise Personalized Ranking for Fast
  Top-N Recommendation with Implicit Feedback}. In
  \bibinfo{booktitle}{\emph{Proceedings of the IJCAI}}.
  \bibinfo{pages}{2159--2165}.
\newblock


\bibitem[Luo et~al\mbox{.}(2024)]%
        {Luo24DLFM}
\bibfield{author}{\bibinfo{person}{Fangyuan Luo}, \bibinfo{person}{Jun Wu},
  {and} \bibinfo{person}{Tao Wang}.} \bibinfo{year}{2024}\natexlab{}.
\newblock \showarticletitle{Discrete Listwise Content-aware Recommendation}.
\newblock \bibinfo{journal}{\emph{ACM Transactions on Knowledge Discovery from
  Data}} \bibinfo{volume}{18}, \bibinfo{number}{1} (\bibinfo{year}{2024}),
  \bibinfo{pages}{7:1--7:20}.
\newblock


\bibitem[Luo et~al\mbox{.}(2023)]%
        {Luo23DeepHash}
\bibfield{author}{\bibinfo{person}{Xiao Luo}, \bibinfo{person}{Haixin Wang},
  \bibinfo{person}{Daqing Wu}, \bibinfo{person}{Chong Chen},
  \bibinfo{person}{Minghua Deng}, \bibinfo{person}{Jianqiang Huang}, {and}
  \bibinfo{person}{Xian{-}Sheng Hua}.} \bibinfo{year}{2023}\natexlab{}.
\newblock \showarticletitle{A Survey on Deep Hashing Methods}.
\newblock \bibinfo{journal}{\emph{ACM Transactions on Knowledge Discovery from
  Data}} \bibinfo{volume}{17}, \bibinfo{number}{1} (\bibinfo{year}{2023}),
  \bibinfo{pages}{15:1--15:50}.
\newblock


\bibitem[Lv et~al\mbox{.}(2007)]%
        {Lv07LSH8}
\bibfield{author}{\bibinfo{person}{Qin Lv}, \bibinfo{person}{William
  Josephson}, \bibinfo{person}{Zhe Wang}, \bibinfo{person}{Moses Charikar},
  {and} \bibinfo{person}{Kai Li}.} \bibinfo{year}{2007}\natexlab{}.
\newblock \showarticletitle{Multi-Probe {LSH:} Efficient Indexing for
  High-Dimensional Similarity Search}. In \bibinfo{booktitle}{\emph{Proceedings
  of the VLDB}}. \bibinfo{pages}{950--961}.
\newblock


\bibitem[Ma et~al\mbox{.}(2024)]%
        {Ma24NegativeSampling}
\bibfield{author}{\bibinfo{person}{Haokai Ma}, \bibinfo{person}{Ruobing Xie},
  \bibinfo{person}{Lei Meng}, \bibinfo{person}{Fuli Feng},
  \bibinfo{person}{Xiaoyu Du}, \bibinfo{person}{Xingwu Sun},
  \bibinfo{person}{Zhanhui Kang}, {and} \bibinfo{person}{Xiangxu Meng}.}
  \bibinfo{year}{2024}\natexlab{}.
\newblock \showarticletitle{Negative Sampling in Recommendation: {A} Survey and
  Future Directions}.
\newblock \bibinfo{journal}{\emph{CoRR}}  \bibinfo{volume}{abs/2409.07237}
  (\bibinfo{year}{2024}).
\newblock


\bibitem[Mao et~al\mbox{.}(2021)]%
        {Mao21SimpleX}
\bibfield{author}{\bibinfo{person}{Kelong Mao}, \bibinfo{person}{Jieming Zhu},
  \bibinfo{person}{Jinpeng Wang}, \bibinfo{person}{Quanyu Dai},
  \bibinfo{person}{Zhenhua Dong}, \bibinfo{person}{Xi Xiao}, {and}
  \bibinfo{person}{Xiuqiang He}.} \bibinfo{year}{2021}\natexlab{}.
\newblock \showarticletitle{SimpleX: {A} Simple and Strong Baseline for
  Collaborative Filtering}. In \bibinfo{booktitle}{\emph{Proceedings of the
  CIKM}}. \bibinfo{pages}{1243--1252}.
\newblock


\bibitem[Marguerite~Frank(1956)]%
        {FW1956}
\bibfield{author}{\bibinfo{person}{Philip~Wolfe Marguerite~Frank}.}
  \bibinfo{year}{1956}\natexlab{}.
\newblock \showarticletitle{An Algorithm for Quadratic Programming}.
\newblock \bibinfo{journal}{\emph{Naval Research Logistics Quarterly}}
  \bibinfo{volume}{3} (\bibinfo{year}{1956}), \bibinfo{pages}{95--110}.
\newblock


\bibitem[Mehrotra et~al\mbox{.}(2020)]%
        {Mehrotra20MultiObjective}
\bibfield{author}{\bibinfo{person}{Rishabh Mehrotra}, \bibinfo{person}{Niannan
  Xue}, {and} \bibinfo{person}{Mounia Lalmas}.}
  \bibinfo{year}{2020}\natexlab{}.
\newblock \showarticletitle{Bandit based Optimization of Multiple Objectives on
  a Music Streaming Platform}. In \bibinfo{booktitle}{\emph{Proceedings of the
  SIGKDD}}. \bibinfo{pages}{3224--3233}.
\newblock


\bibitem[Minar and Naher(2018)]%
        {Matiur18Deep_Survey}
\bibfield{author}{\bibinfo{person}{Matiur~Rahman Minar} {and}
  \bibinfo{person}{Jibon Naher}.} \bibinfo{year}{2018}\natexlab{}.
\newblock \showarticletitle{Recent Advances in Deep Learning: An Overview}.
\newblock \bibinfo{journal}{\emph{CoRR}}  \bibinfo{volume}{abs/1807.08169}
  (\bibinfo{year}{2018}).
\newblock


\bibitem[Motwani et~al\mbox{.}(2007)]%
        {Motwani07LSH5}
\bibfield{author}{\bibinfo{person}{Rajeev Motwani}, \bibinfo{person}{Assaf
  Naor}, {and} \bibinfo{person}{Rina Panigrahy}.}
  \bibinfo{year}{2007}\natexlab{}.
\newblock \showarticletitle{Lower Bounds on Locality Sensitive Hashing}.
\newblock \bibinfo{journal}{\emph{SIAM Journal on Discrete Mathematics}}
  \bibinfo{volume}{21}, \bibinfo{number}{4} (\bibinfo{year}{2007}),
  \bibinfo{pages}{930--935}.
\newblock


\bibitem[Murty(2007)]%
        {Murty07ALM}
\bibfield{author}{\bibinfo{person}{Katta~G. Murty}.}
  \bibinfo{year}{2007}\natexlab{}.
\newblock \showarticletitle{Nonlinear Programming Theory and Algorithms}.
\newblock \bibinfo{journal}{\emph{Technometrics}} \bibinfo{volume}{49},
  \bibinfo{number}{1} (\bibinfo{year}{2007}), \bibinfo{pages}{105}.
\newblock


\bibitem[O'Donnell et~al\mbox{.}(2014)]%
        {ODonnell14LSH6}
\bibfield{author}{\bibinfo{person}{Ryan O'Donnell}, \bibinfo{person}{Yi Wu},
  {and} \bibinfo{person}{Yuan Zhou}.} \bibinfo{year}{2014}\natexlab{}.
\newblock \showarticletitle{Optimal Lower Bounds for Locality-Sensitive Hashing
  (Except When q is Tiny)}.
\newblock \bibinfo{journal}{\emph{ACM Transactions on Computation Theory}}
  \bibinfo{volume}{6}, \bibinfo{number}{1} (\bibinfo{year}{2014}),
  \bibinfo{pages}{5:1--5:13}.
\newblock


\bibitem[Paun(2020)]%
        {Paun20RS_Efficiency}
\bibfield{author}{\bibinfo{person}{Iulia Paun}.}
  \bibinfo{year}{2020}\natexlab{}.
\newblock \showarticletitle{Efficiency-Effectiveness Trade-offs in
  Recommendation Systems}. In \bibinfo{booktitle}{\emph{Proceedings of the
  RecSys}}, \bibfield{editor}{\bibinfo{person}{Rodrygo L.~T. Santos},
  \bibinfo{person}{Leandro~Balby Marinho}, \bibinfo{person}{Elizabeth~M. Daly},
  \bibinfo{person}{Li~Chen}, \bibinfo{person}{Kim Falk}, \bibinfo{person}{Noam
  Koenigstein}, {and} \bibinfo{person}{Edleno~Silva de~Moura}} (Eds.).
  \bibinfo{pages}{770--775}.
\newblock


\bibitem[Qin et~al\mbox{.}(2020)]%
        {Qin20Tanh}
\bibfield{author}{\bibinfo{person}{Haotong Qin}, \bibinfo{person}{Ruihao Gong},
  \bibinfo{person}{Xianglong Liu}, \bibinfo{person}{Mingzhu Shen},
  \bibinfo{person}{Ziran Wei}, \bibinfo{person}{Fengwei Yu}, {and}
  \bibinfo{person}{Jingkuan Song}.} \bibinfo{year}{2020}\natexlab{}.
\newblock \showarticletitle{Forward and Backward Information Retention for
  Accurate Binary Neural Networks}. In \bibinfo{booktitle}{\emph{Proceedings of
  the CVPR}}. \bibinfo{pages}{2247--2256}.
\newblock


\bibitem[Rendle et~al\mbox{.}(2009)]%
        {Rendle09BPR}
\bibfield{author}{\bibinfo{person}{Steffen Rendle}, \bibinfo{person}{Christoph
  Freudenthaler}, \bibinfo{person}{Zeno Gantner}, {and} \bibinfo{person}{Lars
  Schmidt{-}Thieme}.} \bibinfo{year}{2009}\natexlab{}.
\newblock \showarticletitle{{BPR:} Bayesian Personalized Ranking from Implicit
  Feedback}. In \bibinfo{booktitle}{\emph{Proceedings of the UAI}}.
  \bibinfo{pages}{452--461}.
\newblock


\bibitem[Schnabel et~al\mbox{.}(2016)]%
        {Tobias16IPS}
\bibfield{author}{\bibinfo{person}{Tobias Schnabel}, \bibinfo{person}{Adith
  Swaminathan}, \bibinfo{person}{Ashudeep Singh}, \bibinfo{person}{Navin
  Chandak}, {and} \bibinfo{person}{Thorsten Joachims}.}
  \bibinfo{year}{2016}\natexlab{}.
\newblock \showarticletitle{Recommendations as Treatments: Debiasing Learning
  and Evaluation}. In \bibinfo{booktitle}{\emph{Proceedings of the ICML}},
  Vol.~\bibinfo{volume}{48}. \bibinfo{pages}{1670--1679}.
\newblock


\bibitem[Shen et~al\mbox{.}(2018)]%
        {Shen18STE}
\bibfield{author}{\bibinfo{person}{Dinghan Shen}, \bibinfo{person}{Qinliang
  Su}, \bibinfo{person}{Paidamoyo Chapfuwa}, \bibinfo{person}{Wenlin Wang},
  \bibinfo{person}{Guoyin Wang}, \bibinfo{person}{Ricardo Henao}, {and}
  \bibinfo{person}{Lawrence Carin}.} \bibinfo{year}{2018}\natexlab{}.
\newblock \showarticletitle{{NASH:} Toward End-to-End Neural Architecture for
  Generative Semantic Hashing}. In \bibinfo{booktitle}{\emph{Proceedings of the
  ACL}}. \bibinfo{pages}{2041--2050}.
\newblock


\bibitem[Shen et~al\mbox{.}(2015)]%
        {Shen2015Supervised}
\bibfield{author}{\bibinfo{person}{Fumin Shen}, \bibinfo{person}{Chunhua Shen},
  \bibinfo{person}{Wei Liu}, {and} \bibinfo{person}{Heng~Tao Shen}.}
  \bibinfo{year}{2015}\natexlab{}.
\newblock \showarticletitle{Supervised Discrete Hashing}. In
  \bibinfo{booktitle}{\emph{Proceedings of the CVPR}}. \bibinfo{pages}{37--45}.
\newblock


\bibitem[Shrivastava and Li(2014)]%
        {Shrivastava14LSH14}
\bibfield{author}{\bibinfo{person}{Anshumali Shrivastava} {and}
  \bibinfo{person}{Ping Li}.} \bibinfo{year}{2014}\natexlab{}.
\newblock \showarticletitle{Densifying One Permutation Hashing via Rotation for
  Fast Near Neighbor Search}. In \bibinfo{booktitle}{\emph{Proceedings of the
  ICML}}, Vol.~\bibinfo{volume}{32}. \bibinfo{pages}{557--565}.
\newblock


\bibitem[Sileo et~al\mbox{.}(2022)]%
        {Sileo22ZeroShotRS}
\bibfield{author}{\bibinfo{person}{Damien Sileo}, \bibinfo{person}{Wout
  Vossen}, {and} \bibinfo{person}{Robbe Raymaekers}.}
  \bibinfo{year}{2022}\natexlab{}.
\newblock \showarticletitle{Zero-Shot Recommendation as Language Modeling}. In
  \bibinfo{booktitle}{\emph{Proceedings of the ECIR}},
  Vol.~\bibinfo{volume}{13186}. \bibinfo{pages}{223--230}.
\newblock


\bibitem[Singh and Gupta(2022)]%
        {Singh22DeepHash}
\bibfield{author}{\bibinfo{person}{Avantika Singh} {and}
  \bibinfo{person}{Shaifu Gupta}.} \bibinfo{year}{2022}\natexlab{}.
\newblock \showarticletitle{Learning to hash: a comprehensive survey of deep
  learning-based hashing methods}.
\newblock \bibinfo{journal}{\emph{Knowledge and Information Systems}}
  \bibinfo{volume}{64}, \bibinfo{number}{10} (\bibinfo{year}{2022}),
  \bibinfo{pages}{2565--2597}.
\newblock


\bibitem[Tan et~al\mbox{.}(2020)]%
        {Tan20HashGNN}
\bibfield{author}{\bibinfo{person}{Qiaoyu Tan}, \bibinfo{person}{Ninghao Liu},
  \bibinfo{person}{Xing Zhao}, \bibinfo{person}{Hongxia Yang},
  \bibinfo{person}{Jingren Zhou}, {and} \bibinfo{person}{Xia Hu}.}
  \bibinfo{year}{2020}\natexlab{}.
\newblock \showarticletitle{Learning to Hash with Graph Neural Networks for
  Recommender Systems}. In \bibinfo{booktitle}{\emph{Proceedings of the WWW}}.
  \bibinfo{pages}{1988--1998}.
\newblock


\bibitem[van~den Oord et~al\mbox{.}(2017)]%
        {Oord17STE}
\bibfield{author}{\bibinfo{person}{A{\"{a}}ron van~den Oord},
  \bibinfo{person}{Oriol Vinyals}, {and} \bibinfo{person}{Koray Kavukcuoglu}.}
  \bibinfo{year}{2017}\natexlab{}.
\newblock \showarticletitle{Neural Discrete Representation Learning}. In
  \bibinfo{booktitle}{\emph{Proceedings of the NIPS}}.
  \bibinfo{pages}{6306--6315}.
\newblock


\bibitem[Wand and Jones(1995)]%
        {KernelSmoothing}
\bibfield{author}{\bibinfo{person}{Matt~P. Wand} {and}
  \bibinfo{person}{M.~Chris Jones}.} \bibinfo{year}{1995}\natexlab{}.
\newblock \bibinfo{booktitle}{\emph{Kernel Smoothing}}.
\newblock \bibinfo{publisher}{Springer}.
\newblock


\bibitem[Wang et~al\mbox{.}(2022)]%
        {Wang22HCFRec}
\bibfield{author}{\bibinfo{person}{Fan Wang}, \bibinfo{person}{Weiming Liu},
  \bibinfo{person}{Chaochao Chen}, \bibinfo{person}{Mengying Zhu}, {and}
  \bibinfo{person}{Xiaolin Zheng}.} \bibinfo{year}{2022}\natexlab{}.
\newblock \showarticletitle{HCFRec: Hash Collaborative Filtering via Normalized
  Flow with Structural Consensus for Efficient Recommendation}. In
  \bibinfo{booktitle}{\emph{Proceedings of the IJCAI}}.
  \bibinfo{pages}{2270--2276}.
\newblock


\bibitem[Wang et~al\mbox{.}(2019b)]%
        {Wang19DGCNBinCF}
\bibfield{author}{\bibinfo{person}{Haoyu Wang}, \bibinfo{person}{Defu Lian},
  {and} \bibinfo{person}{Yong Ge}.} \bibinfo{year}{2019}\natexlab{b}.
\newblock \showarticletitle{Binarized collaborative filtering with distilling
  graph convolutional networks}. In \bibinfo{booktitle}{\emph{Proceedings of
  the IJCAI}}. \bibinfo{pages}{4802–4808}.
\newblock


\bibitem[Wang et~al\mbox{.}(2019c)]%
        {Wang19ABinCF}
\bibfield{author}{\bibinfo{person}{Haoyu Wang}, \bibinfo{person}{Nan Shao},
  {and} \bibinfo{person}{Defu Lian}.} \bibinfo{year}{2019}\natexlab{c}.
\newblock \showarticletitle{Adversarial Binary Collaborative Filtering for
  Implicit Feedback}. In \bibinfo{booktitle}{\emph{Proceedings of the AAAI}}.
  \bibinfo{pages}{5248--5255}.
\newblock


\bibitem[Wang et~al\mbox{.}(2012)]%
        {Wang12SemiHash}
\bibfield{author}{\bibinfo{person}{Jun Wang}, \bibinfo{person}{Sanjiv Kumar},
  {and} \bibinfo{person}{Shih{-}Fu Chang}.} \bibinfo{year}{2012}\natexlab{}.
\newblock \showarticletitle{Semi-Supervised Hashing for Large-Scale Search}.
\newblock \bibinfo{journal}{\emph{IEEE Transactions on Pattern Analysis and
  Machine Intelligence}} \bibinfo{volume}{34}, \bibinfo{number}{12}
  (\bibinfo{year}{2012}), \bibinfo{pages}{2393--2406}.
\newblock


\bibitem[Wang et~al\mbox{.}(2016)]%
        {Wang16L2H}
\bibfield{author}{\bibinfo{person}{Jun Wang}, \bibinfo{person}{Wei Liu},
  \bibinfo{person}{Sanjiv Kumar}, {and} \bibinfo{person}{Shih{-}Fu Chang}.}
  \bibinfo{year}{2016}\natexlab{}.
\newblock \showarticletitle{Learning to Hash for Indexing Big Data - {A}
  Survey}.
\newblock \bibinfo{journal}{\emph{Proc. IEEE}} \bibinfo{volume}{104},
  \bibinfo{number}{1} (\bibinfo{year}{2016}), \bibinfo{pages}{34--57}.
\newblock


\bibitem[Wang et~al\mbox{.}(2018)]%
        {Wang18L2H}
\bibfield{author}{\bibinfo{person}{Jingdong Wang}, \bibinfo{person}{Ting
  Zhang}, \bibinfo{person}{Jingkuan Song}, \bibinfo{person}{Nicu Sebe}, {and}
  \bibinfo{person}{Heng~Tao Shen}.} \bibinfo{year}{2018}\natexlab{}.
\newblock \showarticletitle{A Survey on Learning to Hash}.
\newblock \bibinfo{journal}{\emph{IEEE Transactions on Pattern Analysis and
  Machine Intelligence}} \bibinfo{volume}{40}, \bibinfo{number}{4}
  (\bibinfo{year}{2018}), \bibinfo{pages}{769--790}.
\newblock


\bibitem[Wang et~al\mbox{.}(2019a)]%
        {Wang19NGCF}
\bibfield{author}{\bibinfo{person}{Xiang Wang}, \bibinfo{person}{Xiangnan He},
  \bibinfo{person}{Meng Wang}, \bibinfo{person}{Fuli Feng}, {and}
  \bibinfo{person}{Tat{-}Seng Chua}.} \bibinfo{year}{2019}\natexlab{a}.
\newblock \showarticletitle{Neural Graph Collaborative Filtering}. In
  \bibinfo{booktitle}{\emph{Proceedings of the SIGIR}}.
  \bibinfo{pages}{165--174}.
\newblock


\bibitem[Wang et~al\mbox{.}(2019d)]%
        {Wang19DR}
\bibfield{author}{\bibinfo{person}{Xiaojie Wang}, \bibinfo{person}{Rui Zhang},
  \bibinfo{person}{Yu Sun}, {and} \bibinfo{person}{Jianzhong Qi}.}
  \bibinfo{year}{2019}\natexlab{d}.
\newblock \showarticletitle{Doubly Robust Joint Learning for Recommendation on
  Data Missing No at Random}. In \bibinfo{booktitle}{\emph{Proceedings of the
  ICML}}, Vol.~\bibinfo{volume}{97}. \bibinfo{pages}{6638--6647}.
\newblock


\bibitem[Wu et~al\mbox{.}(2020)]%
        {Wu2020SDMF}
\bibfield{author}{\bibinfo{person}{Jun Wu}, \bibinfo{person}{Fangyuan Luo},
  \bibinfo{person}{Yujia Zhang}, {and} \bibinfo{person}{Haishuai Wang}.}
  \bibinfo{year}{2020}\natexlab{}.
\newblock \showarticletitle{Semi-discrete Matrix Factorization}.
\newblock \bibinfo{journal}{\emph{IEEE Intelligent Systems}}
  \bibinfo{volume}{35}, \bibinfo{number}{5} (\bibinfo{year}{2020}),
  \bibinfo{pages}{73--83}.
\newblock


\bibitem[Wu et~al\mbox{.}(2024)]%
        {Wu24SSM}
\bibfield{author}{\bibinfo{person}{Jiancan Wu}, \bibinfo{person}{Xiang Wang},
  \bibinfo{person}{Xingyu Gao}, \bibinfo{person}{Jiawei Chen},
  \bibinfo{person}{Hongcheng Fu}, {and} \bibinfo{person}{Tianyu Qiu}.}
  \bibinfo{year}{2024}\natexlab{}.
\newblock \showarticletitle{On the Effectiveness of Sampled Softmax Loss for
  Item Recommendation}.
\newblock \bibinfo{journal}{\emph{ACM Transactions on Information Systems}}
  \bibinfo{volume}{42}, \bibinfo{number}{4} (\bibinfo{year}{2024}),
  \bibinfo{pages}{98:1--98:26}.
\newblock


\bibitem[Wu et~al\mbox{.}(2018)]%
        {Wu18Sqlrank}
\bibfield{author}{\bibinfo{person}{Liwei Wu}, \bibinfo{person}{Cho{-}Jui
  Hsieh}, {and} \bibinfo{person}{James Sharpnack}.}
  \bibinfo{year}{2018}\natexlab{}.
\newblock \showarticletitle{SQL-Rank: {A} Listwise Approach to Collaborative
  Ranking}. In \bibinfo{booktitle}{\emph{Proceedings of the ICML}},
  Vol.~\bibinfo{volume}{80}. \bibinfo{pages}{5311--5320}.
\newblock


\bibitem[Wu et~al\mbox{.}(2023)]%
        {Wu23GNN_Survey}
\bibfield{author}{\bibinfo{person}{Shiwen Wu}, \bibinfo{person}{Fei Sun},
  \bibinfo{person}{Wentao Zhang}, \bibinfo{person}{Xu Xie}, {and}
  \bibinfo{person}{Bin Cui}.} \bibinfo{year}{2023}\natexlab{}.
\newblock \showarticletitle{Graph Neural Networks in Recommender Systems: {A}
  Survey}.
\newblock \bibinfo{journal}{\emph{Comput. Surveys}} \bibinfo{volume}{55},
  \bibinfo{number}{5} (\bibinfo{year}{2023}), \bibinfo{pages}{97:1--97:37}.
\newblock


\bibitem[Wu et~al\mbox{.}(2021)]%
        {Wu21GNN_Survey}
\bibfield{author}{\bibinfo{person}{Zonghan Wu}, \bibinfo{person}{Shirui Pan},
  \bibinfo{person}{Fengwen Chen}, \bibinfo{person}{Guodong Long},
  \bibinfo{person}{Chengqi Zhang}, {and} \bibinfo{person}{Philip~S. Yu}.}
  \bibinfo{year}{2021}\natexlab{}.
\newblock \showarticletitle{A Comprehensive Survey on Graph Neural Networks}.
\newblock \bibinfo{journal}{\emph{IEEE Transactions on Neural Networks and
  Learning Systems}} \bibinfo{volume}{32}, \bibinfo{number}{1}
  (\bibinfo{year}{2021}), \bibinfo{pages}{4--24}.
\newblock


\bibitem[Xu et~al\mbox{.}(2021)]%
        {Xu21Fourier}
\bibfield{author}{\bibinfo{person}{Yixing Xu}, \bibinfo{person}{Kai Han},
  \bibinfo{person}{Chang Xu}, \bibinfo{person}{Yehui Tang},
  \bibinfo{person}{Chunjing Xu}, {and} \bibinfo{person}{Yunhe Wang}.}
  \bibinfo{year}{2021}\natexlab{}.
\newblock \showarticletitle{Learning Frequency Domain Approximation for Binary
  Neural Networks}. In \bibinfo{booktitle}{\emph{Proceedings of the NIPS}}.
  \bibinfo{pages}{25553--25565}.
\newblock


\bibitem[Xu et~al\mbox{.}(2020)]%
        {Xu20MFDCF}
\bibfield{author}{\bibinfo{person}{Yang Xu}, \bibinfo{person}{Lei Zhu},
  \bibinfo{person}{Zhiyong Cheng}, \bibinfo{person}{Jingjing Li}, {and}
  \bibinfo{person}{Jiande Sun}.} \bibinfo{year}{2020}\natexlab{}.
\newblock \showarticletitle{Multi-Feature Discrete Collaborative Filtering for
  Fast Cold-Start Recommendation}. In \bibinfo{booktitle}{\emph{Proceedings of
  the AAAI}}. \bibinfo{pages}{270--278}.
\newblock


\bibitem[Xu et~al\mbox{.}(2023)]%
        {Xu23MDCF}
\bibfield{author}{\bibinfo{person}{Yang Xu}, \bibinfo{person}{Lei Zhu},
  \bibinfo{person}{Zhiyong Cheng}, \bibinfo{person}{Jingjing Li},
  \bibinfo{person}{Zheng Zhang}, {and} \bibinfo{person}{Huaxiang Zhang}.}
  \bibinfo{year}{2023}\natexlab{}.
\newblock \showarticletitle{Multi-Modal Discrete Collaborative Filtering for
  Efficient Cold-Start Recommendation}.
\newblock \bibinfo{journal}{\emph{IEEE Transactions on Knowledge and Data
  Engineering}} \bibinfo{volume}{35}, \bibinfo{number}{1}
  (\bibinfo{year}{2023}), \bibinfo{pages}{741--755}.
\newblock


\bibitem[Xu et~al\mbox{.}(2024)]%
        {Xu24TSGNH}
\bibfield{author}{\bibinfo{person}{Yang Xu}, \bibinfo{person}{Lei Zhu},
  \bibinfo{person}{Jingjing Li}, \bibinfo{person}{Fengling Li}, {and}
  \bibinfo{person}{Heng~Tao Shen}.} \bibinfo{year}{2024}\natexlab{}.
\newblock \showarticletitle{Temporal Social Graph Network Hashing for Efficient
  Recommendation}.
\newblock \bibinfo{journal}{\emph{IEEE Transactions on Knowledge and Data
  Engineering}} \bibinfo{volume}{36}, \bibinfo{number}{7}
  (\bibinfo{year}{2024}), \bibinfo{pages}{3541--3555}.
\newblock


\bibitem[Yang et~al\mbox{.}(2024)]%
        {Yang24DFMR}
\bibfield{author}{\bibinfo{person}{Enyue Yang}, \bibinfo{person}{Weike Pan},
  \bibinfo{person}{Qiang Yang}, {and} \bibinfo{person}{Zhong Ming}.}
  \bibinfo{year}{2024}\natexlab{}.
\newblock \showarticletitle{Discrete Federated Multi-behavior Recommendation
  for Privacy-Preserving Heterogeneous One-Class Collaborative Filtering}.
\newblock \bibinfo{journal}{\emph{ACM Transactions on Information Systems}}
  \bibinfo{volume}{42}, \bibinfo{number}{5} (\bibinfo{year}{2024}),
  \bibinfo{pages}{125:1--125:50}.
\newblock


\bibitem[Yang and Ying(2023)]%
        {Yang23AUC_Survey}
\bibfield{author}{\bibinfo{person}{Tianbao Yang} {and} \bibinfo{person}{Yiming
  Ying}.} \bibinfo{year}{2023}\natexlab{}.
\newblock \showarticletitle{{AUC} Maximization in the Era of Big Data and {AI:}
  {A} Survey}.
\newblock \bibinfo{journal}{\emph{Comput. Surveys}} \bibinfo{volume}{55},
  \bibinfo{number}{8} (\bibinfo{year}{2023}), \bibinfo{pages}{172:1--172:37}.
\newblock


\bibitem[Yi et~al\mbox{.}(2019)]%
        {Yi19sampleingbias}
\bibfield{author}{\bibinfo{person}{Xinyang Yi}, \bibinfo{person}{Ji Yang},
  \bibinfo{person}{Lichan Hong}, \bibinfo{person}{Derek~Zhiyuan Cheng},
  \bibinfo{person}{Lukasz Heldt}, \bibinfo{person}{Aditee Kumthekar},
  \bibinfo{person}{Zhe Zhao}, \bibinfo{person}{Li Wei}, {and}
  \bibinfo{person}{Ed~H. Chi}.} \bibinfo{year}{2019}\natexlab{}.
\newblock \showarticletitle{Sampling-bias-corrected neural modeling for large
  corpus item recommendations}. In \bibinfo{booktitle}{\emph{Proceedings of the
  RecSys}}. \bibinfo{pages}{269--277}.
\newblock


\bibitem[Yilmaz and Robertson(2010)]%
        {Yilmaz10}
\bibfield{author}{\bibinfo{person}{Emine Yilmaz} {and} \bibinfo{person}{Stephen
  Robertson}.} \bibinfo{year}{2010}\natexlab{}.
\newblock \showarticletitle{On the choice of effectiveness measures for
  learning to rank}.
\newblock \bibinfo{journal}{\emph{Information Retrieval}} \bibinfo{volume}{13},
  \bibinfo{number}{3} (\bibinfo{year}{2010}), \bibinfo{pages}{271--290}.
\newblock


\bibitem[Zannettou et~al\mbox{.}(2024)]%
        {Zannettou24Tiktok}
\bibfield{author}{\bibinfo{person}{Savvas Zannettou},
  \bibinfo{person}{Olivia~Nemes Nemeth}, \bibinfo{person}{Oshrat Ayalon},
  \bibinfo{person}{Angelica Goetzen}, \bibinfo{person}{Krishna~P. Gummadi},
  \bibinfo{person}{Elissa~M. Redmiles}, {and} \bibinfo{person}{Franziska
  Roesner}.} \bibinfo{year}{2024}\natexlab{}.
\newblock \showarticletitle{Analyzing User Engagement with TikTok's Short
  Format Video Recommendations using Data Donations}. In
  \bibinfo{booktitle}{\emph{Proceedings of the CHI}}.
  \bibinfo{pages}{731:1--731:16}.
\newblock


\bibitem[Zaslavskiy et~al\mbox{.}(2009)]%
        {ZaslavskiyBV09}
\bibfield{author}{\bibinfo{person}{Mikhail Zaslavskiy},
  \bibinfo{person}{Francis~R. Bach}, {and} \bibinfo{person}{Jean{-}Philippe
  Vert}.} \bibinfo{year}{2009}\natexlab{}.
\newblock \showarticletitle{A Path Following Algorithm for the Graph Matching
  Problem}.
\newblock \bibinfo{journal}{\emph{IEEE Transactions on Pattern Analysis and
  Machine Intelligence}} \bibinfo{volume}{31}, \bibinfo{number}{12}
  (\bibinfo{year}{2009}), \bibinfo{pages}{2227--2242}.
\newblock


\bibitem[Zhang et~al\mbox{.}(2023)]%
        {Zhang23LightFR}
\bibfield{author}{\bibinfo{person}{Honglei Zhang}, \bibinfo{person}{Fangyuan
  Luo}, \bibinfo{person}{Jun Wu}, \bibinfo{person}{Xiangnan He}, {and}
  \bibinfo{person}{Yidong Li}.} \bibinfo{year}{2023}\natexlab{}.
\newblock \showarticletitle{LightFR: Lightweight Federated Recommendation with
  Privacy-preserving Matrix Factorization}.
\newblock \bibinfo{journal}{\emph{ACM Transactions on Information Systems}}
  \bibinfo{volume}{41}, \bibinfo{number}{4} (\bibinfo{year}{2023}),
  \bibinfo{pages}{90:1--90:28}.
\newblock


\bibitem[Zhang et~al\mbox{.}(2016a)]%
        {Zhang2016DCF}
\bibfield{author}{\bibinfo{person}{Hanwang Zhang}, \bibinfo{person}{Fumin
  Shen}, \bibinfo{person}{Wei Liu}, \bibinfo{person}{Xiangnan He},
  \bibinfo{person}{Huanbo Luan}, {and} \bibinfo{person}{Tat~Seng Chua}.}
  \bibinfo{year}{2016}\natexlab{a}.
\newblock \showarticletitle{Discrete Collaborative Filtering}. In
  \bibinfo{booktitle}{\emph{Proceedings of the SIGIR}}.
  \bibinfo{pages}{325--334}.
\newblock


\bibitem[Zhang et~al\mbox{.}(2024)]%
        {zhang2024uncovering}
\bibfield{author}{\bibinfo{person}{Honglei Zhang}, \bibinfo{person}{Shuyi
  Wang}, \bibinfo{person}{Haoxuan Li}, \bibinfo{person}{Chunyuan Zheng},
  \bibinfo{person}{Xu Chen}, \bibinfo{person}{Li Liu},
  \bibinfo{person}{Shanshan Luo}, {and} \bibinfo{person}{Peng Wu}.}
  \bibinfo{year}{2024}\natexlab{}.
\newblock \showarticletitle{Uncovering the Propensity Identification Problem in
  Debiased Recommendations}. In \bibinfo{booktitle}{\emph{Proceedings of the
  ICDE}}. IEEE, \bibinfo{pages}{653--666}.
\newblock


\bibitem[Zhang et~al\mbox{.}(2019)]%
        {Zhang19RS_Survey}
\bibfield{author}{\bibinfo{person}{Shuai Zhang}, \bibinfo{person}{Lina Yao},
  \bibinfo{person}{Aixin Sun}, {and} \bibinfo{person}{Yi Tay}.}
  \bibinfo{year}{2019}\natexlab{}.
\newblock \showarticletitle{Deep Learning Based Recommender System: {A} Survey
  and New Perspectives}.
\newblock \bibinfo{journal}{\emph{Comput. Surveys}} \bibinfo{volume}{52},
  \bibinfo{number}{1} (\bibinfo{year}{2019}), \bibinfo{pages}{5:1--5:38}.
\newblock


\bibitem[Zhang et~al\mbox{.}(2017)]%
        {Zhang17DPR}
\bibfield{author}{\bibinfo{person}{Yan Zhang}, \bibinfo{person}{Defu Lian},
  {and} \bibinfo{person}{Guowu Yang}.} \bibinfo{year}{2017}\natexlab{}.
\newblock \showarticletitle{Discrete Personalized Ranking for Fast
  Collaborative Filtering from Implicit Feedback}. In
  \bibinfo{booktitle}{\emph{Proceedings of the AAAI}}.
  \bibinfo{pages}{1669--1675}.
\newblock


\bibitem[Zhang et~al\mbox{.}(2020)]%
        {Zhang20CGH}
\bibfield{author}{\bibinfo{person}{Yan Zhang}, \bibinfo{person}{Ivor~W. Tsang},
  {and} \bibinfo{person}{Lixin Duan}.} \bibinfo{year}{2020}\natexlab{}.
\newblock \showarticletitle{Collaborative Generative Hashing for Marketing and
  Fast Cold-Start Recommendation}.
\newblock \bibinfo{journal}{\emph{IEEE Intelligent Systems}}
  \bibinfo{volume}{35}, \bibinfo{number}{5} (\bibinfo{year}{2020}),
  \bibinfo{pages}{84--95}.
\newblock


\bibitem[Zhang et~al\mbox{.}(2022)]%
        {Zhang22DPH}
\bibfield{author}{\bibinfo{person}{Yan Zhang}, \bibinfo{person}{Ivor~W. Tsang},
  \bibinfo{person}{Hongzhi Yin}, \bibinfo{person}{Guowu Yang},
  \bibinfo{person}{Defu Lian}, {and} \bibinfo{person}{Jingjing Li}.}
  \bibinfo{year}{2022}\natexlab{}.
\newblock \showarticletitle{Deep Pairwise Hashing for Cold-Start
  Recommendation}.
\newblock \bibinfo{journal}{\emph{IEEE Transactions on Knowledge and Data
  Engineering}} \bibinfo{volume}{34}, \bibinfo{number}{7}
  (\bibinfo{year}{2022}), \bibinfo{pages}{3169--3181}.
\newblock


\bibitem[Zhang et~al\mbox{.}(2018a)]%
        {Zhang18DRMF}
\bibfield{author}{\bibinfo{person}{Yan Zhang}, \bibinfo{person}{Haoyu Wang},
  \bibinfo{person}{Defu Lian}, \bibinfo{person}{Ivor~W. Tsang},
  \bibinfo{person}{Hongzhi Yin}, {and} \bibinfo{person}{Guowu Yang}.}
  \bibinfo{year}{2018}\natexlab{a}.
\newblock \showarticletitle{Discrete Ranking-based Matrix Factorization with
  Self-Paced Learning}. In \bibinfo{booktitle}{\emph{Proceedings of the
  SIGKDD}}. \bibinfo{pages}{2758--2767}.
\newblock


\bibitem[Zhang et~al\mbox{.}(2016b)]%
        {Zhang16NQ}
\bibfield{author}{\bibinfo{person}{Yan Zhang}, \bibinfo{person}{Guowu Yang},
  \bibinfo{person}{Defu Lian}, \bibinfo{person}{Hong Wen}, {and}
  \bibinfo{person}{Jinsong Wu}.} \bibinfo{year}{2016}\natexlab{b}.
\newblock \showarticletitle{Constraint Free Preference Preserving Hashing for
  Fast Recommendation}. In \bibinfo{booktitle}{\emph{Proceedings of the
  GLOBECOM}}. \bibinfo{pages}{1--6}.
\newblock


\bibitem[Zhang et~al\mbox{.}(2018b)]%
        {Zhang2018DDL}
\bibfield{author}{\bibinfo{person}{Yan Zhang}, \bibinfo{person}{Hongzhi Yin},
  \bibinfo{person}{Zi Huang}, \bibinfo{person}{Xingzhong Du},
  \bibinfo{person}{Guowu Yang}, {and} \bibinfo{person}{Defu Lian}.}
  \bibinfo{year}{2018}\natexlab{b}.
\newblock \showarticletitle{Discrete Deep Learning for Fast Content-Aware
  Recommendation}. In \bibinfo{booktitle}{\emph{Proceedings of the ICDM}}.
  \bibinfo{pages}{717--726}.
\newblock


\bibitem[Zhang et~al\mbox{.}(2014)]%
        {Zhang2014PPH}
\bibfield{author}{\bibinfo{person}{Zhiwei Zhang}, \bibinfo{person}{Qifan Wang},
  \bibinfo{person}{Lingyun Ruan}, {and} \bibinfo{person}{Luo Si}.}
  \bibinfo{year}{2014}\natexlab{}.
\newblock \showarticletitle{Preference Preserving Hashing for Efficient
  Recommendation}. In \bibinfo{booktitle}{\emph{Proceedings of the SIGIR}}.
  \bibinfo{pages}{183--192}.
\newblock


\bibitem[Zhou et~al\mbox{.}(2020)]%
        {Zhou20GNN_Survey}
\bibfield{author}{\bibinfo{person}{Jie Zhou}, \bibinfo{person}{Ganqu Cui},
  \bibinfo{person}{Shengding Hu}, \bibinfo{person}{Zhengyan Zhang},
  \bibinfo{person}{Cheng Yang}, \bibinfo{person}{Zhiyuan Liu},
  \bibinfo{person}{Lifeng Wang}, \bibinfo{person}{Changcheng Li}, {and}
  \bibinfo{person}{Maosong Sun}.} \bibinfo{year}{2020}\natexlab{}.
\newblock \showarticletitle{Graph neural networks: {A} review of methods and
  applications}.
\newblock \bibinfo{journal}{\emph{{AI} Open}}  \bibinfo{volume}{1}
  (\bibinfo{year}{2020}), \bibinfo{pages}{57--81}.
\newblock


\bibitem[Zhou and Zha(2012)]%
        {Zhou2012BCCF}
\bibfield{author}{\bibinfo{person}{Ke Zhou} {and} \bibinfo{person}{Hongyuan
  Zha}.} \bibinfo{year}{2012}\natexlab{}.
\newblock \showarticletitle{Learning Binary Codes for Collaborative Filtering}.
  In \bibinfo{booktitle}{\emph{Proceedings of the SIGKDD}}.
  \bibinfo{pages}{498--506}.
\newblock


\bibitem[Zhu et~al\mbox{.}(2021)]%
        {Zhu21CTR}
\bibfield{author}{\bibinfo{person}{Jieming Zhu}, \bibinfo{person}{Jinyang Liu},
  \bibinfo{person}{Shuai Yang}, \bibinfo{person}{Qi Zhang}, {and}
  \bibinfo{person}{Xiuqiang He}.} \bibinfo{year}{2021}\natexlab{}.
\newblock \showarticletitle{Open Benchmarking for Click-Through Rate
  Prediction}. In \bibinfo{booktitle}{\emph{Proceedings of the CIKM}}.
  \bibinfo{pages}{2759--2769}.
\newblock


\bibitem[Zhu et~al\mbox{.}(2023b)]%
        {Zhu23EDCF}
\bibfield{author}{\bibinfo{person}{Lei Zhu}, \bibinfo{person}{Yang Xu},
  \bibinfo{person}{Jingjing Li}, \bibinfo{person}{Weili Guan}, {and}
  \bibinfo{person}{Zhiyong Cheng}.} \bibinfo{year}{2023}\natexlab{b}.
\newblock \showarticletitle{Explainable Discrete Collaborative Filtering}.
\newblock \bibinfo{journal}{\emph{IEEE Transactions on Knowledge and Data
  Engineering}} \bibinfo{volume}{35}, \bibinfo{number}{7}
  (\bibinfo{year}{2023}), \bibinfo{pages}{6901--6915}.
\newblock


\bibitem[Zhu et~al\mbox{.}(2023a)]%
        {Zhu23LLM_Compression}
\bibfield{author}{\bibinfo{person}{Xunyu Zhu}, \bibinfo{person}{Jian Li},
  \bibinfo{person}{Yong Liu}, \bibinfo{person}{Can Ma}, {and}
  \bibinfo{person}{Weiping Wang}.} \bibinfo{year}{2023}\natexlab{a}.
\newblock \showarticletitle{A Survey on Model Compression for Large Language
  Models}.
\newblock \bibinfo{journal}{\emph{CoRR}}  \bibinfo{volume}{abs/2308.07633}
  (\bibinfo{year}{2023}).
\newblock


\end{thebibliography}


\end{document}